\documentclass[sigconf]{acmart}

\usepackage{multirow}
\usepackage{color}
\usepackage{colortbl}

\usepackage{stfloats}

\usepackage{longfbox}

\usepackage{subcaption}
\usepackage{enumitem}
\usepackage{graphicx}
\usepackage{subcaption}
\usepackage{tabularx}

\usepackage{colortbl}
\usepackage{array}

\usepackage{tikz}

\definecolor{quotebackground}{HTML}{EFEFEF}
\definecolor{tableheader}{HTML}{EFEFEF}
\definecolor{tablegrayline}{HTML}{e0e0e0}

\newcommand{\eg}{\textit{e.g.}}

\newcommand{\cf}{\textit{c.f.}}
\newcommand{\etal}{\textit{et al.}}

\newcommand{\sourcecode}{\urlstyle{tt}\url{https://naver-ai.github.io/aacesstalk}}

\newcommand{\sysname}{\textsc{AACessTalk}}
\newcommand{\sysnamehyph}{\textsc{AACess-Talk}}

\newcommand{\labelphantom}[1]{%
  \parbox{0pt}{\phantomsubcaption\label{#1}}%
}

\newcommand{\revised}[1]{#1} 

\newcommand{\rerevised}[1]{\textcolor{blue}{#1}}
\renewcommand{\rerevised}[1]{#1} 

\newcommand{\circledigit}[1]{\textbf{\normalsize{\textsf{\textcircled{\footnotesize{#1}}}}}}

\newcommand{\ipstart}[1]{\vspace{1mm} \noindent{\textbf{\textit{#1.}}}}

\makeatletter
\newdimen\@tempdimd
\makeatother

\definecolor{aaccardcolor}{HTML}{f0f0f0}

\newfboxstyle{boxparam}{padding=1.5pt, padding-left=2pt, padding-right=2pt, height=7.5pt, background-color=aaccardcolor, border-radius=2pt, border-right-width=1pt, border-bottom-width=1pt, border-color=gray}

\newcommand{\aaccard}[1]{\lfbox[boxparam]{\textsf{#1}}}
\newcommand{\aaccardtable}[1]{\lfbox[boxparam, height=6pt, border-bottom-width=0.5pt, border-right-width=0.5pt]{\textsf{#1}}}

\newcommand{\rectwrapsmall}[2]{\lfbox[boxparam, border-radius=0pt, padding-left=2pt, padding-right=2pt, height=5.5pt, border-width=0pt, background-color=#1]{\sffamily{\textcolor{white}{#2}}}}

\newcommand{\blackrectsmall}[1]{\rectwrapsmall{darkgray}{#1}}

\usepackage{newfloat}
\DeclareFloatingEnvironment[fileext=fml,placement={H},name=Dialogue]{dialogue}
\copyrightyear{2025}
\acmYear{2025}
\setcopyright{rightsretained}
\acmConference[CHI '25]{CHI Conference on Human Factors in Computing Systems}{April 26-May 1, 2025}{Yokohama, Japan}
\acmBooktitle{CHI Conference on Human Factors in Computing Systems (CHI '25), April 26-May 1, 2025, Yokohama, Japan}\acmDOI{10.1145/3706598.3713792}
\acmISBN{979-8-4007-1394-1/25/04}

\acmSubmissionID{1243}
\begin{document}

\title[\sysname{}: AI-driven Communication Mediation System for Minimally Verbal Autistic Children and Parents]{\sysname{}: Fostering Communication between Minimally~Verbal Autistic Children and Parents with Contextual~Guidance and Card Recommendation}

\settopmatter{authorsperrow=3}

\author{Dasom Choi}
\orcid{0000-0002-3392-6243}
\authornote{Dasom Choi conducted this work as a research intern at NAVER AI Lab.}
\affiliation{%
  \institution{KAIST}
  \country{Republic of Korea}
}
\email{dasomchoi@kaist.ac.kr}

\author{SoHyun Park}
\orcid{0000-0001-8703-0584}
\affiliation{%
  \institution{NAVER Cloud}
  \country{Republic of Korea}}
\email{sohyun@snu.ac.kr}

\author{Kyungah Lee}
\affiliation{%
  \institution{Dodakim Child Development Center}
  \country{Republic of Korea}
}
\email{hiroo6900@hanmail.net}

\author{Hwajung Hong}
\orcid{0000-0001-5268-3331}
\affiliation{%
  \institution{KAIST}
  \country{Republic of Korea}
}
\email{hwajung@kaist.ac.kr}

\author{Young-Ho Kim}
\orcid{0000-0002-2681-2774}
\affiliation{%
  \institution{NAVER AI Lab}
  \country{Republic of Korea}
}
\email{yghokim@younghokim.net}

\begin{abstract}
As minimally verbal autistic (MVA) children communicate with parents through few words and nonverbal cues, parents often struggle to encourage their children to express subtle emotions and needs and to grasp their nuanced signals. We present \sysname{}, a tablet-based, AI-mediated communication system that facilitates meaningful exchanges between an MVA child and a parent. \sysnamehyph{} provides real-time guides to the parent to engage the child in conversation and, in turn, recommends contextual vocabulary cards to the child. Through a two-week deployment study with 11 MVA child-parent dyads, we examine how \sysname{} fosters everyday conversation practice and mutual engagement. Our findings show high engagement from all dyads, leading to increased frequency of conversation and turn-taking. \sysname{} also encouraged parents to explore their own interaction strategies and empowered the children to have more agency in communication. We discuss the implications of designing technologies for balanced communication dynamics in parent-MVA child interaction.



\end{abstract}



\begin{teaserfigure}
    \centering
    \includegraphics[width=\textwidth]{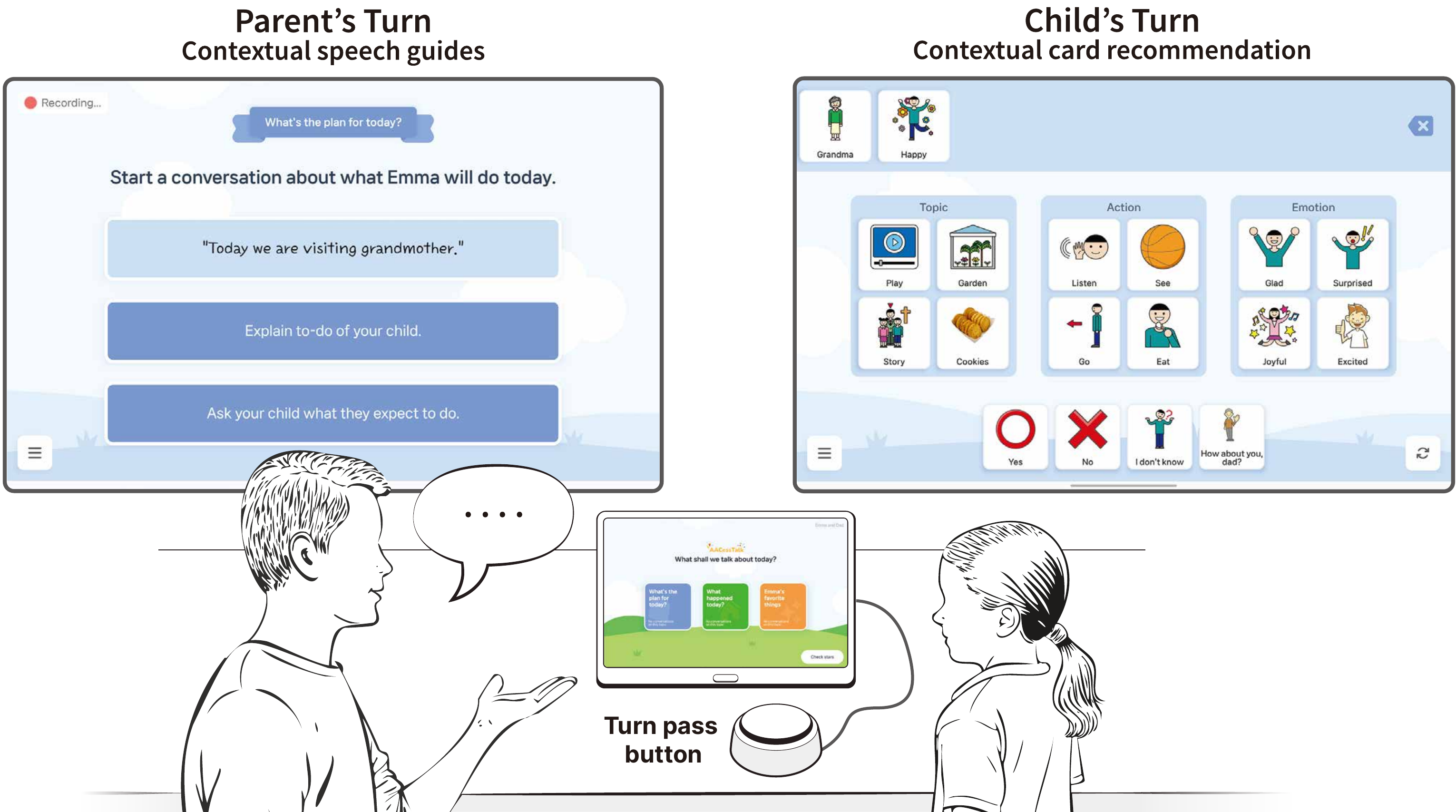}
    \caption[An illustration depicting the use of AACessTalk by a parent and a minimally verbal autistic child (referred to as MVA children). AACessTalk is a communication mediation system that operates on a tablet device complemented by a hardware button. On the left, the Parent’s Turn interface displays contextual speech guides with sample dialogues such as, 'Today we are visiting grandmother,' alongside prompts designed to engage the child. On the right, the Child’s Turn interface features a contextual vocabulary card recommendation system with categories including Emotion, Action, and Topic, represented by symbols like 'Happy', 'Listen', and 'Garden'.]{\sysname{} is a communication mediation system that runs on a tablet device accompanied by a hardware button (Bottom). \sysname{} shapes a turn-taking conversation, and the button is used to switch turns. On the parent's turn, the system displays guide messages and example utterances that the parent can refer to when responding to the child (Left). On the child's turn, the system recommends a set of vocabulary cards that the child can select to convey their message (Right).}
    \label{fig:interface:teaser}
\end{teaserfigure}

\maketitle

\section{Introduction}
Minimally verbal autistic (MVA)\footnote{In this work, we choose to use identity-first language (\eg{}, autistic children) rather than person-first language (\eg{}, child with autism). This decision aligns with the preferences of autistic individuals~\cite{Lorcan2016} and recent academic trends~\cite{Brian2002}.} children engage in communication. mainly with nonverbal cues, non-speech vocalizations, or a small repertoire of words or fixed phrases~\cite{Kasari2013}. Regardless of the form, communication with parents is just as essential for autistic children's emotional and social development as it is for non-autistic children~\cite{Bernard2000}. However, parents often shoulder the significant responsibility of driving this communication---initiating conversations by asking questions, interpreting nuanced communicative signals, and constantly responding to encourage the child's participation~\cite{Damiao2023, McCauley2022}. As parents manage these interactions with little cooperation from their children, they experience feelings of isolation and frustration~\cite{del2018thorn}.

To encourage the child's participation in communication, many parents have adopted Augmentative and Alternative Communication (AAC) systems, such as symbol cards, that children can use instead of verbalizing their message~\cite{beukelman2020}. While these tools enable MVA children to express themselves by selecting visualized symbols of words, they are often criticized for failing to capture the child’s authentic thoughts and feelings as the symbols tend to prioritize functional communication over the user's personal context and true intent ~\cite{Cynthia2018, Janna2022}. Most AAC systems are pre-programmed with ``necessary words'' chosen by service providers and, in most cases, by parents~\cite{assistiveware, Neimy2022}. As a result, children's messages are often limited to the boundaries set by parents rather than reflecting children's unique experiences and perspectives. 

Previous studies have examined how MVA children can take a more active role in communication with their parents. Parent training, often conducted with therapists in face-to-face settings~\cite{Elaine2013, Gerald2003}, remotely~\cite{Baharav2010, suess2014}, or self-guided by video materials~\cite{DAI201836, hong2018}, have equipped parents with strategies to encourage their children to communicate~\cite{INGERSOLL2007163, Oono2013}. Research in special education and HCI also proposed conversational storyboards based on behavioral data~\cite{black2010, TINTAREV20161}, or customizable AAC systems that generate vocabularies based on geographical data~\cite{Epp2012, patel2007} or photographs~\cite{Fontana2022, Vargas2024}. However, previous attempts have primarily considered education-oriented language-learning contexts that target either parents or children, lacking an integrated approach to empower both parties in everyday communication on the fly.

In this work, we aim to support the agency of both parents and MVA children in reciprocal communication by designing and developing an intelligent system that mediates their conversational interaction\footnote{\revised{In this work, we use the term \textit{communication} to denote the holistic interaction between parents and MVA children, embracing not only exchanges of words and utterances but also emotional connections and behaviors~\cite{watzlawick2011, knapp1978}. Also, we use the term \textit{conversation} to denote a specific subset of communication characterized by a structured and reciprocal exchange of messages between two individuals~\cite{sacks1974}.}}. To that end, we conducted formative interviews with nine autism experts and five parents of MVA children to better understand communication challenges between parents and MVA children and glean insights from professional practices to enhance communication. The interviews revealed that parents exert significant control over the structure and topics of conversation, underscoring the need to support turn-taking between parents and children around mutually engaging subjects.

Based on these insights, we designed and developed \sysname{} (\autoref{fig:interface:teaser}), a communication mediation system that fosters meaningful exchanges of ideas and emotions through mutual contribution between MVA children and their parents. \sysname{} runs on a tablet accompanied by a hardware button (\autoref{fig:interface:teaser}, bottom) for taking turns, providing an environment for explicit turn-taking conversations. On the parent's turn, the app provides guide messages on what and how to respond to the child (\autoref{fig:interface:teaser}, left). On the child's turn, the app curates a set of vocabulary cards relevant to the conversational context (\autoref{fig:interface:teaser}, right).
To accommodate a range of in-situ, daily topics, \sysname{} leverages large language models (LLMs) to generate parental guidance and card curation, based on the parent's voice input transcribed in text and the child's input in selected cards.

To examine how parents and MVA children interact with \sysnamehyph{} and how the system impacts their communication, we conducted a two-week home deployment study with 11 parent-child dyads in South Korea. Our results show a considerable level of participant engagement in \revised{conversation} through \sysname{}, carrying out 232 conversation sessions in total. During these interactions, parents incorporated \sysname{}'s parental guidance in 78\% of their turns, and children selected a total of 2,244 AAC cards recommended by the system. The debriefing interviews and daily surveys revealed that \sysname{} alleviated parental pressure to have MVA children produce complete sentences and provided the children chances to express their communicative intent. Furthermore, using \sysname{} motivated parents to depart from brief and instructive dialogue, turning conversations with their MVA children into daily routines focused on empathy and understanding. The contribution of this work is fourfold:
\begin{enumerate}[leftmargin=*, itemsep=4pt, topsep=0pt]
\item Findings from a formative study with nine autism experts and five parents of MVA children, revealing the unique challenges in communication between MVA children and their parents, along with the current best practices in the field.
\item The design and implementation of \sysname{}, an AI-driven communication mediation system that supports turn-taking conversations, offering parental guidance to parents and recommending AAC cards to the child, encouraging mutual participation in parent-child communication. The source code of \sysname{} is publicly available at \sourcecode{}.
\item Empirical findings from a two-week deployment study involving 11 parent-MVA child dyads, demonstrating how parents and children leveraged \sysname{} for communication and how this experience influenced parents' perceptions towards communicating with MVA children.
\item Design considerations for communication technologies that foster parental reflection on interactions with MVA children, thereby promoting the inclusion of neurodiverse children as active conversational partners.
\end{enumerate}
\section{Related Work}
In this section, we cover the related work in the areas of (1) technology support for communication with neurodiverse children, (2) parenting technology support, and (3) AAC for MVA children.

\subsection{Technology Support for Communication with Neurodiverse Children}
In parent-child communication, a child’s self-expression is fundamental in establishing a shared understanding, which allows parents to address the child’s unique needs and challenges effectively~\cite{lin2023}. However, forming this shared understanding is often challenging within neurodiverse dyads—comprising MVA children and non-autistic parents~\cite{Zolyomi2023}. This is because, first, MVA children often have lower intrinsic motivation to communicate~\cite{schuler1997}, and thus reveal less of their inner selves to their parents. Second, they tend to express their needs and opinions through various modalities beyond verbal language, such as touch, sound, pointing, and gestures~\cite{Verónica2020}. Even their nonverbal vocalizations can hold emotional and self-expressive information~\cite{johnson2023}, yet parents often struggle to interpret these subtle communication cues~\cite{Damiao2023}. While communication difficulties in these neurodiverse dyads stem from differences in how they perceive and express the world, many educational and technological efforts have focused on teaching and supporting verbal communication to MVA children~\cite{Rachel2022, kientz2014}. \revised{For example, systems that visualize audio in real-time to promote vocal imitation~\cite{Hailpern2010, Hailpern2012, Vera1999} and virtual tutors to train spoken word production in autistic children~\cite{Dominic2006} have been developed and assessed for use in clinical speech-therapy settings.}

The field of HCI has contributed numerous assistive communication technologies aimed at empowering MVA children to have a ``voice.'' Importantly, these recent efforts redefine having a ``voice'' not just as the ability to speak, but as enabling MVA children to represent themselves by having a communicative agency~\cite{alper2017}. This paradigm shift also aligns with the Ability-Based Design approach~\cite{Wobbrock2011, Wobbrock2017}, which focuses on individuals' abilities rather than limitations, encouraging MVA children to engage in communication on their own terms. For example, Wilson \etal{} developed a personalized interactive dictionary that taps into MVA children's special interests to motivate self-expression~\cite{Wilson2018}. Researchers have also proposed tangible devices that enable MVA children to express themselves through sound, light, touch, and body movement~\cite{Singh2024, Wilson2020}, AAC tools with symbolic images or photos~\cite{Fontana2022, Vargas2024}, and \revised{VR/AR letterboards~\cite{Nazari2024, Alabood2024, Alabood2022}}. Some have even involved the children as design partners in creating these technologies~\cite{Wilson2019, Vargas2024}. 

While previous studies have facilitated more independent self-expression in MVA children, there has been relatively less focus on how parents engage in communication with MVA children. Communication inherently involves \textit{interdependent} actions that require mutual effort from all parties. Bennett \etal{}~\cite{Bennett2018} applied the concept of \textit{interdependence} from Disability Studies to the design of assistive technology (AT), proposing that AT needs to support collaborative efforts among people with disabilities, those they interact with, and their surrounding environments. Through the lens of interdependence, current communication technologies for MVA children-parent dyads predominantly focus on facilitating access for MVA children. This unilateral targeting can inadvertently overlook neurotypical parents' accessibility to engage in conversations with their neurodiverse child, potentially leading to frustration and feelings of alienation~\cite{del2018thorn}. Thus, we see a critical opportunity in designing AT that assist communication by accommodating the needs of both MVA child and the parent. Specifically, our work integrates previously separated communication supports for parents and MVA children into a unified system to mediate real-time interactions among them.

\subsection{Parenting Technology Support}
There has been a growing emphasis on family-centered support for autistic children, highlighting the need for parents to actively engage in setting goals, making decisions, and implementing interventions at home to improve interactions with their children~\cite{bindlish2018}. This led parents to learn new modes of communication, adapt their interaction styles, and modify their home environments~\cite{alli2015, american2006}. Moreover, for the successful integration of those changes and interventions, including AAC, parents are required to have specialized training to learn effective strategies~\cite{Michael2013, Matthew2014, Gadberry2011, Erinn2009, Jennifer2013, Debora2007}. Despite these efforts, the visibility of improvements or responses from autistic children to these interventions can vary significantly~\cite{Paynter2018}, making it challenging for parents to reflect on and adjust their parenting and communication strategies.

The HCI researchers have explored real-time interventions to assist parents in interacting with their children. Several studies aim to facilitate parental reflection by providing direct access to interaction data. For example, the Dyadic Mirror~\cite{Wonjung2020}, a wearable mirror worn around a child’s neck, shows parents live views of their own face, increasing their awareness of emotional states. SpecialTime~\cite{Huber2019} utilizes a predefined set of guidelines derived from Parent–Child Interaction Therapy (PCIT) to track parent interactions and provide auto-labeled feedback after conversation. Another tool, TalkLime~\cite{Song2016}, employs real-time visualizations to analyze the number of utterances, initiation ratios, and turn-taking between parents and children. 

Building on advances in multimodal sensors and AI automation, recent studies offer more direct feedback and contextual guidance by analyzing conversation dynamics between parents and children. TalkBetter~\cite{Hwang2014} monitors conversations between parents and language-delayed children, triggering alerts when parents show undesirable language habits, such as interrupting, speaking too quickly, or not waiting. Captivate!~\cite{Kwon2022} focuses on play sessions with language-delayed children, identifying objects of joint focus and displaying related phrases on a tablet. TIPS~\cite{Hossain2022} offers context-responsive recommendations of American Sign Language (ASL) for hearing parents of Deaf and Hard of Hearing (DHH) children. 
\revised{While numerous technological attempts have demonstrated the potential of at-home parental guidance, most focus on general or behavioral-level support, such as improving social interactions or communicative intent. For MVA children, however, effective guidance demands detailed speech-level support, grounded in expert linguistic and interaction strategies, to help parents navigate conversations tailored to their unique communication characteristics~\cite{KASARI2014635, Megan2011}. 
Thus, our work extends this line of research by developing parental guides that incorporate professional strategies tailored to MVA children.}

\subsection{AAC for MVA children}
Symbol-based AAC tools, which allow users to select symbolic images representing specific objects or concepts, have been widely adopted by MVA children to facilitate their self-expression and social participation~\cite{Janice2012}. While the effectiveness of AAC in learning language and requesting needs has become evident~\cite{Teresa2016}, several barriers persist in its implementation for everyday conversation~\cite{Cynthia2018}. Low-tech AAC in the form of paper cards, often used in the initial stages of adoption, is time-consuming for parents to prepare~\cite{CHIEN201579} and has limited portability when carrying many images~\cite{hayes2010}. As for high-tech AAC devices, such as tablets, children often struggle to navigate numerous cards to find those suitable for specific communication purposes~\cite{chung2015, Rhonda2014}. Although these tools commonly support configuring a preset of the AAC vocabulary in advance for that reason, the presets are not flexible to cover serendipitous topics that arise in real-time, everyday conversations~\cite{CHIEN201579, Gunilla2007, thunberg2011}. Moreover, the vocabulary presets are often selected by others, mostly parents, which makes it uncertain whether they truly reflect the child's intentions~\cite{assistiveware, Neimy2022}.

A large body of research on AAC technologies has focused on recommending appropriate vocabulary or symbols, primarily aimed at enabling AAC users to participate in communication more efficiently. Early work employed rule-based algorithms to predict relevant words or similar symbols based on the user's input of alphabets or symbols~\cite{Trnka2008, voros2014}. Subsequently, several works analyzed the speech of a conversation partner with Natural Language Processing and suggested context-appropriate noun phrases to users~\cite{Bruce2008, Bruce2009}. Recently, the introduction of AI and Large Language Models (LLMs) has allowed for the recommendation of relevant sentences based on users' abbreviated text entry~\cite{Valencia2023, cai2023, Valencia2024}. Furthermore, Vargas~\etal{}~\cite{Fontana2024, Fontana2022} have developed AI-driven AAC boards, which create narrative stories with vocabulary cards based on photos provided by users. Neamtu~\etal{}~\cite{neamtu2019} also proposed LIVOX, an AI-based system that recommends pictograms based on the user's geographic location and time data.

Although existing systems have shown a promising avenue for AI-driven AAC technologies, they rarely target communication with MVA children, which calls for consideration beyond the quick navigation of cards. Many of these children tend to exhibit a significant gap between receptive and expressive language~\cite{Chen2024}, which means they may know specific words and phrases but struggle to recall them when needed. In this context, building on prior research on AI-based AAC recommendations, we aim to explore how AI can act as a stimulus to prompt expressions in MVA children, potentially  their vocabulary.
\section{Formative Study}
\begin{sloppypar}To inform the design of \sysname{}, we conducted 14 semi-structured interviews with autism experts and parents of MVA children. We aimed to understand the challenges in communication between MVA children and their parents and identify professional practices and parental efforts to enhance communication. We tailored the interview protocols for each group---experts and parents---and conducted the interviews separately to better elicit their unique experiences and perspectives. Both interview studies were approved by the Institutional Review Board.\end{sloppypar}

\subsection{Procedure and Analysis}
The interviews were conducted either in-person or remotely depending on the participant's convenience and availability.

\subsubsection{Interviews with Autism Experts}
We recruited nine autism experts (E1--9) by distributing flyers to local private psychiatry hospitals and child development centers. The participants included two child psychiatrists, two child development specialists, two speech-language pathologists, one child psychotherapist, one elementary special education teacher, and one behavioral therapist. The experts had an average of 15.56 years of experience (ranged 8--25 years), and all of them had clinical experiences with MVA children in promoting their language and behavioral development.

We first asked about the communication challenges MVA children and their parents have, the patterns of AAC adoption and usage within these families, and the strategies experts used to understand the intentions of MVA children and facilitate reciprocal conversations. Following this, we presented a video prototype to prompt experts to grasp the concept of an AI-driven communication mediation system. 
\revised{This prototype served as an early proof of concept for integrating parental guidance and AAC technology to facilitate meaningful conversations between parents and MVA children. The video depicted a scenario in which a parent and an MVA child discuss ``what happened today," with the AI providing conversational guides to the parent and AAC symbols to the child. The scenario was developed based on the counsel of one of the authors, an expert in autism communication strategies, drawing from her extensive counseling and clinical experience.} We asked the experts about the clinically desired direction of conversations, principles of parental guidance, potential opportunities for AI integration, and any potential risks.  The interviews lasted about 1 to 1.5 hours. We offered a 100,000 KRW (approx. 80 USD) gift card as compensation.

\subsubsection{Interviews with Parents of MVA Children}
We recruited five parents (F1--5) of MVA children by snowball sampling and the internal network of one researcher, who is both an autism specialist and a parent-child counselor for autistic families. The participants included four mothers and one father, with an average age of 44.2 years (ranged 36--51, $SD = 5.11$). Four of the parents had sons, and one had a daughter, with their average age being 9 years (ranged 6--14, $SD = 2.83$). Three families actively used low-tech AAC with paper symbol cards at home and in educational settings, while the other two had been advised to adopt AAC but not yet tried.

\begin{figure*}[b]
    \centering
     \begin{subfigure}[t]{0.37\textwidth}
         \centering
         \includegraphics[height=3.7cm]{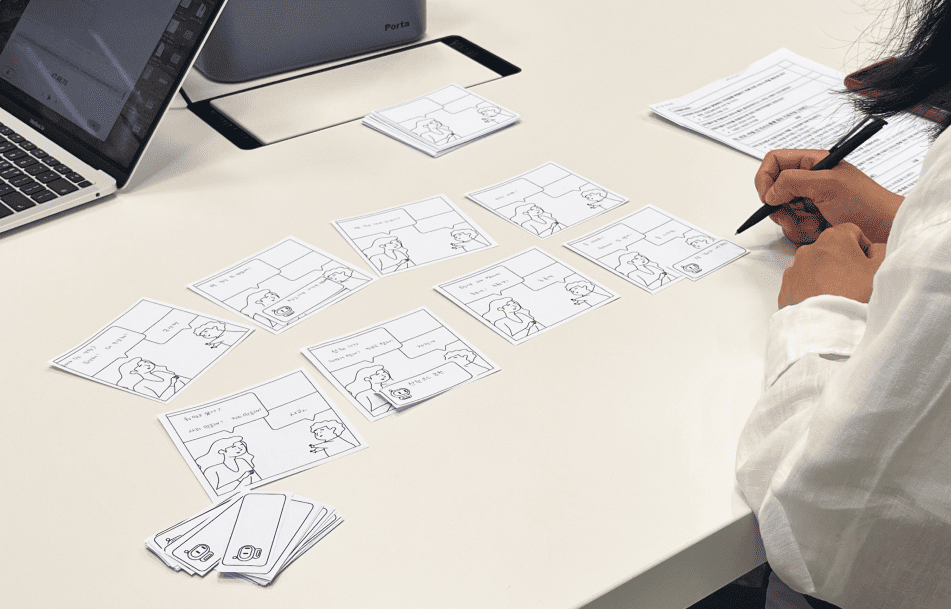}
         \caption{Recalling conversations on comic strips}         \label{fig:formative:photos:comic}
     \end{subfigure}
     \hspace{6mm}
     \begin{subfigure}[t]{0.53\textwidth}
         \centering
         \includegraphics[height=3.7cm]{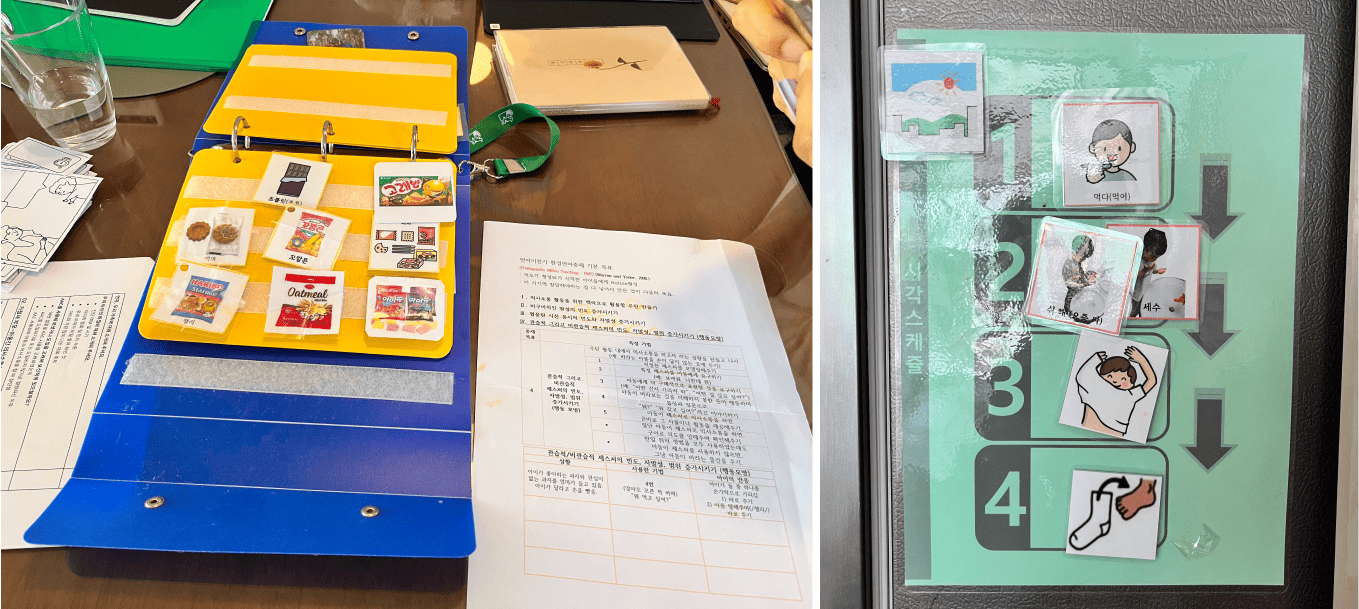}
         \caption{Parental guidance and AACs participants are using}
         \label{fig:formative:photos:aac}
     \end{subfigure}
   \caption[Photos from parent interviews in a formative study: (a) shows a parent writing down a series of comic strips on a desk, detailing conversations with their child. (b) displays an array of parental guidance materials and coated paper-based AAC cards used at home, spread out on a table. These include therapy session handouts, various communication cards, and a clipboard with notes.]{Photos taken from the parent interviews from formative study: (a) A parent writing down conversations they had with their child on the comic strip, and (b) parental guidance materials received from the therapy sessions and coated paper-based AAC cards used at home.}
    \label{fig:formative:photos}
\end{figure*}

The interviews, which lasted about an hour, began by discussing the communication characteristics of MVA children and the parents' experiences with AAC adoption and application. To better capture the nature of current interactions between parents and their MVA children in everyday conversation, we provided a comic strip (See \autoref{fig:formative:photos:comic}) with images of a parent and a child and blank speech bubbles. Parents were asked to fill in the bubbles with examples of enjoyable or challenging conversations they had with their child. We then asked about the efforts parents had made to enhance communication with their child, \rerevised{including the resources and tools used at home (See \autoref{fig:formative:photos:aac})}, the strategies that led to successful outcomes, and any difficulties they encountered in the process. Finally, we explained the basic concept and capabilities of LLMs and invited parents to share their expectations regarding the potential role of AI in supporting parent-child communication. We offered a 50,000 KRW (approx. 40 USD) gift card as compensation.

\subsubsection{Analysis}
All interviews were audio-recorded and later transcribed. Applying Thematic Analysis~\cite{Braun2006ThematicAnalysis}, one researcher open-coded the transcript to identify emerging themes. The entire research team then finalized the themes through multiple rounds of discussions. In the following, we present the findings from the formative study.

\subsection{Finding 1: Parent-Led Conversations}
Both our expert and parent participants pointed out that conversations between MVA children and their parents are often led by parents, where the parent's level of engagement is markedly higher than that of the child. This was particularly evident in two aspects: the conversation structure and the content of the conversation.

\subsubsection{Control over the Conversation} 
We found that most parents had a significant level of control over the entire conversation with their MVA children. For example, all conversations were initiated by the parent, as MVA children often have challenges in making spontaneous communication attempts~\cite{sigafoos1993}. The parent also determined when to conclude the conversation upon noting nonverbal signals from the child getting overwhelmed or disengaged. Even during conversations, when the child remained silent or slow to respond, parents would steer the interaction by asking continuous questions to prevent the conversation from coming to a halt. As a result, most interactions involved minimal turn-taking, typically only 0 to 1 exchange.

When conversing with children unfamiliar with self-expression and social interaction, it is natural and encouraged for the parent to guide the conversation~\cite{Choi2020}. However, some experts pointed out that such asymmetric dynamics could limit the child’s opportunities to learn that communication is an effective means to convey their intentions, leaving them as passive participants. Experts suggested that for meaningful conversations that enhance parent-child bonding, the system should support reciprocal exchanges between the child and parent. E7 remarked: \textit{``Both the parent and the child need to learn that a conversation isn't something you do alone. It’s back-and-forth. But because one person (the parent) is speaking and the other (the MVA child) isn't, they participate in asymmetrical ways, so they've had fewer chances to really build this skill.''}

\subsubsection{Teaching Speech instead of Communication}
Parent participants reflected that their conversations with their MVA children at home often focused on repeatedly teaching essential words for daily life, with an emphasis on producing correct sentences. F1 mentioned, \textit{``As my child grows older and has to go to preschool or school without me, they need to at least know how to say things like `bathroom' or `water.' It feels urgent to make sure they can say those basic words.''} Experts noted that parents can become so absorbed in honing their child’s speech that they mistakenly believe a few spoken words mean they’re having a conversation. E4 said: ``\textit{Autistic kids often repeat the same word over and over throughout the day. So the parents take that as a chance to ask questions. But the kid just keeps saying the same word. In the end, there’s nothing shared between them.}'' They propose that instead of teaching what the parent believes necessary, it's important to establish a shared interest between the parent and child and provide rich resources and stimuli that can enable the child to express their thoughts and feelings.

\subsection{Finding 2: Lack of Actionable Parental Guidance}
Regarding parents' challenges in improving conversations with their MVA children, most parent participants mentioned difficulty in applying key principles and strategies from parent training in day-to-day conversations. Considering the broad spectrum of characteristics displayed by autistic children, the communication guides currently available to parents are often too general and impractical. F3 remarked: ``\textit{After my child was diagnosed with autism, I bought these huge, thick books, watched every YouTube video by doctors, and got all sorts of pamphlets from the therapy sessions on how to guide conversations. I know what's important now. But all that material is really broad. When I try to use it with my child, I just freeze, and the words don’t come out right.}'' Similarly, F4 recalled, ``\textit{My child’s speech therapist once told me, `Most of your conversations with your child are just questions. You need to diversify your interactions.' It was a huge revelation. But then I realized I have no idea what to say if I’m not asking a question.}''

Some parents (F2 and F4) closely observed how the therapist conversed with their child during clinic sessions and then practiced the therapist's phrases at home. However, they pointed out that it’s impossible to collect examples for every situation and topic. Additionally, secondary caregivers, who spend less time with the child and have fewer opportunities to observe examples from the clinic, often lack confidence in guiding interactions. It sometimes led secondary caregivers to avoid conversations with their children, ultimately weakening their bond.

\subsection{Reflection: Opportunities for AIs as Communication Mediators}
To address the challenges mentioned above, parent participants have implemented various professional supports and workarounds. Most of them (F1, F3, F4, and F5) attended parent training sessions led by therapists, and some (F1, F2, and F3) created customized AAC cards using photos of objects, people, and activities from their child's daily life. However, they noted that these approaches are not always readily available and impose significant time and cost burdens, raising concerns about their long-term sustainability. This led participants to express the expectation that AI could provide low-cost, accessible at-home guidance for parents and AAC card curation for MVA children. F5 remarked, ``\textit{I heard AI can make drawings now. The first thing I thought was, it’d be great if it could make AAC cards for my kid. My child’s growing up, but I can’t keep up with making the cards. So lately, I feel like they’re not really useful anymore.}'' Experts also recognized the potential of AI to broaden the range of expression and interaction in parent-child conversations. However, they emphasized that the parent and child should remain the primary drivers of the conversation, with AI's suggestions serving only as a point of reference.

\section{\sysname{}}
\sloppy{
Informed by the formative study, we designed and developed \sysname{}, a communication mediation system for both parent and MVA child. In this section, we discuss our design rationales from formative interviews and literature, user interface and system components of \sysname{}, description of generative pipelines, and implementation details.}

\subsection{Design Rationales}
\ipstart{DR1. Structure Turn-taking through Expressive Cues}\phantomsection{}\label{sec:dr1}
From the formative interviews, experts stressed the importance of MVA children having more control throughout the conversation process, thereby inviting them to engage in back-and-forth exchanges with their parents. Given that autistic children are often visual learners~\cite{Paula2003}, we decided to provide visual and behavioral cues to make them better understand the flow of conversation and effectively express their intentions through actions. To this end, we set the system on a tablet device so that both the parent and child can visually track the conversation while jointly focusing on the screen. When starting a conversation, the parent and child are guided to select a topic together. Finally, a shared turn pass button (\autoref{fig:interface:teaser}, bottom) serves as a turn-switching signal, reinforcing the concept of turn-taking in a tangible way.

\ipstart{DR2. Contextualize Parental Guidance}\phantomsection{}\label{sec:dr2}
Considering the parents' needs for immediately applicable conversation guides, we \revised{designed two types of context-aware guidance based on the formative interviews with experts on general principles of parental guidance. First, \textbf{conversational guides} provided in general conversations assist parents in diversifying the forms of interaction. Drawing on strategies from the Hanen More Than Words program~\cite{Elaine2013}, we identified 12 parent response types to promote communication with MVA children (see \autoref{tab:guides}). These guides are provided along with relevant examples. Second, \textbf{feedbacks} alert parents to be aware of negative interaction dynamics. Based on the principles of an evidence-based intervention, Parent-Child Interaction Therapy (PCIT) \cite{eyberg2011}, we outlined three types of negative conversational patterns that parents should avoid (see \autoref{tab:feedback}). When such responses are detected during conversations, the corresponding feedback is shown in the next turn to avoid disrupting the conversation between parents and children. Such delayed delivery was designed to stimulate parents to reflect on the direction of the conversation themselves. Therefore, rather than instructing parents not to make specific remarks, the feedback informs parents about the potential impact and risks on conversations with their child. The categories and detailed instructions for LLM prompting for both conversational guides and feedback were developed based on formative interviews and reviewed by an expert author.}

Upon the experts' concerns about parents' overreliance on AI suggestions, we implemented two safeguards. First, we limited the number of guides presented to three, allowing parents their own room to ideate proactively. Second, the example utterances would not appear automatically; instead, parents must tap guides to reveal the examples. In doing so, we intended to encourage parents to internalize the guidance and implement it in their own words. All guides, examples, and feedback are presented in a concise, glanceable format to minimize disruption during conversation.

\ipstart{DR3. Prioritize MVA Children's Vocabulary}\phantomsection{}\label{sec:dr3}
MVA children often struggle to retrieve the appropriate words, even when they have something to express~\cite{williams2006}. Generating a variety of AAC cards using LLMs can help the MVA child to recognize unmet needs and boost self-expression. Considering the age and vocabulary of MVA children, we simplified the AAC board into a more accessible format. We structured the AAC board into four categories: \textbf{Topic, Action, Emotion}, and \textbf{Core}. This setup provides balanced exposure to various linguistic components. Topic and Action words are contextually generated by the LLM, whereas Emotion words are curated from a predefined set of 12 basic emotions---which were crafted in consultation with experts---specifically aimed at autistic children who struggle to grasp complex and subtle emotions~\cite{american1994}. 
\revised{Existing AAC boards typically provide \textbf{Core} words---a fixed set of vocabularies frequently used to express users' basic needs---that are consistently displayed on the screen for easy access~\cite{beukelman2013, Karen2001}. We employed \aaccard{Yes} and \aaccard{No} as Core words, both of which are highly frequent in existing core vocabulary lists and are easily understood by MVA children~\cite{Emily2020}. Based on an expert's feedback, w}e expanded them to include \aaccard{I don't know}, which allows children to avoid habitual, non-meaningful responses, and \aaccard{How about you, mom/dad?} to facilitate mutual exchange in conversations.
To reduce cognitive load, each category presents only four cards at a time, resulting in a total of 12 AI-recommended cards and four default options on the screen. A refresh button is also provided, allowing children to request new recommendations if the initial ones do not meet their needs, ensuring they are not forced to choose a word that doesn't align with their intent. Moreover, for children who struggle with symbol recognition, photos of familiar people, places, and objects are pre-uploaded and used as custom AAC symbols.


\begin{figure*}
    \centering
    \includegraphics[width=0.95\textwidth]{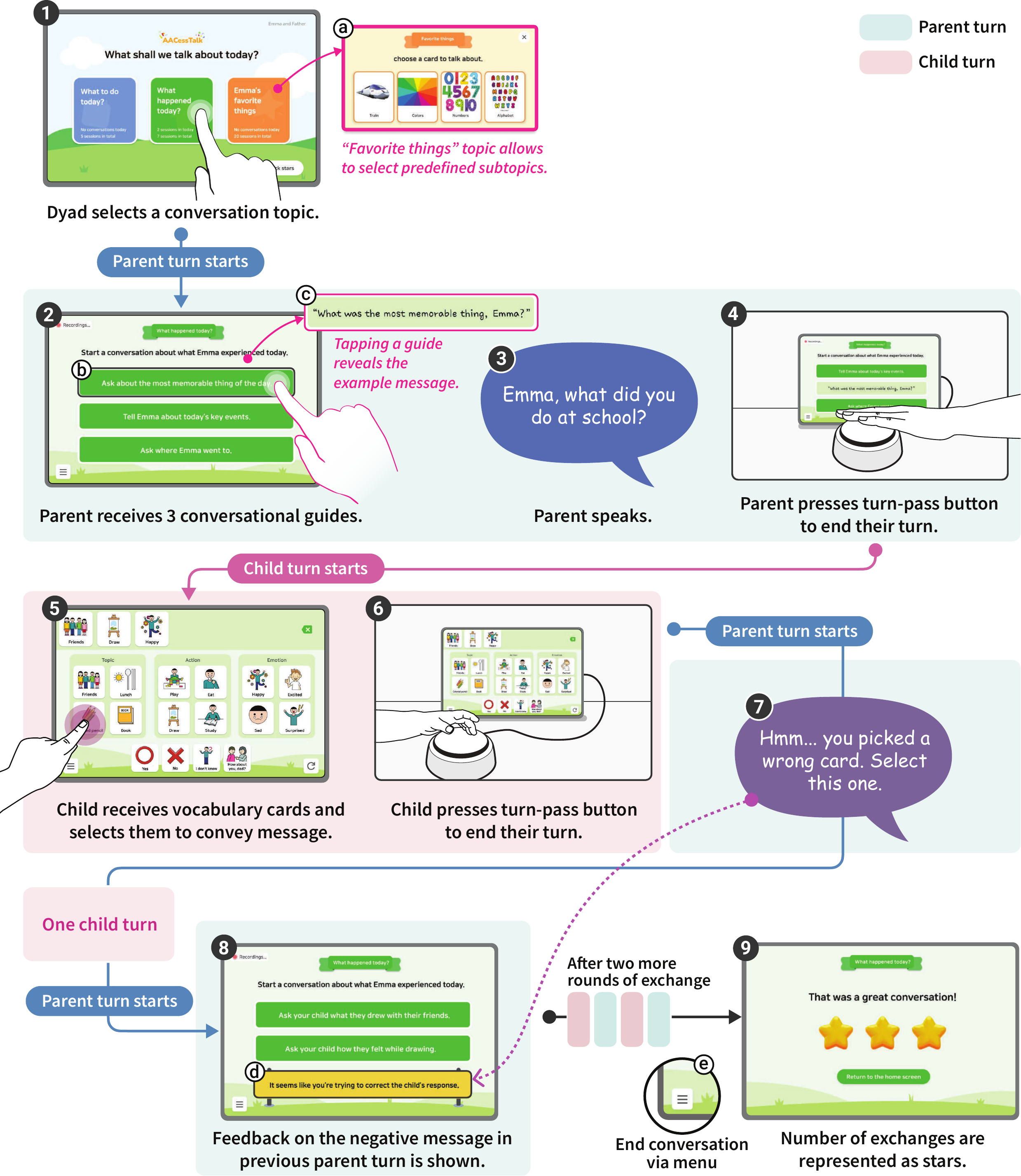}
    \caption[Figure 3 depicts the main screens and interaction flow of AACessTalk, a system designed to facilitate guided conversations between parents and children. First, Home Screen: The first screen shows options for selecting conversation topics, including "What to do today?", "What happened today?", and "Emma's favorite things". The "Favorite things" topic allows the selection of predefined subtopics. Second, Parent’s Turn Screen: During the parent’s turn, the screen provides three conversation guides (e.g., "Ask about the most memorable thing of the day"), with example messages revealed upon tapping the guide. The parent uses this guidance to initiate dialogue and speaks aloud to the child. A "turn-pass" button allows the parent to signal the end of their turn. Third, Child’s Turn Screen: The child’s screen displays vocabulary cards grouped into categories such as "Topic," "Action," "Emotion," and "Core". The child selects cards to construct their response and ends their turn by pressing the turn-pass button. Last, Feedback and Conversation End: If a negative conversational pattern is detected during the parent's previous turn, feedback is shown on the parent’s turn screen, helping improve interaction. At the end of the conversation, the final screen celebrates the interaction with a "That was a great conversation!" message, displaying three stars and a button to return to the home screen. The figure also illustrates the iterative process of turn-taking between parent and child, supported by AACessTalk's features, until the conversation concludes via the menu.]{{Main screens and usage flow of \sysname{}: \circledigit{1} The home screen, where users select conversation topics. Selecting `Plan' (Blue) or `Recall' (Green) topics immediately begins the conversation, while `Interest' (Orange) presents a pop-up \circledigit{a} with the child’s pre-uploaded interests for further selection. \circledigit{2} The parent's turn screen, which provides \circledigit{b} conversation guides and \circledigit{c} example phrases for the parent. \revised{\circledigit{3} The parent speaks to the child and then \circledigit{4} presses the turn-pass button to end their turn. \circledigit{5} The child's turn screen, which offers vocabulary cards for the child. \circledigit{6} The child presses the turn-pass button to end their turn. \circledigit{7} If a negative conversational pattern is detected in the parent’s previous turn, \sysname{} shows \circledigit{d} feedback during the next parent turn \circledigit{8}. After several iterations of turn-taking like \circledigit{2}-\circledigit{6}, either the parent or the child ends the conversation \circledigit{9} via menu ~\circledigit{e}.}}}
    \vspace{-1mm}
    \label{fig:interface:flow}
\end{figure*}

\subsection{User Interface and Interaction with \sysname{}}
\autoref{fig:interface:flow} illustrates the overall use scenario of the \sysname{}. Users first select a conversation topic (\autoref{fig:interface:flow}-\circledigit{1}), then the interface moves to the parent's turn, displaying \revised{three \textit{Conversational guides}} on the screen (\autoref{fig:interface:flow}-\circledigit{2}). By starting with the parent's turn, we aimed to gather contextual hints from the parent's speech to recommend suitable AAC cards. 
Once the parent completes their turn, it transitions to the child's turn, showing the AAC board (\autoref{fig:interface:flow}-\circledigit{5}). \revised{This process repeats with several exchanges between the parent and child. Additionally, if a negative conversational pattern is detected in parent's speech, \sysname{} provides \textit{Feedback} to the parent during their next turn, alongside two \textit{conversational guides} (\autoref{fig:interface:flow}-\circledigit{8}). Finally, the parent or the child can conclude the conversation session whenever they want \autoref{fig:interface:flow}-\circledigit{9}).}

Here, we illustrate a hypothetical scenario of a MVA child-parent communication with \sysname{}. 
    Every evening, Daniel and his 5-year-old autistic daughter, Emma, share a special moment to talk about their day. Tonight, they sit at Emma's bedside table and launch the \sysname{} app on a tablet with a set of turn pass button in front of them. The \sysname{} displays three topic cards on the main screen (\autoref{fig:interface:flow}-\circledigit{1}): (1) \textbf{What to do today?} (Plan)---sharing the day’s schedule to reduce the child's anxiety about unexpected events~\cite{Shari1998}, (2) \textbf{What happened today?} (Recall)---discussing the day's experiences to support the parent and child in processing and reflecting on what happened~\cite{Charlotte2013}, and (3) \textbf{Emma's favorite things} (Interest)---talking about a topic the child enjoys to increase their motivation to participate in the conversation~\cite{Mary2007}. Daniel is curious about Emma's experience during the job exploration field trip at kindergarten. So, among the three conversation topics on the main screen, he selects the ``What happened today'' topic.


    The \sysname{} then transitions to the parent turn screen (~\autoref{fig:interface:flow}-\circledigit{2}). When the parent's turn begins, recording starts automatically to transcribe Daniel's speech, with an animated recording indicator appearing in the top left corner. In the center of the screen, three conversation guides generated by LLMs are presented as cards to help Daniel to start the discussion about today's event. Daniel skims through the options and picks one that starts with: ``Remind Emma of the important events from today.' He pauses briefly, considering how to make it engaging for Emma. Then he taps on the guide, and an example phrase pops up. Inspired, Daniel says, \textit{``Emma, did you see the firefighter you like at the job trip today?''} and then presses the turn pass button to signal the end of his turn.


    \sysname{} now switches to Emma's turn (\autoref{fig:interface:flow}-~\circledigit{5}). In the center of the screen, each of the three sections, labeled Topic, Action, and Emotion, displays four AAC cards related to firefighters. Emma carefully looks through the cards and taps on the \aaccard{Firetruck} card. Then the card appears in the selected card area at the top of the screen, and the app voices, ``firetruck'' Emma then touches the refresh button in the bottom right corner to receive new card suggestions. She then selects \aaccard{Ride}, \aaccard{Fire hose}, and \aaccard{Happy}. After a pause, she looks up at Daniel. Daniel says, \textit{``Emma, if you're done, do you want to press the button? Then it’ll be Daddy's turn.''} When Emma presses the turn pass button, \sysname{} switches back to the parent screen.
    
    After a couple more exchanges, Emma's turn arrives again. But instead of selecting a card, she repeatedly presses the turn pass button. Daniel smiles and says, \textit{``Emma, that was so much fun. How about we call it a day here?''} Emma agrees by selecting the \aaccard{Yes} card on the screen. Daniel then \revised{ends the conversation session through the menu button} (\autoref{fig:interface:flow}-~\circledigit{e}) on the app. The system displays a cheerful message and also shows how many turns they have had with star icons, celebrating their achievement before going to bed (\autoref{fig:interface:flow}-~\circledigit{9}).


\subsection{Generative Pipelines}

The system generates contextual \revised{guidance} for both parents (\autoref{fig:system:pipeline:parent}) and children's (\autoref{fig:system:pipeline:child}) turns, \revised{considering the dialogue messages as an input.} To build text-based dialogue data (\eg, \circledigit{A} in \autoref{fig:system:pipeline:parent} and \circledigit{a} in \autoref{fig:system:pipeline:child}), the system treats the transcript of the parents' utterances as the parent messages, and the list of the selected AAC card labels as the child messages.

\subsubsection{Generating Parental Guidelines}
\begin{figure*}[h]
    \centering
    \includegraphics[width=\textwidth]{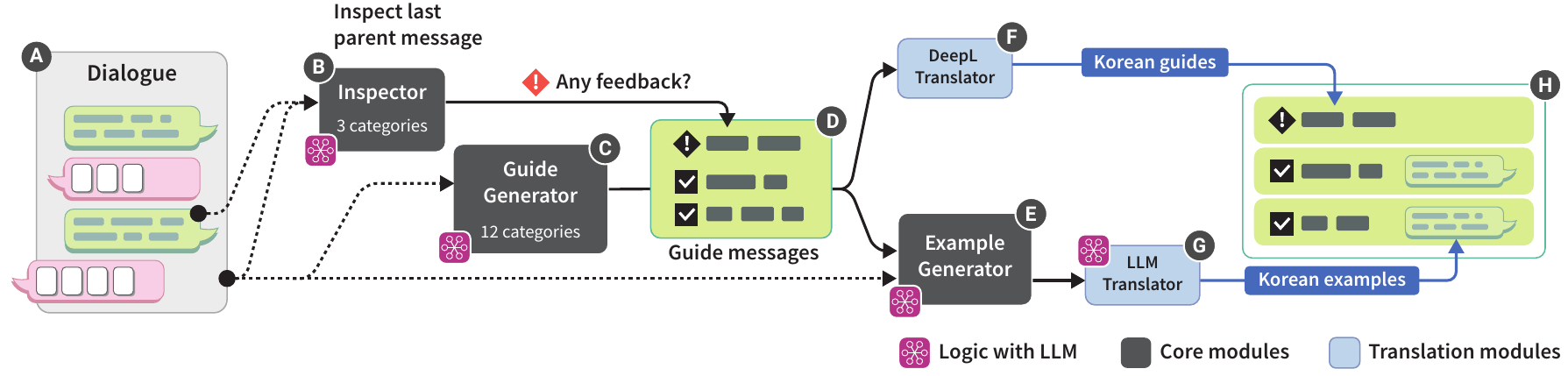}
    \caption[Figure 4: Generative pipeline for parental guidelines. The diagram visualizes the system's workflow beginning with the inspection of the last parent message (A). The inspector module (B) analyzes this message to identify if it falls into any of three \revised{feedback} categories: blame, correction, or complexity. If identified, feedback is generated. Next, the guide generator (C) reviews the entire dialogue and produces up to three guide statements based on the absence or presence of \revised{feedback} categories. These guides are derived from 12 predefined directions chosen to suit the dialogue context. Subsequently, the example generator (D) uses these guides to create example messages for parents, which are practical suggestions on how to respond. Finally, the system translates the guides and examples into Korean using both the DeepL translator (E) for formal translations and an LLM translator (F) for informal tone, aligning with typical parent-child communication styles in South Korea.]{Generative pipeline for parental guidelines. The pipeline analyzes the current dialogue~\circledigit{A} and generates parental guidelines with example messages~\circledigit{H}.}
    \label{fig:system:pipeline:parent}
\end{figure*}

When a parent turn starts, the \textbf{inspector} (\circledigit{B} in \autoref{fig:system:pipeline:parent}) first checks the previous parent message if it falls under any of the three \revised{feedback} categories: (1) \textit{blame} (the parent criticizes or negatively evaluates the child), (2) \textit{correction} (the parent is compulsively correcting the child's response or pointing out that the child is wrong), and (3) \textit{complex} (the dialogue contains more than one goal or intent). If the message is tagged with any one of these categories, the inspector generates the feedback message.
The \textbf{guide generator} (\circledigit{C} in \autoref{fig:system:pipeline:parent}) analyzes the entire dialogue and generates guide statements for the next parent response. To simplify the information to be shown to the parent, we limited the number of guides to two when there is inspector feedback or three when there is none (See \circledigit{D} in \autoref{fig:system:pipeline:parent}). To avoid generating duplicate guides, we provided 12 predefined guide directions---\textit{Ask for
Elaboration}, \textit{Show
Encouragement}, \textit{Suggest
Choices}, \textit{Encourage
Self-Disclosure}, \textit{Ask for
Intentions}, \textit{Extend Topic}, \textit{Open Up}, \textit{Show
Empathy}, \textit{Pique Interest}, \textit{Provide Clues}, \textit{Suggest
Coping Strategies}, and \textit{Wrap Up} (See \autoref{tab:guides} for descriptions)---and prompted the model to choose the most appropriate direction for each guide. 
\revised{The feedback categories and guide directions were derived from both formative interviews with experts and widely recognized intervention programs~\cite{Elaine2013, eyberg2011}, and then reviewed by a child psychotherapist.}

The guide statements are then passed onto the \textbf{example generator} (\circledigit{E} in \autoref{fig:system:pipeline:parent}), which generates example messages that the parent can try. For instance, an example message \textit{``Emma, how do you feel when running on the soccer field?''} can be generated from the guide ``Ask Emma how she feel when playing soccer.''

As we planned to deploy \sysname{} to parent-MVA child dyads in South Korea, we also incorporated translation steps in the end (\circledigit{F} and \circledigit{G} in \autoref{fig:system:pipeline:parent}). \revised{We used the DeepL translation API~\cite{DeepL} to translate the guide statements due to their formal nature. DeepL has been shown to outperform other AI-based machine translation systems, including ChatGPT-3.5, in English-to-Korean translation quality evaluations~\cite{liu2024corpus}. For the translation of example messages, we used \texttt{GPT-4o} to apply an informal style of voice, which is typical in parent-to-child communication in Korean. To ensure that the model generates translation in consistent quality and style, we employed few-shot prompting~\cite{zhang2023prompting}, where the model input includes three translation samples most similar to the input English text, retrieved through a similarity search. For the sample database, we crafted 18 English-Korean translation references inspired by the cases observed during internal testing.}

\subsubsection{Generating Card Decks for Children}
\begin{figure*}[h]
    \centering
    \includegraphics[width=\textwidth]{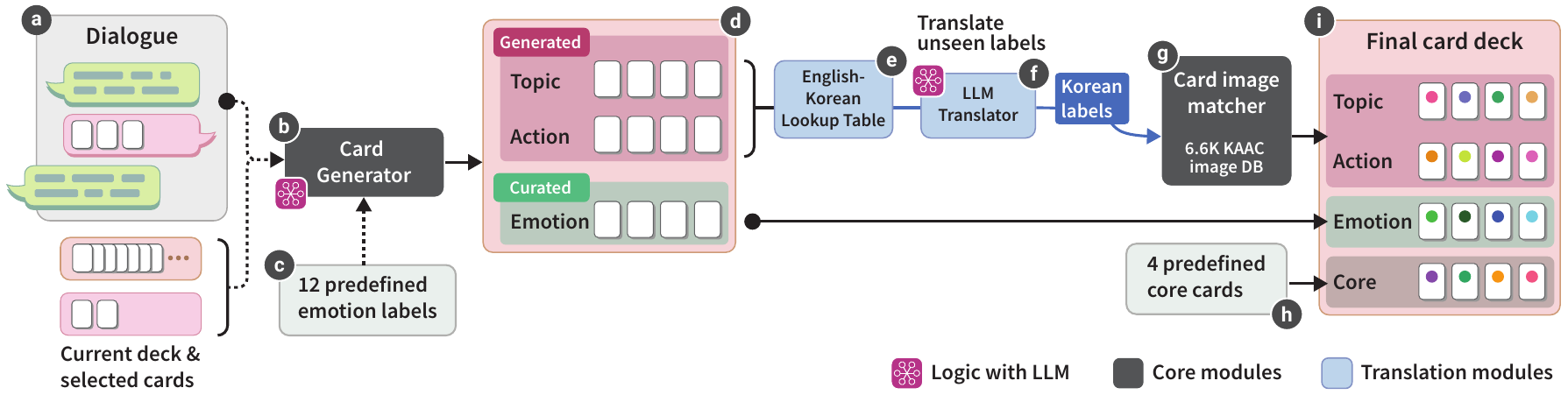}
    \caption[Generative pipeline for curating card decks for children. The process begins by analyzing the current dialogue, card deck, and selected cards (a). The card generator (b) then produces labels for Topic and Action cards and curates four Emotion cards from a predefined list of 12 possible emotions. The generated and curated cards undergo a translation phase (e and f), where they are converted into Korean using a lookup table and LLM-based translation, ensuring cultural and linguistic appropriateness. Finally, the card image matcher (g) uses a database of 6.6K AAC symbol and label pairs to visually match each card label with an appropriate symbol, resulting in the final card deck (i), which includes four predefined core cards and generated Topic, Action, and Emotion cards.]{Generative pipeline for curating card decks for children. The pipeline analyzes the current dialogue and card information~\circledigit{a} and generates a deck of cards in four categories~\circledigit{i} along with symbol illustration retrieved from the KAAC database~\circledigit{g}.}
    \label{fig:system:pipeline:child}
\end{figure*}

The current dialogue, card deck, and selected cards (\circledigit{a} in \autoref{fig:system:pipeline:child}) are considered when generating the next card deck. The \textbf{card generator} (\circledigit{b} in \autoref{fig:system:pipeline:child}) generates four card labels in \textit{topic} and \textit{action} categories respectively. \revised{For the \textit{emotion} category, the LLM curates four \textit{emotion} cards by selecting from a predefined list of 12 emotions} (\circledigit{c} in \autoref{fig:system:pipeline:child})---\textit{joyful}, \textit{glad}, \textit{happy}, \textit{excited}, \textit{sad}, \textit{angry}, \textit{upset}, \textit{scared}, \textit{afraid}, \textit{surprised}, \textit{amazed}, and \textit{bored}. We derived this set of emotions by referring to Plutchik’s wheel of emotions~\cite{plutchik1980general} and based on previous research on an LLM-driven chatbot for children~\cite{seo2024chacha}. A child psychotherapist reviewed the list, excluding emotions that children with autism may rarely understand. \revised{For the \textit{core} category, a fixed set of four cards (\aaccard{Yes}, \aaccard{No}, \aaccard{I don't know}, and \aaccard{How about you, mom/dad?}) was consistently used.}

The generated topic and action cards (\circledigit{d} in \autoref{fig:system:pipeline:child}) go through a translation process combining a cached lookup table~(\circledigit{e} in \autoref{fig:system:pipeline:child}) and an LLM translator~(\circledigit{f} in \autoref{fig:system:pipeline:child}). 
\revised{The lookup table, indexed by pairs of categories and English labels, contains 305 translation references based on cases collected from multiple rounds of internal testing. The LLM translator translates only labels not found in the lookup table, but it also retrieves the five most similar references from the table for each label to use in few-shot prompting.}

The translated Korean card labels are then matched with appropriate symbol images by \textbf{card image matcher} (\circledigit{g} in \autoref{fig:system:pipeline:child}). We used 6,658 AAC symbol and label pairs in the KAAC Symbols Search\footnote{\url{http://symbol.ksaac.or.kr/}} database~\cite{Parl2016KAAC}. The card image matcher performs a similarity search with the card labels across the captions describing the symbol illustrations, generated by GPT-4~\cite{openai2023gpt4} and Gemini~\cite{google2024gemini} in advance, to find the best match. For all cards, if the user has registered their own card image for the same label, the system prioritizes the custom images.

\subsection{Implementation}
We implemented the core system of \sysname{} in Python using a FastAPI~\cite{FastAPI} server that serves REST APIs. The generative pipelines incorporate OpenAI~\cite{openai}'s ChatCompletion APIs to run the underlying LLM inferences. We used \texttt{gpt-4-0613} model for generative tasks (\eg, \circledigit{B}, \circledigit{C}, and \circledigit{E} in \autoref{fig:system:pipeline:parent} and \circledigit{b} in \autoref{fig:system:pipeline:child}) as the model showed promising performance in complex instruction follow-ups and language generation tasks in previous research~(\eg,~\cite{kim2024mindfuldiary, shin2023planfitting, seo2024chacha}). To minimize peaking latency in the user interface caused by the English-to-Korean translation overhead, we used faster \texttt{gpt-3.5-turbo} and \texttt{gpt-4o} models for translation tasks (\eg, \circledigit{G} in \autoref{fig:system:pipeline:parent} and \circledigit{f} in \autoref{fig:system:pipeline:child}). The interaction and conversation logs were stored in the SQLite database.

The client tablet app was implemented in TypeScript~\cite{TypeScript} on React Native~\cite{ReactNative} as a cross-platform running on Android and iPad. The client communicates with the server via REST API.
On parent turns, the utterance is audio-recorded and sent to the server. The recording is then transcribed using CLOVA Speech Recognition API\footnote{\url{https://api.ncloud-docs.com/docs/en/ai-naver-clovaspeechrecognition}}, which is widely applied to Korean automatic speech recognition tools.
On child turns, we also used CLOVA Voice API\footnote{\url{https://api.ncloud-docs.com/docs/en/ai-naver-clovavoice}} to generate voice-over audios for card labels.
\section{Deployment Study}
We conducted a two-week field deployment study with 11 dyads of MVA children and their parents. We aimed to examine how MVA children and parents interact with \sysname{} to engage in reciprocal conversations and how the use of \sysname{} influences the everyday conversation and interaction dynamics of these dyads. The study protocol was approved by the Institutional Review Board.

\subsection{Participants}

\newcommand{\listparbox}[1]{\parbox{0.37\textwidth}{#1}}

\newenvironment{listenum}{\begin{itemize}[leftmargin=*, label=\textbullet, itemsep=0.1em, labelsep=0.2em, itemjoin={\newline}]}{\end{itemize}}

\begin{table*}[b]

\sffamily
\smaller
	\def\arraystretch{1.0}\setlength{\tabcolsep}{0.2em}
		    \centering
\caption[Table 1 presents the demographics of participants in a study involving minimally verbal autistic children and their parents. The table is organized into columns listing the parent’s age, parental role (such as Mother (primary) or Father (secondary)), the child’s alias, age, gender, a description of their speech and cognition status, their use of AAC, presence of siblings, and personal interests. Each row corresponds to a different child-parent pair, detailing the diverse speech capabilities and family dynamics.]{Demographics of our study participants. It includes the age and parental role of the parent participants, as well as the age and gender of the MVA children. Additionally, it features a parent-reported description of the children's speech and cognition status, their use of AAC, the presence of siblings, and personal interests.}~\label{tab:demographic}

\begin{tabular}{|c!{\color{lightgray}\vrule}c!{\color{lightgray}\vrule}>{\centering\arraybackslash}m{0.075\textwidth}!{\color{gray}\vrule}c!{\color{lightgray}\vrule}c!{\color{lightgray}\vrule}c!{\color{tablegrayline}\vrule}>{\raggedright\arraybackslash}m{0.37\textwidth}!{\color{tablegrayline}\vrule}c!{\color{tablegrayline}\vrule}m{0.09\textwidth}!{\color{tablegrayline}\vrule}m{0.14\textwidth}|}
\hline
\rowcolor{lightgray}\multicolumn{3}{|l!{\color{gray}\vrule}}{\textbf{Parents}}      & \multicolumn{7}{l|}{\textbf{MVA Children}}                                                                                                                                                                                                    \\
\rowcolor{tableheader}\textbf{Alias} & \textbf{Age} & \textbf{Type}               & \textbf{Alias} & \textbf{Age} & \textbf{Gender} & \textbf{Description of speech/cognition} & \textbf{AAC use}         & \textbf{Siblings}                                                  & \textbf{Interests}                                                                                 \\\hline
\cellcolor{tableheader}\textbf{P1}    & 35  & Mother\newline{}(primary)   & \cellcolor{tableheader}\textbf{C1}   & 6   & Boy    &  \listparbox{\begin{listenum} \item Produces speech with 1-2 words for requests. \item Understands familiar expressions but struggles with\newline{}verbal responses. \end{listenum}}                & -           & -           & \footnotesize{Clay, Train, Animals,\newline{}Whale, Shark,\newline{}KATURI (animation),\newline{}Baby Shark (animation)}             \\ \arrayrulecolor{lightgray}\hline
\cellcolor{tableheader}\textbf{P2}    & 38  & Mother\newline{}(primary)   & \cellcolor{tableheader}\textbf{C2}   & 6   & Boy    & \listparbox{\begin{listenum} \item Produces echolalia with around 10 words. \item Uses PECS cards to request. \item Understands basic instructions (e.g., sit, stand, no). \end{listenum}}                 & Very freq. & Non-autistic\newline{}twin brother                                 & \footnotesize{Train, Vehicles,\newline{}Numbers, Alphabet,\newline{}Colors, English}                                       \\\hline
\cellcolor{tableheader}\textbf{P3}    & 47  & Mother\newline{}(primary)   & \cellcolor{tableheader}\textbf{C3}   & 8   & Boy    & \listparbox{\begin{listenum} \item Produces speech with simple 3-word phrases\newline{}for requests. \item Verbalizes interest-related words. \item Pronunciation is unclear due to articulation issues. \item Communicates through writing. \end{listenum}}                & Very freq. & -                                                         & \footnotesize{Elevator, Automatic\newline{}restroom door,\newline{}Water slide} \\\hline
\cellcolor{tableheader}\textbf{P4}    & 43  & Mother\newline{}(primary)   & \cellcolor{tableheader}\textbf{C4}   & 5   & Boy    & \listparbox{\begin{listenum} \item Produces speech using nouns. \item Verbalizes favorite characters. \item Understands simple instructions. \end{listenum}}                & -           & Two non-\newline{}autistic\newline{}brothers and \newline{}one autistic\newline{}twin brothers & \footnotesize{Firefighter,\newline{}Pororo (animation),\newline{}Dinosaur}                                                 \\\hline
\cellcolor{tableheader}\textbf{P5}    & 51  & Father\newline{}(secondary) & \cellcolor{tableheader}\textbf{C5}   & 13  & Girl   & \listparbox{\begin{listenum} \item Produces speech in sentences for requests about\newline{}areas of interest. \item Can read and write. \end{listenum}}                & Very freq. & -                                                         & \footnotesize{Catch! Teenieping\newline{}(animation), Coloring,\newline{}Ramen, Pizza}                                     \\\hline
\cellcolor{tableheader}\textbf{P6}    & 47  & Mother\newline{}(primary)   & \cellcolor{tableheader}\textbf{C6}   & 15  & Boy    & \listparbox{\begin{listenum} \item Produces speech with 2-3 word sentences for requests. \item Expressive language is limited compared to\newline{}receptive language. \item Pronunciation is unclear. \item Understands symbolic images and written language well. \end{listenum}}                 & Very freq. & Non-autistic\newline{}sister                                       & \footnotesize{Soccer, Running,\newline{}Magformers (toy),\newline{}Puzzle, iPad, Songs,\newline{}Todo Hangul\newline{}(education material)}  \\\hline
\cellcolor{tableheader}\textbf{P7}    & 42  & Mother\newline{}(primary)   & \cellcolor{tableheader}\textbf{C7}   & 6   & Boy    & \listparbox{\begin{listenum} \item Produces speech with 2-word phrases to convey\newline{}requests. \item Understands symbolic images well. \end{listenum}}                 & Sometimes       & -                                                         & \footnotesize{Bus, Bread Barber\newline{}Shop (animation),\newline{}Duda \& Dada\newline{}(animation),\newline{}Swimming}                      \\\hline
\cellcolor{tableheader}\textbf{P8}    & 42  & Mother\newline{}(primary)   & \cellcolor{tableheader}\textbf{C8}   & 10  & Boy    & \listparbox{\begin{listenum} \item Produces speech with 2-word phrases to convey\newline{}requests. \item Pronunciation is unclear. \item Understands and learns symbolic images well. \end{listenum}}                 & Very freq. & Two non-\newline{}autistic\newline{}brothers                                 & \footnotesize{Car, Train, Spaceship}                                                           \\\hline
\cellcolor{tableheader}\textbf{P9}    & 37  & Mother\newline{}(primary)   & \cellcolor{tableheader}\textbf{C9}   & 8   & Girl   & \listparbox{\begin{listenum} \item Frequent echolalia and non-meaningful vocalizations. \item Understands parents' speech to some extent. \item Recently shows signs of reading words without images. \end{listenum}}                 & -           & -                                                         & \footnotesize{Pororo (animation),\newline{}Puzzle, Scribbling}                                                    \\\hline
\cellcolor{tableheader}\textbf{P10}   & 47  & Mother\newline{}(primary)   & \cellcolor{tableheader}\textbf{C10}  & 14  & Boy    & \listparbox{\begin{listenum} \item Produces speech with 2-word phrases for requests. \end{listenum}} & Sometimes       & -                                                         & \footnotesize{YouTube, Toy cars}                                                                         \\\hline
\cellcolor{tableheader}\textbf{P11}   & 34  & Mother\newline{}(primary)   & \cellcolor{tableheader}\textbf{C11}  & 6   & Boy    &  \listparbox{\begin{listenum} \item Produces echolalia, but no spontaneous speech. \item Pronunciation is unclear. \item Expresses requests through pointing. \item Understands parents' instructions to some extent. \end{listenum}} & -           & -                                                         & \footnotesize{Pinkfong (animation),\newline{}MyChew (candy),\newline{}Bread}                \\ \arrayrulecolor{black}\hline 
\end{tabular}
\end{table*}

We recruited MVA child-parent dyads by ensuring that both the child and the family met the following specific criteria: For the child, (1) a diagnosis of Autism Spectrum Disorder classified as Level 2 autism per CDC guidelines, (2) an ability to understand vocabulary to some extent but with challenges in verbal expression, 
\revised{(3) no difficulties with hand coordination, ensuring they could interact with a touchscreen tablet without assistance.}
We did not impose an age restriction on the children in the study, as the cognitive and communicative abilities of autistic children are not strictly dependent on age~\cite{Odermatt2022}. For the family: (1) one parent who can consistently use the \sysname{} to communicate with the child over the two-week period, and (2) a stable home Wi-Fi connection. The flyers were distributed to potential parent participants who met our criteria through a local child development center in South Korea, where one of the authors is affiliated, along with snowball sampling. A total of 25 dyads expressed interest in participation.

We involved two additional screening steps to ensure the system would not cause discomfort or difficulties to the children. First, we asked parents to provide a written description of their child's verbal and cognitive characteristics, as well as their usual interaction patterns with the parent, before the study. An autism expert then assessed whether the system would be suitable for each dyad and whether there were any potential negative outcomes associated with its use. Second, we provided parents with a demo video explaining how to use \sysname{} and asked for their feedback on whether it could be used with their child. We specifically requested that parents focus on whether their child could understand and engage with the system's flow rather than on their ability to use it fluently in conversation. After this process, three parents decided not to participate in the study.

Finally, a total of 11 parent-child dyads (P1--11, C1--11; a parent and a child with the same number denote the same dyad) gave their informed written consent and participated in the study, with no dropouts. While we designed the recruiting and screening questionnaires to target potential parent participants according to IRB guidelines, these questionnaires also asked parents to confirm their children's willingness to participate. \autoref{tab:demographic} summarizes the demographics of the participants, including the child’s verbal and cognitive characteristics as reported by the parents. The MVA child participants aged from 5 to 15 ($avg.=8.8$, $SD = 3.63$). Among the 11 children, ten were boys and two were girls. This gender distribution aligns with previous studies, which report a male-to-female ratio of 4--7.38:1~\cite{Simon2005,whiteley2010gender}. \revised{Three children primarily used echolalia, and the remaining eight were reported to verbalize 1–3 words mainly for requests, with little to no spontaneous speech in daily communication with their parents.}
The average age of parent participants was 42 years (ranged 43–51, $SD = 5.58$); 10 self-identified as the primary caregiving mother, and 1 as the secondary caregiving father. All parents were aware of AAC, with 3 dyads reporting its frequent use, 1 dyad reporting occasional use, and 7 dyads reporting no prior experience with AAC. Additionally, all dyads with previous AAC experience had used low-tech AAC, such as physical cards made by parents, rather than high-tech AAC on tablets or digital tools.

For compensation, we rewarded participants up to 200,000 KRW (approx. 150 USD) per their participation. We gave 20,000 KRW (approx. 15 USD) for attending the introductory session, 160,000 KRW (approx. 120 USD) for completing the deployment period, and an additional 20,000 KRW for a debriefing interview. To meet the minimum requirement for completing the deployment period, participants were advised to use the \sysname{} for at least nine out of the 14 days. This allowed participants to miss up to four days, providing some room for dyads to take time in getting used to the system. Additionally, we did not make it a mandatory condition to meet a certain number of conversations to encourage and observe naturally occurring conversations.

\subsection{Study Instrument}
We deployed a Samsung Galaxy Tab S9 tablet and a hardware button from Taotao Technology at each dyad's home. The tablet has an 11-inch AMOLED display with a 1600 $\times$ 2560 resolution (274 PPI). The button is round and has a diameter of 90mm with a height of 40mm. It is connectable to the tablet using a USB cable. To minimize the time for the initial setup, we configured the tablet and installed the \sysname{} app in advance. 

\subsection{Procedure}
Our two-week deployment study consisted of four phases: (1) pre-study preparation, (2) an introductory session, (3) the deployment, and (4) debriefing interviews.

\begin{figure*}[t]
    \centering
    \includegraphics[width=\textwidth]{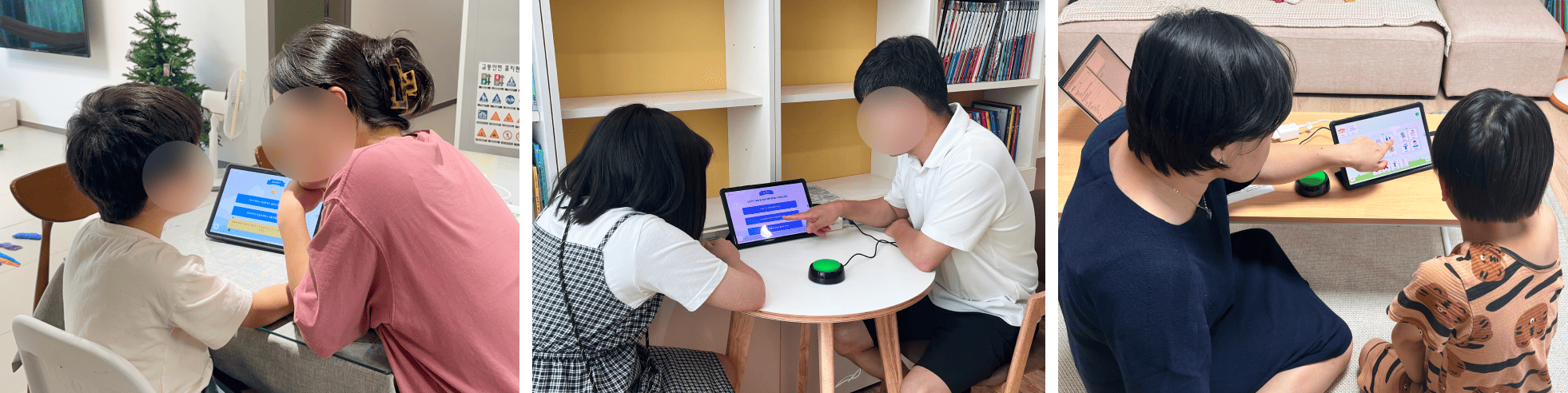}
    \caption[Three images showing parent and child participants engaging in conversation using AACessTalk during an introductory session.]{Parents and MVA child participants are engaging in conversation using the \sysname{} during the introductory session.}
    \label{fig:method:deploy}
\end{figure*}

\ipstart{Pre-study Preparation}
We first sent parents a video outlining our study goal, the overall procedure of the study, and the usage flow of \sysname{}. Then, through a remote survey link, we asked parents to share their child's interests to be included in the system's Interests section. We also requested photos of people, places, and objects familiar to the child to be used as custom AAC symbols. 

\ipstart{Introductory Session}
A few days after the remote survey, one researcher visited participants' homes to connect the tablet device to the home Wi-Fi and set up the device (see \autoref{fig:method:deploy}). To ensure participants fully understood how to use the system, the parent and MVA child participated in a pilot session, freely engaging in conversation using \sysname{}. During this session, the researcher ensured parent participants that \sysname{} is not a teaching aid but a communication aid, emphasizing not to pressure the child to use precise words, complete sentences, or to force the child to use it by physically placing the child's hand to make AAC card selection. \revised{We also obtained parents’ informed consent for speech recording and transcription, with transcriptions accessed by researchers only to verify recording accuracy.} The introductory session lasted about 30 minutes.

\ipstart{Deployment}
From the day of the introductory session, participants began using \sysname{} for the two-week period. \revised{We designed the deployment environment to establish a conversational routine at home, making it a natural and accessible part of daily life. Therefore,} we informed participants that they could use \sysname{} at any time of the day and have multiple conversations in a single day. They were also free to choose conversation topics and were not required to use each topic equally. \revised{Importantly, to prevent parents from becoming overly dominant when initiating conversations, we instructed them to have conversations only when both were prepared and comfortable. During conversations using \sysname{}, recognizing that AAC can facilitate a gradual transition from touch-based communication to verbal expression~\cite{Ralf2008}, we encouraged parents not to prevent their child's spontaneous speech or echolalia but to address it naturally, as they would in their usual interactions.} We collected all data related to participants' interactions with \sysname{}, including parents' voice input and the LLMs' output for parental guides and AAC cards.

Each morning, we sent a text reminder to remind them of their participation. If \sysname{} was used in the day, we sent the conversation logs in the evenings in plain text and asked parents to review the interactions through a survey. The survey included 5-point Likert scale questions on overall satisfaction with the conversation, the \revised{smoothness} of back-and-forth exchanges, and the child’s level of engagement, as well as open-ended questions for parents' self-reflection.

To see if \sysname{} may play a role in parents' sense of efficacy in parenting and their perceived burden in communicating with their child, we asked participants to complete an online survey three times: the day before the experiment, after Week 1, and after the deployment. The survey utilized the Parenting Sense of Competence Scale (PSOC)~\cite{gibaud1978} and the Family Empowerment Scale (FES)~\cite{Nirbhay1995}, each consisting of eight items on a 5-point Likert scale.

\ipstart{Debriefing}
The day after the 2-week deployment period, we visited each household to conduct a debriefing session that included completing surveys and a semi-structured interview with parent participants. The survey consisted of a total of 8 items on a 7-point Likert scale, including five items based on the Technology Acceptance Model (TAM)~\cite{Venkatesh2008} and three items assessing the appropriateness of the AI outcomes (AAC words, symbols, and parental guidance). To gather the voices of MVA children regarding their experience with \sysname{}, we provided a simplified version of the TAM survey on a 3-point scale (Yes, Maybe, and No)~\cite{Kory2019}. However, since we did not receive valid responses from all children participants, we do not include these results in the analysis.

Following this, parents participated in an hour-long interview with one researcher. The interview questions covered topics including the overall usability of \sysname{} and its impact on conversation patterns, the influence of the system's parental guidance and AAC recommendations, challenges encountered while using the system, and any changes in attitudes or perceptions of both parents and children in conversations. All interviews were audio-recorded and anonymized for analysis.

\subsection{Data Analysis}
Before analysis, we pre-processed the parent messages transcribed by the system by correcting typos and removing erroneous texts. We referred to the audio recordings of the parent turns to confirm the transcript.
We conducted a descriptive statistical analysis from the collected dialogue dataset to investigate the usage patterns of participants with \sysname{}. This includes the total number of conversation sessions and exchanges---defined as turn-taking counts of parent and child at a time---per session, the duration of each turn for parents and children, and the distribution of topics discussed. Following this, we assessed the types and frequencies of parental guides and feedback recommended by the system. To measure the adoption rate of these guides by the parents, one researcher confirmed whether each parent's utterance during their turn reflected the system's guides or example phrases. Another researcher reviewed these analyses for consistency, and discrepancies were discussed and resolved collaboratively. Additionally, we analyzed the usage patterns of AI-recommended vocabulary cards by the child participants, examining the frequency of selection by categories and the number of unique vocabularies recommended to each child. We also conducted an overlap coefficient analysis on the top 20 vocabulary words recommended to each child to examine the degree of personalization in the recommendations.

To assess the shift in the quality of conversations over time, we employed a mixed-effect model~\cite{Pinheiro2000MixedEffects} using the \texttt{lmer} package~\cite{lmer} in R to analyze three key indicators from daily surveys: overall satisfaction, the \revised{smoothness} of back-and-forth exchanges, and the child’s level of engagement. Additionally, we examined changes in parental self-efficacy using the Friedman test on survey data collected at pre-deployment, one-week, and post-deployment.

We further analyzed debriefing interviews and parents' reflections on conversations recorded in daily surveys using open coding and Thematic Analysis~\cite{Braun2006ThematicAnalysis}. All qualitative data were digitized using ATLAS.ti~\cite{atlas}, and the first author generated initial code themes. Another researcher then read the classified initial themes and quotes and responded with feedback. Based on this, the entire research team discussed any disagreements and iteratively revised the themes. Additionally, we applied open coding to dialogue of parent-child conversations to identify recurring patterns and interactions. These themes were used to support the analysis of the debriefing interviews. Through our comprehensive qualitative analysis, we found the impacts of \sysname{} on MVA child-parent conversation patterns and changes in parents' perceptions of conversation with their MVA children.
\section{Findings}
We present our findings on: (1) the overall use of \sysname{}; (2) conversational experiences with \sysname{}; and (3) changes in parents' perceptions of conversations with their MVA children over \sysname{} use.

\begin{figure*}[t]
    \centering     
        \includegraphics[width=\textwidth]{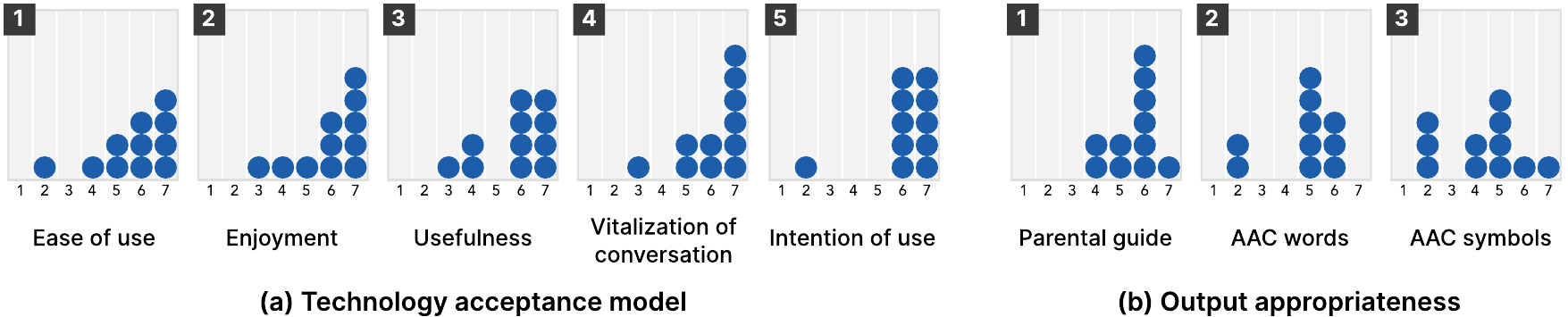}

        \labelphantom{fig:results:surveys:tam}
        \labelphantom{fig:results:surveys:output}
        \vspace{-4mm}
        \caption[Figure 6 displays two sets of bar graphs labeled (a) Technology Acceptance Model and (b) Output Appropriateness. Part (a) contains five bar graphs representing different aspects: Ease of Use, Enjoyment, Usefulness, Vitalization of Conversation, and Intention of Use, each numbered from 1 to 5 respectively. Part (b) includes three bar graphs titled Parental Guide, AAC Words, and AAC Symbols. Each graph has vertical columns numbered from 1 to 7, representing the rating scale, with blue circles plotted at various points within these columns to indicate individual parent ratings. For all scales, the higher number indicates a more positive rating.]{Distribution of parent participants' post-study ratings for the technology acceptance model (a) and output appropriateness surveys (b). Each circle represents the rating score of an individual parent participant. For all scales, the higher number indicates a more positive rating.}         \label{fig:results:surveys}
\end{figure*}

\subsection{Overall System Usage}
Over the two-week period, participants actively used \sysname{} for daily conversations. Five dyads out of 11 used the system every single day, and no dyad skipped more than three consecutive days. According to the 7-point scale TAM survey that measured acceptance of \sysname{}, parent participants gave positive evaluations, with scores of 5.64 ($SD = 1.57$) for ease of use, 5.91 ($SD = 1.38$) for enjoyment of use, and 5.73 ($SD = 1.42$) for the system’s usefulness in facilitating conversations (see \autoref{fig:results:surveys:tam}, \blackrectsmall{1}--\blackrectsmall{3}). Notably, parent participants rated the system favorably on whether the system vitalized their conversations and whether they are willing to continue using it, both giving a score of 6.09 ($SD = 1.30$ and $SD = 1.45$, respectively; see \autoref{fig:results:surveys:tam}, \blackrectsmall{4} and \blackrectsmall{5}). In this section, we describe how participants interacted with the \sysname{} based on system usage logs.

\begin{figure*}[b]
    \centering     
        \includegraphics[width=\textwidth]{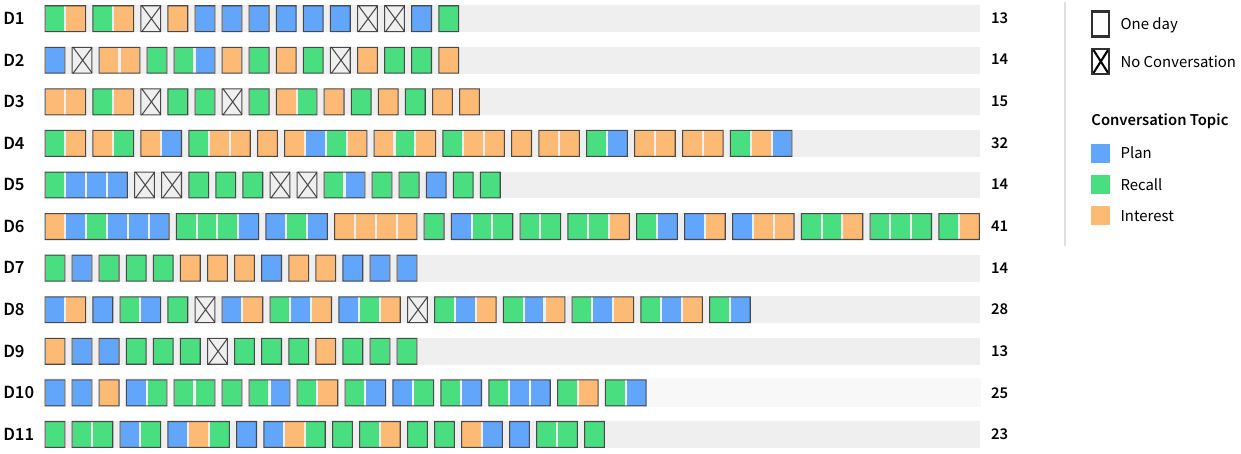}

        \labelphantom{fig:results:overview}
        \vspace{-3mm}
        \caption[Figure 7 displays a horizontal bar chart showing daily conversation sessions over a two-week period for 11 dyads, labeled D1 through D11. Each dyad's row consists of colored squares, each representing a day. Days with conversations are encased in black-bordered squares, and days without conversations are indicated by a black-bordered square containing an 'X'. The squares are colored blue, green, or orange, corresponding to conversation topics 'Plan', 'Recall', and 'Interest', respectively. On the right side of each row, the total number of conversation sessions is numerically displayed]{Overview of participant engagement with \sysname{}. The colored squares represent daily conversation sessions attempted by each dyad. Days with conversations are enclosed in squares with black borders, while days without conversations are marked with an 'X' inside the bordered squares. The color of each square indicates the topic of the conversation. On the right side of the chart, the total number of conversation sessions for each dyad is numerically displayed.}         
        \label{fig:results:timeline}
\end{figure*}

\subsubsection{Dialogue}
\autoref{fig:results:timeline} shows participants' daily engagement with the \sysname{} over the two-week period. The colored squares represent the conversation sessions attempted by each dyad, and the color of the square indicates the topic of the conversation. Participants engaged in a total of 232 conversation sessions, with an average of 21.09 sessions per participant ($min = 13$ [D1 and D9], $max = 41$ [D6]). On average, participants engaged in 1.55 conversations per day ($SD = 0.70$), with a maximum of 6 sessions in a single day [D6]. The conversation topics were fairly evenly distributed, with 63 sessions focused on ``Plan'', 99 on ``Recall'', and 70 on ``Interest''. The average duration of each session was 248 seconds (around 4 minutes) ($SD = 91.87 seconds$). Participants averaged 2.66 exchanges per session, with notable variance across dyads ($SD = 1.16$, $min = 1.27$ [D8], $max = 4.21$ [D6]). The highest number of exchanges in a single session was 8.5, recorded by D6. Given the few conversational exchanges participants had before the deployment, it appears that \sysname{} positively influenced both the duration of conversation and the number of exchanges. 

During the deployment period, parent participants took a total of 690 turns with speech and child participants took 608 turns using AAC cards. The average duration of a parent's turn was 31 seconds ($SD = 14.68$), whereas the child's turn averaged 63 seconds ($SD = 32.89$), indicating children spent more than twice as much time per turn as their parents. The average syllable count for parents' turns was 67.46 ($SD = 49.25$). The analysis of children's AAC-based communication will be discussed in Section 6.1.3.

\begin{table*}[]

\sffamily
\smaller
	\def\arraystretch{1.1}\setlength{\tabcolsep}{0.2em}
		    \centering
\caption[Table displaying 12 types of parental guidance, descriptions, and examples along with the total number of recommendation and its acceptance rate by parents. The types include 'Ask for Elaboration', 'Show Encouragement', 'Suggest Choices', 'Encourage Self-Disclosure', 'Ask for Intentions', 'Extend Topic', 'Open Up', 'Show Empathy', 'Pique Interest', 'Provide Clues', 'Suggest Coping Strategies', and 'Wrap Up'.]{Guide types and descriptions for parental guidance, including examples of recommended guides and sample phrases from the deployment period, along with the total number of recommendations and the parents' acceptance rate.}
~\label{tab:guides}

\begin{tabular}{|m{0.12\textwidth}!{\color{lightgray}\vrule}m{0.15\textwidth}m{0.58\textwidth}!{\color{lightgray}\vrule}m{0.04\textwidth}!{\color{tablegrayline}\vrule}c|}
\hline
\rowcolor{tableheader}
\textbf{Guide type} & \textbf{Definition} & \textbf{Example} & \textbf{Total} & \textbf{Accepted} \\ \hline
\textbf{Ask for \newline{} Elaboration} & Ask about ``what"\newline{}to specify the event. & \textbf{P6}: The Olympics have a lot of sports. Right now, archery is on TV. \newline{}Let's watch the archery first, then we can check out table tennis later. \newline{} Sound good? Can you cheer for Korea? \newline{}\textbf{C6}: \aaccardtable{Archery} \aaccardtable{Cheer up}\newline{}\newline{}\textbf{Ask your child what Olympic sport they like to watch.} \newline{} ``\textit{What sport do you like watching in the Olympics?}'' & 835 (11Ps) & 310 (37.1\%) \\ \hline

\textbf{Show\newline{}Encouragement} & Encourage the child's\newline{}actions or emotions. & \textbf{P9}: [child], at kindergarten, you made amazing things. Do you want to make more? \newline{} \textbf{C9}: \aaccardtable{Go} \aaccardtable{Eat} \aaccardtable{Go} \aaccardtable{Draw}\newline{}\newline{} \textbf{Show excitement about what the child plans to create.} \newline{} ``\textit{I can't wait to see what you’re gonna make!}'' & 236 (11Ps) & 59 (25.0\%) \\ \hline

\textbf{Suggest \newline{} Choices} & Provide choices for\newline{}children to select their answers. & \textbf{P2}: [child], wanna see ABCD? \newline{} \textbf{C2}: \aaccardtable{Alphabet}\newline{} \newline{} \textbf{Suggest some alphabets and ask which one the child likes.} \newline{} ``\textit{Which letter do you like the most, [child]? E or F?}''& 214 (11Ps) & 57 (26.6\%) \\ \hline

\textbf{Encourage\newline{}Self-Disclosure} & Ask about the child's\newline{}feelings and emotions. & \textbf{P11}: Now that you brushed your teeth and took a bath, should we go to bed? \newline{} \textbf{C11}: \aaccardtable{How about you, Mom?} \aaccardtable{How about you, Mom?}\newline{}\newline{} \textbf{Ask the child how they feel after the bath.} \newline{} ``\textit{How do you feel after the bath, [child]?}'' & 194 (11Ps) & 74 (38.1\%) \\ \hline

\textbf{Ask for \newline{} Intentions} & Check the intention\newline{}behind the child's\newline{}response and ask back. & \textbf{P11}:  Today, you're gonna meet your teacher with me, and then go study, okay?\newline{} \textbf{C11}: \aaccardtable{Mom} \aaccardtable{No}\newline{}\newline{}\textbf{Check if the child doesn't feel like studying today.} \newline{} ``\textit{Don't feel like studying today?}'' & 124 (11Ps) & 29 (23.4\%) \\ \hline

\textbf{Extend Topic} & Induce an expansion or\newline{}change of the conver-\newline{}sation topic. & \textbf{P3}: [child], why did you think of Dubai? Was it because it was your first time seeing it? \newline{} \textbf{C3}: \aaccardtable{Travel} \aaccardtable{Song} \aaccardtable{Airplane}\newline{}\newline{} \textbf{Talk with the child about other places childs might want to travel to.} \newline{} ``\textit{Is there anywhere else you’d like to go? Maybe Paris or London?}'' & 110 (11Ps) & 11 (10\%) \\ \hline

\textbf{Open Up} & Share the parent's\newline{}emotions and thoughts\newline{}in simple language. & \textbf{P3}: How does the automatic bathroom door sound? \newline{} \textbf{C3}: \aaccardtable{Sensor}\newline{}\newline{} \textbf{Talk about your experience with automatic bathroom doors.} \newline{} ``\textit{I get surprised by the automatic bathroom doors too!}'' & 90 (11Ps) & 19 (21.1\%) \\ \hline

\textbf{Show\newline{}Empathy} & Empathize with\newline{}the child's feelings. & \textbf{P5}: [child], where did you go with mom after lunchtime? Went to the hospital, right?\newline{} \textbf{C5}: \aaccardtable{Hospital} \aaccardtable{Doctor} \aaccardtable{Nurse} \aaccardtable{Scary} \aaccardtable{Bed} \aaccardtable{Afraid}\newline{}\newline{}\textbf{Let the child know it’s okay to be scared of the hospital.} \newline{} ``\textit{The hospital can be scary, right? You were so brave.}'' & 90 (10Ps) & 16 (17.8\%) \\ \hline

\textbf{Pique Interest} & To stimulate the\newline{}child's interest,\newline{}present information\newline{}that contradicts\newline{}what is known. & \textbf{P4}: Wanna talk about Pororo?\newline{} \textbf{C4}: \aaccardtable{Pororo} \aaccardtable{Play} \aaccardtable{Poby} \aaccardtable{Crong} \aaccardtable{Pororo} \aaccardtable{Yes}\newline{}\newline{} \textbf{Tell the child that Poby is taller than Pororo and ask if they noticed.} \newline{} ``\textit{Did you see that Poby is taller than Pororo?}'' & 48 (9Ps) & 9 (18.8\%) \\ \hline

\textbf{Provide Clues} & Give clues that can be\newline{}answered based on\newline{}previously known\newline{}information. & \textbf{P7}: Wanna talk about buses? \newline{} \textbf{C7}: \aaccardtable{Bus} \aaccardtable{Ticket} \aaccardtable{Tayo} \aaccardtable{Bus Stop} \aaccardtable{Open} \aaccardtable{No} \aaccardtable{Yes}\newline{}\newline{} \textbf{Mention the last time you took the bus together.} \newline{} ``\textit{Remember when we took the bus to the park last week?}'' & 29 (7Ps) & 11 (37.9\%) \\ \hline

\textbf{Suggest\newline{}Coping\newline{}Strategies} & Suggest coping\newline{}strategies for specific\newline{}situations to the child. & \begin{tabular}[c]{@{}l@{}}\textbf{P10}: [child], are you gonna watch the iPad tonight for study? \\ \textbf{C10}: \aaccardtable{Study} \aaccardtable{Read} \aaccardtable{Sad}\\ \\ \textbf{Suggest strategies to make studying more enjoyable.}\\ ``\textit{How about we read a book together instead?}''\end{tabular} & 9 (5Ps) & 4 (44.4\%) \\ \hline

\textbf{Wrap Up} & \multicolumn{2}{l!{\color{lightgray}\vrule}}{Inquire about the desire to end the conversation.} & 0 & 0 \\ \hline
\end{tabular}
\end{table*}
\begin{table*}[t]

\sffamily
\small
	\def\arraystretch{1.2}\setlength{\tabcolsep}{0.5em}
		    \centering
\caption[A table showing 3 types of feedback used in parental guidance, categorized into 'Complex', 'Blaming', and 'Overcorrection'. Each type includes a description, an example of suggested feedback during deployment study, and the total instances each type was provided. ]{Feedback types and descriptions for parental guidance, including examples from the deployment period and the total number of feedback provided.}~\label{tab:feedback}

\begin{tabular}{|l!{\color{lightgray}\vrule}m{0.25\textwidth}m{0.53\textwidth}!{\color{lightgray}\vrule}c|}
\hline
\rowcolor{tableheader}
\textbf{Feedback type} & \textbf{Description} & \textbf{Example} & \textbf{Total} \\ \hline
\textbf{Complex} & When a parent's dialogue contains\newline{}more than one goal or intent. & \textbf{P9}: It's Saturday! We're going to draw, visit the zoo, and have ice cream.\newline{}Do you know what we should talk about today? Should we start with the zoo?\newline{}\newline{}\textbf{Feedback}\newline{}\textit{Focus on one topic at a time and guide the child clearly\newline{}so they can easily understand and respond.} & 62 \\ \hline

\textbf{Blaming} & When the parent criticizes or negatively evaluates the child's response, like\newline{}reprimanding or scolding. & \textbf{P6}: Hmm, no... take a closer look. This isn't light blue. Look carefully and answer.\newline{}\newline{}\textbf{Feedback}\newline{}\textit{Create an environment where the child feels understood and supported.} & 39 \\ \hline

\textbf{Overcorrection} & When the parent is compulsively\newline{}correcting the child's response or\newline{}pointing out that the child is wrong. & \textbf{P5}: Let's do that again. We had chicken for dinner. Next time, pick chicken. \newline{}\newline{}\textbf{Feedback}\newline{}\textit{It seems like you're trying to correct the child's response. \newline{}Please support the child in expressing themselves in their own way.} & 29 \\ \hline
\end{tabular}
\end{table*}

\subsubsection{Parental Guides}
Parent participants received a total of 1,979 speech guides over two weeks ($M=179.82$, $SD=130.02$ per participant). \autoref{tab:guides} summarizes the guide types with example snippets and the acceptance ratio. On average, parents adopted 78\% of the LLM generated parental guides during their turns ($SD = 14$; $min=47\%$ [P1], $max=95\%$ [P7]). ``\textit{Ask for elaboration}'' was the most commonly provided guide type (835 out of 1,979; 42\%), followed by ``\textit{Show encouragement}'' (236 out of 1,979; 12\%) and ``\textit{Suggest choices}'' (214 out of 1,979; 11\%). 
During the deployment, the ``\textit{Wrap up}'' guide did not appear, potentially because that guide type was triggered only after three or more exchanges were made, but only 74 out of 232 conversations met this criterion. 
In debriefing, parent participants also mentioned that they often concluded the conversation upon the child's behavioral signals.

As for the dialogue inspection feedback, parent participants received a total of 92  feedback alerts ($M = 8.36$, $SD = 8.62$ per participant; $min=0$ [P2], $max=16$ [P6]). \autoref{tab:feedback} summarized the feedback types, example snippets, and the number of occurrences. The feedback was most frequently related to complex language (47\%), followed by blame (30\%) and correction (29\%). Additionally, parent participants rated the guides and examples as useful for conversations, with an average score of 5.55 ($SD = 0.93$) on a 7-point scale survey (see \autoref{fig:results:surveys:output}-\blackrectsmall{1}).

\subsubsection{Children's Use of Cards}
Child participants selected a total of 2,244 vocabulary cards recommended by \sysname{} to engage in conversations. On average, they picked 9.74 cards per session, with significant individual differences ($SD = 4.73$, $min = 2.21$, $max = 17.23$). Children used Topic cards for more than half of their conversations (51.92\%). The remaining categories—Emotion (18.63\%), Core (14.80\%), and Action (14.66\%)—were fairly evenly distributed, though usage patterns varied significantly between participants, as shown in \autoref{tab:result:AAC}. Among the Core words, \aaccard{Yes} and \aaccard{No} accounted for about half (55.42\%) of the usage, followed by \aaccard{I don't know} (25\%) and \aaccard{How about you, Mom/Dad?} (19.58\%).

\begin{table}[b]

\sffamily
\small
\def\arraystretch{1.3}\setlength{\tabcolsep}{0.25em}
\centering
\caption[A table showing the number of vocabulary cards used by each MVA child participant, divided into four categories: Topic, Action, Emotion, and Core. Columns labeled C1 through C11 represent different participants, with the total usage summarized at the end of each row. The cells are color-coded to indicate the proportion of word category usage, with shades ranging from light to dark green corresponding to 0\% to 100\% usage.]{The number of vocabulary cards used by each MVA child participant for each word category. The intensity of the cell color indicates the ratio of each word category used by the participants.}~\label{tab:result:AAC}
\begin{tabular}{|l!{\color{lightgray}\vrule}c!{\color{lightgray}\vrule}c!{\color{lightgray}\vrule}c!{\color{lightgray}\vrule}c!{\color{lightgray}\vrule}c!{\color{lightgray}\vrule}c!{\color{lightgray}\vrule}c!{\color{lightgray}\vrule}c!{\color{lightgray}\vrule}c!{\color{lightgray}\vrule}c!{\color{lightgray}\vrule}c|c|}

\hline
\rowcolor{tableheader}
\textbf{Categories} & \textbf{C1} & \textbf{C2} & \textbf{C3} & \textbf{C4} & \textbf{C5} & \textbf{C6} & \textbf{C7} & \textbf{C8} & \textbf{C9} & \textbf{C10} & \textbf{C11} & \textbf{Total} \\
\hline
\cellcolor{white}\textbf{Topic} & \cellcolor[HTML]{3A9F7D}45 & \cellcolor[HTML]{359A7B}21 & \cellcolor[HTML]{3FA47F}42 & \cellcolor[HTML]{3EA37F}165 & \cellcolor[HTML]{389C7C}128 & \cellcolor[HTML]{309578}280 & \cellcolor[HTML]{51B587}63 & \cellcolor[HTML]{34997A}252 & \cellcolor[HTML]{ABDDC4}33 & \cellcolor[HTML]{4EB386}76 & \cellcolor[HTML]{95D5B6}40 & \cellcolor{white}1,165 \\
\cellcolor{white}\textbf{Action} & \cellcolor[HTML]{B7E2CD}9 & \cellcolor[HTML]{FFFFFF}0 & \cellcolor[HTML]{AADDC4}11 & \cellcolor[HTML]{D5EEE2}21 & \cellcolor[HTML]{9AD7B9}40 & \cellcolor[HTML]{DAF0E5}25 & \cellcolor[HTML]{9BD7BA}30 & \cellcolor[HTML]{DDF1E7}22 & \cellcolor[HTML]{7AC9A2}52 & \cellcolor[HTML]{4AAF84}85 & \cellcolor[HTML]{A5DBC1}34 & \cellcolor{white}329 \\
\cellcolor{white}\textbf{Emotion} & \cellcolor[HTML]{D7EFE4}5 & \cellcolor[HTML]{DAF0E6}2 & \cellcolor[HTML]{7BCAA3}17 & \cellcolor[HTML]{B1E0C9}39 & \cellcolor[HTML]{9DD8BB}39 & \cellcolor[HTML]{B4E1CB}50 & \cellcolor[HTML]{6CC499}44 & \cellcolor[HTML]{B3E1CA}48 & \cellcolor[HTML]{3DA27E}131 & \cellcolor[HTML]{B5E1CC}25 & \cellcolor[HTML]{D0ECDE}18 & \cellcolor{white}418 \\
\cellcolor{white}\textbf{Core} & \cellcolor[HTML]{97D5B7}13 & \cellcolor[HTML]{6BC398}8 & \cellcolor[HTML]{E0F3EA}4 & \cellcolor[HTML]{85CEAA}61 & \cellcolor[HTML]{FDFEFE}1 & \cellcolor[HTML]{D4EEE1}29 & \cellcolor[HTML]{8AD0AE}35 & \cellcolor[HTML]{BEE5D2}41 & \cellcolor[HTML]{EBF7F1}8 & \cellcolor[HTML]{EBF7F1}7 & \cellcolor[HTML]{3EA37F}125 & \cellcolor{white}332 \\\arrayrulecolor{darkgray}\hline
\rowcolor{white}\textbf{Total} & 72 & 31 & 74 & 286 & 228 & 384 & 172 & 363 & 224 & 193 & 217 & 2,244 \\
\arrayrulecolor{black}\hline
\end{tabular}
\null\hfill\includegraphics[width=3.5cm]{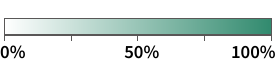}
\end{table}

Excluding the Emotion and Core categories with fixed word pools, the \sysname{} recommended a total of 4,324 unique vocabulary cards to the children, with 2,673 unique Topic cards and 1,700 unique Action cards. On average, each conversation session featured 11.9 unique Topic cards ($SD = 2.22$) and 7.7 unique Action cards ($SD = 1.31$), exposing children to an average of 19.3 unique cards ($SD = 3.13$) per session. The results of the overlap coefficient analysis of the top 20 cards recommended to each child revealed an average overlap of 33\% on the first turn. This indicates that while the recommendations started relatively generally, they still showed a significant degree of personalization, likely influenced by parents' initial input. As the turns progressed, the overlap decreased to 25\% on the second turn, 14\% on the third turn, and 10\% on the fourth turn. This declining trend suggests that the system could tailor its recommendations as the conversation unfolded, resulting in more contextualized selections and less overlap among participants.

Nevertheless, parent participants rated the appropriateness of the recommended words at an average of 4.73 ($SD = 1.42$; see \autoref{fig:results:surveys:output}-\blackrectsmall{2}) and the appropriateness of the symbols at an average of 4.27 ($SD = 1.68$; see \autoref{fig:results:surveys:output}-\blackrectsmall{3}), which are relatively lower scores compared to the overall satisfaction with the system. In the debriefing interviews, parents attributed these lower ratings to inaccuracies in speech recognition, which led to incorrect word recommendations, and suggestions that didn’t align with the Korean cultural context. They also highlighted the need for more individualized AAC recommendations that reflect the child’s cognitive and developmental levels, as well as their daily events.

\subsection{Conversational Experiences with \sysname{}}
\begin{figure*}[]
    \centering
     \begin{subfigure}[t]{0.35\textwidth}
         \centering
         \includegraphics[height=4.8cm]{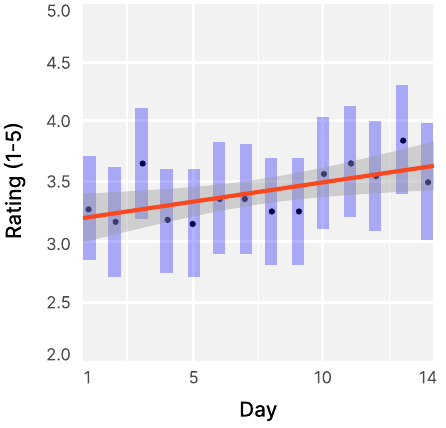}
         \caption{Overall satisfaction \revised{with the conversation}}
     \end{subfigure}
     \begin{subfigure}[t]{0.3\textwidth}
         \centering
         \includegraphics[height=4.8cm]{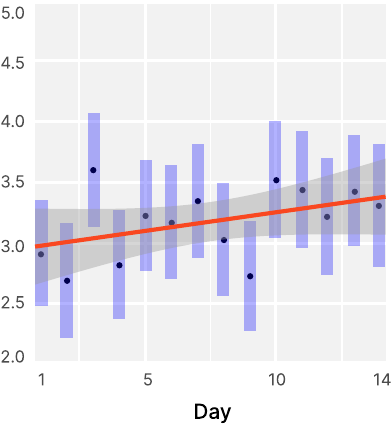}
         \caption{\revised{Smoothness of t}urn-taking}
     \end{subfigure}
     \begin{subfigure}[t]{0.3\textwidth}
         \centering
         \includegraphics[height=4.8cm]{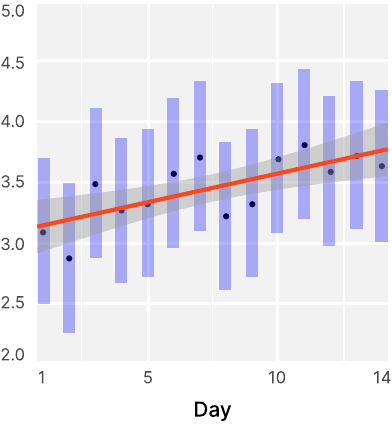}
         \caption{\revised{Level of c}hild engagement}
     \end{subfigure}
   \caption[Three line graphs depicting daily trends over a 14-day period in (a) overall satisfaction, (b) turn-taking, and (c) child engagement. Each graph displays purple bars to visualize the data, with a red trend line generated using a linear model to highlight the overall trend. The x-axis labels the days from 1 to 14, while the y-axis ranges from 2.0 to 5.0.]{Daily trends in \revised{the estimated marginal means of parent's self-evaluations of} (a) overall satisfaction, (b) \revised{the smoothness of} turn-taking, and (c) \revised{the level of} child engagement over a 14-day period\revised{, after controlling the random effect of individuals using mixed-effect models. The blue bars indicate the 95\% confidence interval.}}
    \label{fig:dailysurvey}
\end{figure*}

The daily conversation survey results from parent participants showed that the MVA child-parent conversation improved not only in frequency but also in the quality of their interactions. Over the two-week period, overall satisfaction with conversations ($p = 0.006^{**}$), \revised{the smoothness of} turn-taking ($p = 0.028^{*}$), and especially the \revised{level of} childs' engagement ($p < 0.0001^{***}$) all showed statistically significant increases (see~\autoref{fig:dailysurvey}). Based on the conversation logs and interview analysis, we illustrate how \sysname{} facilitated parent-child conversations.

\subsubsection{Enriched Conversational Patterns}
\sysname{}’s guides provided parents with ideas for responding in varied ways, even in repetitive situations. All parent participants mentioned that their previous daily conversations with their child had become routine and fixed. P5 described, \textit{``My child’s routine is pretty much the same every day, so I always end up asking, `What did you do at school today?' or `What did you eat for lunch?' And my child always answers, `It was fun' or `Pasta.' But sometimes that meal wasn’t even served. They’re just saying what they’ve memorized.''} The parental guides led parents to move beyond the habitual ``what'' questions, encouraging them to explore the various response strategies. In doing so, parents gradually developed a way to communicate for their children to easily understand and enjoy engaging with. P8 remarked, \textit{``I'd read about praising and empathizing in books, but this was my first time trying it. I thought the conversation would just end if I said something like that. But when I said, `Wow, that’s awesome,' my child’s face lit up, and they got more engaged. So I kept reacting, the conversation kept going, and it felt like we were really talking.''}

The increased self-expression of the child through AI-driven AAC was another factor that made the conversations more vibrant. Unlike previous AAC experiences with the limited words provided by parents, \sysname{} continuously introduced new words to the children. While children didn't understand every word, they often used words in context that their parents had assumed they wouldn’t know. This naturally led to a wider variety of conversation topics and more dynamic responses from parents. P6 mentioned, \textit{``I kept saying, `You know that word?' When he went on a field trip, he saw a drone and pressed} \aaccard{scared}. \textit{I didn’t think he knew the word `scared.' But then, remembering that my child is usually afraid of heights, they pressed the} \aaccard{scared} \textit{button with an uncomfortable expression. He knew what it meant! When I realized he knew more than I thought, I suddenly had so much more to talk about.''} (see \autoref{dial:newword}) \revised{Several parents (P3, P4, P6, P9, and P10) even reported that by the later stages of using \sysname{}, their children began repeating the AI-generated voice after selecting a card, significantly increasing their imitative speech.}

\begin{dialogue}[h]

\sffamily
\small
\def\arraystretch{1.4}\setlength{\tabcolsep}{0.5em}
\centering

\begin{tabular}{cl}
\hline
\arrayrulecolor{gray!50} 
\multicolumn{1}{l}{} & {[}...{]} \\ \hline
\textbf{P6} & The drone flew up into the sky. How did you feel while flying it? \\ \hline
\textbf{C6} & \aaccardtable{Balloon} \aaccardtable{Sky} \aaccardtable{Button} \aaccardtable{Sky} \aaccardtable{Scared} \\ \hline
\textbf{P6} & Scared? Were you scared flying the drone, {[}child{]}? \\ 
\arrayrulecolor{black}
\hline
\end{tabular}

\caption{Dialogue snippet from P6 and C6.}~\label{dial:newword}
\end{dialogue}

\subsubsection{Serendipitious Parent-child Interaction}
Some parents (P1, P2, P7, P8, and P11) occasionally questioned if their children's expression through \sysname{} included communicative intention. They mentioned that the child seems to treat the system as an engaging activity or game, with words and symbols continually appearing, and chooses AAC cards to listen to and mimic the sounds. Early on, parents perceived their children were not fully engaging in conversation. However, they began to see the \sysname{} sessions as opportunities to initiate interaction. For example, some parents searched for and explained the words their child selected (P3, P4, and P9) or incorporated the device into playtime, like discussing fruits with \sysname{} while playing with fruit-shaped clay (P1; see \autoref{dial:play}).

\begin{dialogue}[h]

\sffamily
\small
\def\arraystretch{1.4}\setlength{\tabcolsep}{0.5em}
\centering

\begin{tabular}{cm{0.9\columnwidth}}
\hline
\arrayrulecolor{gray!50} 
\textbf{P1} & {[}child{]}, let’s play with the fruits today. What’s the red fruit? And the yellow fruit like lemons, mangoes, and pears.Yellow corn, purple grapes, purple sweet potato, purple eggplant, and purple onion. \\ \hline
\textbf{C1} & \aaccardtable{Fruit} \aaccardtable{Grapes} \aaccardtable{Red fruit} \aaccardtable{Fruit game} \aaccardtable{Purple fruit}\newline{} \aaccardtable{Orange fruit} \aaccardtable{Broccoli} \aaccardtable{Yellow fruit} \aaccardtable{Purple fruit} \\ \hline
\textbf{P1} & It's so juicy! {[}child{]}, let's sort them by color. \\ \hline
\textbf{C1} & \aaccardtable{Yes} \\ \hline
\textbf{P1} & Wow, you sorted the fruits and veggies so well. Look at that! These are the fruit friends, and those are the veggie friends. Is this a fruit or a veggie? Cabbage? Is cabbage a veggie? {[}child{]}, which fruit do you like the most? \\ \hline
\textbf{C1} & \aaccardtable{Fruit} \aaccardtable{Fruit shop} \aaccardtable{Fruit friends} \\ 
\arrayrulecolor{black}
\hline
\end{tabular}

\caption{Dialogue snippet from P1 and C1.}~\label{dial:play}
\end{dialogue}

Moreover, the unexpected AAC use by children provided parents with a sense of enjoyment in their conversation. During our introductory session, every parent mentioned that their child likely would not use the \aaccard{How about you, Mom/Dad?} card, as they had never asked such a question before. However, all children except C5 used this card in their conversations, a total of 65 times ($M = 5.9, SD = 9.67$). Although parents were uncertain about children’s intent in using the card, they were all pleasantly surprised by the experience of being asked a question by their child and happily shared their own feelings and thoughts (see \autoref{dial:enjoyment}): \textit{``Usually, I'm the one asking all the questions, and it gets pretty lonely. But when [child] used the \aaccard{How about you, Mom?} card, it was so sweet and touching. It really felt like [child] cared. I got all excited and, for the first time, shared my own story with my child. It felt like we were in conversation `together,'  like we were connected. [P11]''} 


\begin{dialogue}[h]

\sffamily
\small
\def\arraystretch{1.4}\setlength{\tabcolsep}{0.5em}
\centering

\begin{tabular}{cm{0.87\columnwidth}}
\hline
\arrayrulecolor{gray!50} 
\textbf{P11} & [child], you just take a bath with mommy. How do you feel? \\ \hline
\textbf{C11} & \aaccardtable{How about you, mom?} \aaccardtable{How about you, mom?} \\ \hline
\textbf{P11} & How about me? I feel so happy because I took a bath with you. \\ \hline
\textbf{C11} & \aaccardtable{Feelings} \aaccardtable{No} \\ \hline
\textbf{P11} & You're doing great. You're even talking with mommy, [child]. Mommy is so happy to take a bath with you like this. Give me a kiss. Love you so much. \\ \hline
\textbf{C11} & \aaccardtable{How about you, mom?} \aaccardtable{Happy} \aaccardtable{How about you, mom?}\newline{} \aaccardtable{Happy} \\ 
\arrayrulecolor{black}
\hline
\end{tabular}

\caption{Dialogue snippet from P11 and C11.}~\label{dial:enjoyment}
\end{dialogue}

\subsubsection{Reflection in Conversation}
Parent participants valued the AI-generated feedback on their speech as it allowed them to reflect on their conversational patterns and consciously adjust them in the next turn. Most parent participants (all except P5) did not have experience receiving feedback on their dialogue with their child from experts or others. P4 remarked, \textit{``Speech therapists usually talk a lot about the child but not so much about ME. Most of the parents are already doing their best, so criticizing them might just overwhelm them. It can feel like a stab to the heart and could even lead to emotional issues.''} Parents also appreciated that the feedback was provided by AI rather than a person, as it enabled them to digest feedback with less emotional burden and make their own decision on whether to accept it.

Some parents (P4, P6, and P10) even reflected on their speech patterns through the AAC cards that the AI recommended to their child: \textit{ ``After a few days, I realized that to make [child] express himself better, I need to ask specific questions. If I ask something broad like, `What do you remember most about today?' his cards show sorts of unrelated stuff. Seeing the cards, I realized my questions made it tough for him to think of responses. So I made it more specific, like, `What did you play with at the pool?'\null''} The AI's recommendation of AAC cards mirrored the parents' speech patterns and served as implicit feedback, leading parents to self-assess and adjust their speech.

\begin{table*}[b]
\sffamily\small
\def\arraystretch{1.5}\setlength{\tabcolsep}{0.2em}
\centering
\caption[Questions 1 through 4 are positively rated with higher numbers, whereas Questions 5 through 7 are negatively impacted by higher numbers. Significant changes are marked with asterisks next to the p-values, indicating statistically significant differences.]{Parental efficacy survey questions with average ratings from parents at Pre-study, After Week 1, and Post-study periods. Results of the Friedman test are reported as p-values. Q1-Q4 indicate that a higher number represents a positive rating, whereas for Q5-Q7, a lower number signifies a positive rating.}~\label{tab:result:efficacy}

\begin{tabular}{
    |>{\centering\arraybackslash}p{0.03\textwidth}  
    p{0.55\textwidth}                                
    >{\centering\arraybackslash}p{0.08\textwidth}   
    >{\centering\arraybackslash}p{0.10\textwidth}   
    >{\centering\arraybackslash}p{0.08\textwidth}   
    >{\centering\arraybackslash}p{0.07\textwidth}   
    >{\centering\arraybackslash}p{0.04\textwidth}|  
}

\hline
\rowcolor[HTML]{ECECEC} 
\cellcolor[HTML]{ECECEC}\textbf{\#} & \cellcolor[HTML]{ECECEC}\textbf{Questionnaire} & \textbf{Pre-study} & \textbf{After Week 1} & \textbf{Post-study} & \textbf{\textit{p}-value} & \textbf{sig.} \\ \hline
Q1 & I feel confident in supporting my child's growth and development. & 3.09 & 3.9 & 4.00 & \textbf{0.0259} & * \\ \hline
Q2 & I can effectively resolve issues related to my child. & 3.18 & 3.82 & 3.91 & 0.0545 & - \\ \hline
Q3 & I make efforts to learn new approaches to support my child's growth and development. & 4.45 & 4.55 & 4.73 & 0.1738 & - \\ \hline
Q4 & I understand my child’s challenges better than anyone else. & 3.91 & 4.09 & 4.00 & 1.0000 & - \\ \hline
Q5 & I feel frustrated because my child does not follow my guidance and instruction. & 4.27 & 3.36 & 2.82 & \textbf{0.0027} & ** \\ \hline
Q6 & I worry that I might not be a good parent. & 3.00 & 2.36 & 2.27 & 0.0545 & - \\ \hline
Q7 & Talking to my child makes me feel anxious and tense. & 2.64 & 2.09 & 2.18 & 0.4578 & - \\ \hline
\end{tabular}
\end{table*}

\subsubsection{Shared Control Between Parent and MVA Child}
\sysname{} served as a channel for MVA children to take the lead and express their intentions within the conversation flow. Half of the children (C3, C4, C6, C8, C9, and C11) initiated communication by handing the tablet with their preferred conversation topic open to their parents. Some gently expressed their desire to end the conversation by finding and pressing the hidden ``End Conversation'' button (C4, C5, C10) or by closing the tablet case and saying  \textit{``End''} (C6). Notably, the visually distinct screens for the parent's and child's turns make the children better understand the concept of turn-taking by showing them when to wait and when to express themselves.

We found that the parents also developed a better sense of turn-taking through \sysname{}. By observing their children carefully read and select AAC cards, parents could recognize that the child's turn was not yet over. P3 noted, \textit{ ``Normally, if my child didn’t answer right away, I’d just jump in with the answer. But using \sysname{}, I noticed [child] was actually thinking what to say. And when he didn’t pick cards right away, I went to press the button, but he stopped me. That’s when I realized, `Oh, he's talking. I just need to wait.' It made me see I need to respect my kid's turn.''}

\subsection{Parent Perceptions of Conversations with \sysname{}}
Based on the analysis of parents' debrief interviews, we observed that the consistent use of \sysname{} over two weeks positively influenced parents' perceptions and attitudes toward conversations with their MVA children.

\subsubsection{It's Okay to Have Imperfect Conversations}
The results of the parenting efficacy surveys conducted before the deployment, on day 7, and on day 14 showed a general trend of positive change (see \autoref{tab:result:efficacy}). Notably, there was a statistically significant increase in parents’ confidence in supporting their child’s development ($p = 0.026^{*}$) and a statistically significant decrease in feelings of frustration when the child did not follow their guidance as expected ($p = 0.003^{**}$). While it is difficult to directly attribute these changes to the short two-week system usage, we also found evidence of shifts in parenting efficacy and parents' perspectives on conversations through the debrief interviews.

Using \sysname{}, parents experienced a release from the pressure and burden they once had in conversations with their children. Most parents felt responsible for guiding their children to perfectly match subjects and predicates and use contextually appropriate language in conversations. Parents' emphasis on refining their child’s expressions was noticeable in the early stages of using \sysname{}. P1 pointed out, \textit{``The screen has topics, actions, and emotions on it. I felt like my child needed to pick them in order. So I guided his hand to press the cards, but he didn’t like it and stopped using it (\sysname{}). I backed off and let him press what he wanted. Then, [child] mostly picked words from the topic section, but I could still understand what he meant. [...] If we both get it and have fun, that's the conversation. It's just important to keep going.''} Parents realized there is no right or wrong in conversation and tried to make it a natural, enjoyable part of daily routine.

\subsubsection{Moving Beyond Autism Labels}
As parents reviewed the daily conversation logs each evening, many of them reflected on the stereotypical thoughts they had previously held that conversations with autistic children need to be short and simple. P11 remarked, \textit{``Funny how I realized the things I've been saying here [\sysname{}] are just like what I used to tell my child every day before the autism diagnosis. He was just a baby then, but I talked to him so much. But after the diagnosis, I stopped and only gave simple commands, thinking that conversations like this wouldn't be possible. But when we tried, it turned out we could do it. I was the one who was trapped in this mindset while my child was growing in their own way all along.''} Parents who also have non-autistic siblings (P4 and P6) or twins (P2 and P8) were surprised to notice that they had never spoken to their autistic child in the same way they did with their non-autistic children. This reflection led parents to resolve to provide their MVA children with a wider range of topics and stimuli, breaking from self-imposed limits on daily conversations.
\section{Discussions}
In this section, we reflect on how parents and MVA children adopted \sysname{} as a mediating tool for conversation through a collaborative meaning-making process. We also discuss how the system usage log can be utilized to better understand MVA children. We then propose broader design implications for supporting long-term interaction. We also share the challenges and strategies of running a deployment study with MVA children. Lastly, we discuss the study limitations that could impact the generalization of the findings.

\subsection{Collaborative Meaning-making about Conversation Tool}
\sysname{} provided a scaffolded conversation structure through visual and behavioral cues that helped both MVA children and their parents naturally develop key communication skills such as initiating conversations, expressing themselves, taking turns, waiting for others to speak, and signaling when they want to stop. MVA children, in particular, used various components of the system to actively express their communicative intent. For example, they would bring the tablet or choose a favorite topic to initiate conversation, use the \aaccard{I don't know} card to indicate confusion, and press the ``End Conversation'' button to conclude the conversation. Beyond these explicit expressions of intent, parents also noticed implicit cues--such as the child carefully examining vocabulary cards or their facial expressions while selecting one--which encouraged them to respect the child's agency in the conversation.

More importantly, each dyad developed unique routines for using the system and assigned personalized meanings to its elements to express communicative intent. 
\revised{In the early stages of deployment, parental involvement played a significant role in initiating conversations using \sysname{}, as is common in parent-child studies~\cite{siller2002}. However, the flexible conversation settings, locations, and features of \sysname{} allowed each dyad to collaboratively shape the system's role over time and identify environments that matched their mutual conversational needs.} For example, some used it for bedtime reflection, while one dyad incorporated it during playtime, and others used it to study new words that the child showed interest in. Also, some dyads employed the turn pass button for end their turn, while others had the child prevent the parent from pressing the button as a way to express refusal. 

In the adoption of technology by autistic children, Spiel et al. pointed out that neurotypical adults often define the meaning and purpose of the technology in a one-sided manner, limiting children's ability to articulate their thoughts within narrow boundaries set by the adults~\cite{Spiel2019}. They argue that autistic children should have agency in deciding which forms of interaction are meaningful to them~\cite{alper2017}. Similarly, in our study, we observed some parents initially guiding the child's system use by directing their hands to select specific cards or pressing the turn-taking button on their behalf. While this parent-directed approach often led to the child refusing to engage, it prompted parents to reflect and begin respecting the child's intent. However, the risk of parents unconsciously imposing the use of the communication tool remains. Thus, we emphasize the need for a space where children can freely explore the tool and identify interactions that suit them in a self-directed manner, especially during the early stages of its introduction. In doing so, parents should not act as instructors but as equal communication partners, collaboratively defining which forms of use foster meaningful and mutually satisfying conversations.

\subsection{Toward Better Understanding of MVA Children through Data}
We found that \sysname{} allows parents to better understand their MVA children's language use characteristics and capabilities. By leveraging the generative power of an LLM, \sysname{} continuously provides children with diverse language input across various categories of topics, actions, and emotions. This stimulates children to use vocabulary they may know but have had little opportunity to express. Parents are often surprised by their child's unexpected language use like, \textit{``Do you know this word (or emotion)?''} This, in turn, encourages parents to introduce richer topics and vocabulary, setting new goals for language development.

Parents often struggle with a lack of objective data to assess the communicative and language capacities of MVA children~\cite{Adam2023}. The assessment of these abilities varies greatly, depending on the perspectives and methods used by individual professionals~\cite{Daniela2016}. Moreover, clinical feedback typically emphasize a child's weaknesses, as the primary goal is to improve communication~\cite{Matthew2017}. In this context, \sysname{} has helped parents gain a clearer understanding of the words and symbols their child comprehends and uses, as well as the types of words and combinations they are skilled at expressing through the card selection process. Our analysis of children's vocabulary card usage showed significant variation in the categories of words used for communication. While some children had over 50\% of their expression on topic-related words, others used all four categories more evenly, and some expressed emotions in over 50\% of their interactions.

Building on these findings, we envision that a future version of \sysname{} can empower parents to better understand MVA children through data. By providing parents with insight into the words their child is exposed to and patterns in their word usage, \sysname{} could offer a more objective view of the child’s communicative abilities. This would allow parents to identify their child's strengths and characteristics, enabling them to set more informed interaction and educational goals in collaboration with experts. Furthermore, gaining deeper insight into the child's unique traits can increase the sense of parental efficacy by empowering parents to better support their child's growth and development.

However, as our findings revealed, not all words were fully comprehended or deliberately used by the MVA children. As a result, parents often inferred their child's communicative intent through their facial expressions or familiar interests. By reviewing conversation logs through the daily survey, parents reflected on their child's expressions within the broader context. They also evaluated the effectiveness of their own communication strategies and parenting approaches, identifying ways to better understand and support their child's expression. Building on this, we suggest introducing systemic nudges that encourage parents to reflect on both their children's word usage and their own communication styles, fostering more meaningful interactions.

\subsection{Design Considerations for Long-term Interaction}
Our findings showed that \sysname{} increased the frequency of conversations and turn-taking between parents and MVA children, enabling more meaningful interactions. Furthermore, by addressing the unique needs of both parties, the system reduced the parents' burden while encouraging children's self-expression, fostering mutual engagement. Notably, changes in daily conversation practice and parents' attitudes resulted in 10 out of 11 parents expressing their desire to continue using the system. However, our findings also revealed potential challenges that need to be addressed to support long-term conversations as MVA children grow.

First, the cognitive and language stages of MVA children, as well as their pace of development, vary significantly between individuals~\cite{Odermatt2022}. 
While we customize LLM prompts based on the characteristics of MVA children to set appropriate vocabulary levels, many parents articulate a need for more personalized vocabulary recommendations tailored to their child's language capabilities. A discrepancy between MVA children's developmental stage and the recommended vocabulary can reduce motivation and engagement, potentially hindering genuine self-expression. To address this, the system could continuously update the vocabulary recommendations by regularly assessing each child’s communicative capabilities using clinical measures, such as ``MacArthur-Bates Communicative Development Inventories~\cite{fenson2007}'' or the ``Expressive Vocabulary Test~\cite{williams1997}'' In addition to these clinical measures, the system could analyze previous conversation sessions to extract specific data, such as topics, content, and the frequency of specific words used, on communication characteristics of MVA children. This information could be stored in the LLM’s long-term memory~\cite{Bae2022, Wang2023} and used in future sessions to offer tailored recommendations to each child’s language use. \revised{To implement such approaches while considering user privacy issues, measures such as providing options for users to exclude sensitive topics or contexts from being recorded or analyzed can be included. Additionally, while our study involved transcription to allow researchers to verify recording accuracy afterward, future systems could adopt technologies such as speaker separation for more accurate real-time speech recognition~\cite{Subakan2021}, ensuring that no sensitive information is unnecessarily stored.}

Secondly, 
previous research has highlighted that AAC systems designed for autistic users often rely on the communicative norms and practices of neurotypical frameworks~\cite{Janna2022, Mankoff2010}. In our study, the LLM used to generate parental guidance and vocabulary cards primarily reflected neurotypical perspectives, which constitute the majority in our society~\cite{Gadiraju2023}. This raises concerns that the LLM's recommendations may potentially enforce neurotypical standards and normative behaviors on MVA children, limiting their communicative agency~\cite{Choi2024}. Therefore, further studies are needed to consider the unique characteristics of neurodiverse populations throughout the entire LLM-mediated communication process. For example, this could involve fine-tuning or integrating recommendation logic trained on neurodiverse communication datasets. Additionally, the system could broaden communication modalities to include sensory, physical, and tangible elements, integrating more natural signals from MVA children into conversations.


\subsection{Conducting a Deployment Study with MVA Children}
\rerevised{Since our study involved MVA children---who are often unfamiliar with both technology and conversational interactions---we had to address several challenges in conducting a home-based deployment study with minimal researcher guidance. Here, we share the challenges we encountered and expert-guided strategies we implemented to address them.}

\rerevised{First, during the recruitment phase, we found that many parents of MVA children viewed our AAC-integrated system as a language education tool, as suggested by our formative study. To align parents' expectations of \sysname{} with our goal of fostering natural conversations, we conducted two phone calls with applicants to clarify the study’s purpose in detail and ensure their understanding and agreement.
}

\rerevised{Second, providing a detailed system walkthrough during the introductory session was impractical due to the involvement of MVA children in a home setting. To keep the session concise, we provided pre-recorded instructional videos to parents in advance. Additionally, we installed TeamViewer\footnote{https://www.teamviewer.com/en/} remote control app on the tablets, enabling researchers to remotely troubleshoot technical issues during deployment (\eg, installing an updated version of the app with bug fixes).} 

\rerevised{Lastly, in our formative study, both parents and experts pointed out the lack of motivation among MVA children to engage with \sysname{}, which may create a burden for both parents and children as they attempt to use the system. To motivate children, we incorporated reward-based features that could incentivize autistic children to participate in conversations~\cite{Scott2010}. Specifically, dyads could earn stars based on the number of completed conversational turns (see \autoref{fig:interface:flow}-\circledigit{9}). During the introductory session, we instructed parents to implement their own motivational strategies upon the star system at their discretion, tailoring their approach to their child’s individual characteristics.}

\rerevised{By navigating these challenges, we contribute to a broader understanding of how technology can be effectively deployed to support communication in MVA children. We hope this study provides valuable guidance for researchers conducting deployment studies with similar populations.}
\subsection{Limitations and Future Work}
In this section, we discuss the limitations of our study that could impact the generalization of the findings. 
First, we used an LLM trained on datasets dominated by Western languages and cultures to support parent-child conversations in a Korean context. Although the LLM-driven recommendations did not pose significant issues in exchanging intentions between parents and MVA children, some participants reported that the recommendations did not adequately reflect Korean cultural sensitivities (\cf{},~\cite{naous2024CulturalBiasLLM}). Running a study with LLMs aligned with the primary language and cultural context (\eg{}, using HyperCLOVA X~\cite{yoo2024hyperclovaxtechnicalreport} for Korean) may improve the quality of dialogue understanding and thus further impact the engagement.

Second, we sourced participants for our deployment study through a single expert. As an early step in designing an LLM-driven communication mediation tool for parents and MVA children, we selectively sampled dyads that could employ \sysname{} without discomfort or difficulties in a home setting. Although we sought diversity in parent roles, genders, and children’s speech and cognitive statuses, our participants are not representative samples of these populations. Therefore, further investigation is needed across populations with diverse backgrounds and characteristics.

Lastly, we faced challenges ensuring that the MVA child's \textit{voice} was authentically represented through the system, without being influenced by parental guidance or lacking conversational intent. To understand the child’s behaviors and intentions, we relied on parents' recalls and secondary interpretations. 
In addition, we could not obtain evaluations directly from the MVA children regarding their satisfaction with the system. As highlighted in previous discussions~\cite{Wilson2019, Wilson2020, Spiel2019}, there is a need for a participatory approach in designing, developing, and evaluating systems for MVA children that incorporates their perspectives.

\section{Conclusion}
In this paper, we designed \sysname{} to enrich conversations between MVA children and their parents through reciprocal engagement. Drawing on formative interviews with professionals and parents of MVA children, \sysname{} leverages AI to provide parents with actionable speech guides that encourage participation of MVA children in conversation, while offering the children context-relevant vocabulary cards to expand their expressive range. Our deployment study with 11 parent-MVA child dyads over two weeks demonstrated that \sysname{} supported mutual participation, significantly increasing the frequency of conversations and turn-taking. Parents felt relieved from the pressure to lead flawless conversations, and increased self-expression by MVA children led parents to respect their child’s agency. They also discovered that conversations could become enjoyable interactions that deepen mutual understanding and empathy. We hope this research contributes to the development of inclusive communication technologies that bridge the communicative gap between the neurodiverse and neurotypical populations. 

\begin{acks}
\rerevised{We thank our participants from the formative interviews and the deployment study for their time and efforts. We are also grateful to Eunjoo Kim at Yonsei University for helping us recruit professionals for our formative study. We also thank Suhyeon Yoo for providing feedback on our early version of the draft.
This work was supported through a research internship at NAVER AI Lab of NAVER Cloud and in part by National Research Foundation of Korea (NRF) grant funded by the Korea government (MSIT) (RS-2024-00458557 and NRF-2024S1A5B5A19043580).}
\end{acks}

\bibliographystyle{ACM-Reference-Format}
\bibliography{bibliography}


\begin{thebibliography}{141}


\ifx \showCODEN    \undefined \def \showCODEN     #1{\unskip}     \fi
\ifx \showDOI      \undefined \def \showDOI       #1{#1}\fi
\ifx \showISBNx    \undefined \def \showISBNx     #1{\unskip}     \fi
\ifx \showISBNxiii \undefined \def \showISBNxiii  #1{\unskip}     \fi
\ifx \showISSN     \undefined \def \showISSN      #1{\unskip}     \fi
\ifx \showLCCN     \undefined \def \showLCCN      #1{\unskip}     \fi
\ifx \shownote     \undefined \def \shownote      #1{#1}          \fi
\ifx \showarticletitle \undefined \def \showarticletitle #1{#1}   \fi
\ifx \showURL      \undefined \def \showURL       {\relax}        \fi
\providecommand\bibfield[2]{#2}
\providecommand\bibinfo[2]{#2}
\providecommand\natexlab[1]{#1}
\providecommand\showeprint[2][]{arXiv:#2}

\bibitem[Alabood et~al\mbox{.}(2024)]%
        {Alabood2024}
\bibfield{author}{\bibinfo{person}{Lorans Alabood}, \bibinfo{person}{Travis Dow}, \bibinfo{person}{Kaylyn~B Feeley}, \bibinfo{person}{Vikram~K. Jaswal}, {and} \bibinfo{person}{Diwakar Krishnamurthy}.} \bibinfo{year}{2024}\natexlab{}.
\newblock \showarticletitle{From Letterboards to Holograms: Advancing Assistive Technology for Nonspeaking Autistic Individuals with the HoloBoard}. In \bibinfo{booktitle}{\emph{Proceedings of the CHI Conference on Human Factors in Computing Systems}} (Honolulu, HI, USA) \emph{(\bibinfo{series}{CHI '24})}. \bibinfo{publisher}{Association for Computing Machinery}, \bibinfo{address}{New York, NY, USA}, Article \bibinfo{articleno}{71}, \bibinfo{numpages}{18}~pages.
\newblock
\showISBNx{9798400703300}
\urldef\tempurl%
\url{https://doi.org/10.1145/3613904.3642626}
\showDOI{\tempurl}


\bibitem[Alabood et~al\mbox{.}(2022)]%
        {Alabood2022}
\bibfield{author}{\bibinfo{person}{Lorans Alabood}, \bibinfo{person}{Evan Krul}, \bibinfo{person}{Ali Shahidi}, \bibinfo{person}{Vikram~K. Jaswal}, \bibinfo{person}{Diwakar Krishnamurthy}, {and} \bibinfo{person}{Mea Wang}.} \bibinfo{year}{2022}\natexlab{}.
\newblock \showarticletitle{HoloType-CR: Cross Reality Communication Training for Minimally Verbal Autistic Persons}. In \bibinfo{booktitle}{\emph{2022 IEEE International Symposium on Mixed and Augmented Reality Adjunct (ISMAR-Adjunct)}}. \bibinfo{pages}{187--190}.
\newblock
\urldef\tempurl%
\url{https://doi.org/10.1109/ISMAR-Adjunct57072.2022.00042}
\showDOI{\tempurl}


\bibitem[Alli et~al\mbox{.}(2015)]%
        {alli2015}
\bibfield{author}{\bibinfo{person}{Aaishah Alli}, \bibinfo{person}{Shabnam Abdoola}, {and} \bibinfo{person}{Anniah Mupawose}.} \bibinfo{year}{2015}\natexlab{}.
\newblock \showarticletitle{Parents' journey into the world of autism}.
\newblock \bibinfo{journal}{\emph{South African Journal of Child Health}} \bibinfo{volume}{9}, \bibinfo{number}{3} (\bibinfo{year}{2015}), \bibinfo{pages}{81--84}.
\newblock


\bibitem[Alper(2017)]%
        {alper2017}
\bibfield{author}{\bibinfo{person}{Meryl Alper}.} \bibinfo{year}{2017}\natexlab{}.
\newblock \bibinfo{booktitle}{\emph{Giving voice: Mobile communication, disability, and inequality}}.
\newblock \bibinfo{publisher}{MIT Press}.
\newblock


\bibitem[American Psychiatric~Association et~al\mbox{.}(1994)]%
        {american1994}
\bibfield{author}{\bibinfo{person}{A American Psychiatric~Association} {et~al\mbox{.}}} \bibinfo{year}{1994}\natexlab{}.
\newblock \bibinfo{title}{Diagnostic and statistical manual of mental disorders: DSM-IV}.
\newblock
\newblock


\bibitem[AssistiveWare(2024)]%
        {assistiveware}
\bibfield{author}{\bibinfo{person}{AssistiveWare}.} \bibinfo{year}{2024}\natexlab{}.
\newblock \bibinfo{booktitle}{\emph{Giving AAC users the power of independent communication}}.
\newblock
\urldef\tempurl%
\url{https://www.assistiveware.com/blog/giving-aac-users-the-power-of-independent-communication}
\showURL{%
Retrieved Sep 10, 2024 from \tempurl}


\bibitem[Association et~al\mbox{.}(2006)]%
        {american2006}
\bibfield{author}{\bibinfo{person}{American Speech-Language-Hearing Association} {et~al\mbox{.}}} \bibinfo{year}{2006}\natexlab{}.
\newblock \showarticletitle{Principles for speech-language pathologists in diagnosis, assessment, and treatment of autism spectrum disorders across the life span. doi: 10.1044/policy}.
\newblock \bibinfo{journal}{\emph{TR2006-00143}} (\bibinfo{year}{2006}).
\newblock


\bibitem[atlas.ti(2023)]%
        {atlas}
\bibfield{author}{\bibinfo{person}{atlas.ti}.} \bibinfo{year}{accessed 2023}\natexlab{}.
\newblock \bibinfo{booktitle}{\emph{atlas.ti}}.
\newblock
\urldef\tempurl%
\url{https://atlasti.com}
\showURL{%
Retrieved Sep 10, 2024 from \tempurl}


\bibitem[Bae et~al\mbox{.}(2022)]%
        {Bae2022}
\bibfield{author}{\bibinfo{person}{Sanghwan Bae}, \bibinfo{person}{Donghyun Kwak}, \bibinfo{person}{Soyoung Kang}, \bibinfo{person}{Min~Young Lee}, \bibinfo{person}{Sungdong Kim}, \bibinfo{person}{Yuin Jeong}, \bibinfo{person}{Hyeri Kim}, \bibinfo{person}{Sang-Woo Lee}, \bibinfo{person}{Woomyoung Park}, {and} \bibinfo{person}{Nako Sung}.} \bibinfo{year}{2022}\natexlab{}.
\newblock \showarticletitle{Keep Me Updated! Memory Management in Long-term Conversations}. In \bibinfo{booktitle}{\emph{Findings of the Association for Computational Linguistics: EMNLP 2022}}, \bibfield{editor}{\bibinfo{person}{Yoav Goldberg}, \bibinfo{person}{Zornitsa Kozareva}, {and} \bibinfo{person}{Yue Zhang}} (Eds.). \bibinfo{publisher}{Association for Computational Linguistics}, \bibinfo{address}{Abu Dhabi, United Arab Emirates}, \bibinfo{pages}{3769--3787}.
\newblock
\urldef\tempurl%
\url{https://doi.org/10.18653/v1/2022.findings-emnlp.276}
\showDOI{\tempurl}


\bibitem[Baharav and Reiser(2010)]%
        {Baharav2010}
\bibfield{author}{\bibinfo{person}{Eva Baharav} {and} \bibinfo{person}{Carly Reiser}.} \bibinfo{year}{2010}\natexlab{}.
\newblock \showarticletitle{Using Telepractice in Parent Training in Early Autism}.
\newblock \bibinfo{journal}{\emph{Telemedicine and e-Health}} \bibinfo{volume}{16}, \bibinfo{number}{6} (\bibinfo{year}{2010}), \bibinfo{pages}{727--731}.
\newblock
\urldef\tempurl%
\url{https://doi.org/10.1089/tmj.2010.0029}
\showDOI{\tempurl}
\showeprint{https://doi.org/10.1089/tmj.2010.0029}
\newblock
\shownote{PMID: 20583950}.


\bibitem[Baron-Cohen et~al\mbox{.}(2005)]%
        {Simon2005}
\bibfield{author}{\bibinfo{person}{Simon Baron-Cohen}, \bibinfo{person}{Rebecca~C. Knickmeyer}, {and} \bibinfo{person}{Matthew~K. Belmonte}.} \bibinfo{year}{2005}\natexlab{}.
\newblock \showarticletitle{Sex Differences in the Brain: Implications for Explaining Autism}.
\newblock \bibinfo{journal}{\emph{Science}} \bibinfo{volume}{310}, \bibinfo{number}{5749} (\bibinfo{year}{2005}), \bibinfo{pages}{819--823}.
\newblock
\urldef\tempurl%
\url{https://doi.org/10.1126/science.1115455}
\showDOI{\tempurl}
\showeprint{https://www.science.org/doi/pdf/10.1126/science.1115455}


\bibitem[Bates et~al\mbox{.}(2015)]%
        {lmer}
\bibfield{author}{\bibinfo{person}{Douglas Bates}, \bibinfo{person}{Martin M{\"a}chler}, \bibinfo{person}{Ben Bolker}, {and} \bibinfo{person}{Steve Walker}.} \bibinfo{year}{2015}\natexlab{}.
\newblock \showarticletitle{Fitting Linear Mixed-Effects Models Using {lme4}}.
\newblock \bibinfo{journal}{\emph{Journal of Statistical Software}} \bibinfo{volume}{67}, \bibinfo{number}{1} (\bibinfo{year}{2015}), \bibinfo{pages}{1--48}.
\newblock
\urldef\tempurl%
\url{https://doi.org/10.18637/jss.v067.i01}
\showDOI{\tempurl}


\bibitem[Bennett et~al\mbox{.}(2018)]%
        {Bennett2018}
\bibfield{author}{\bibinfo{person}{Cynthia~L. Bennett}, \bibinfo{person}{Erin Brady}, {and} \bibinfo{person}{Stacy~M. Branham}.} \bibinfo{year}{2018}\natexlab{}.
\newblock \showarticletitle{Interdependence as a Frame for Assistive Technology Research and Design}. In \bibinfo{booktitle}{\emph{Proceedings of the 20th International ACM SIGACCESS Conference on Computers and Accessibility}} (Galway, Ireland) \emph{(\bibinfo{series}{ASSETS '18})}. \bibinfo{publisher}{Association for Computing Machinery}, \bibinfo{address}{New York, NY, USA}, \bibinfo{pages}{161–173}.
\newblock
\showISBNx{9781450356503}
\urldef\tempurl%
\url{https://doi.org/10.1145/3234695.3236348}
\showDOI{\tempurl}


\bibitem[Bernard-Opitz et~al\mbox{.}(2000)]%
        {Bernard2000}
\bibfield{author}{\bibinfo{person}{V Bernard-Opitz}, \bibinfo{person}{A Chen}, \bibinfo{person}{AJ Kok}, {and} \bibinfo{person}{N Sriram}.} \bibinfo{year}{2000}\natexlab{}.
\newblock \showarticletitle{Analysis of pragmatic aspects of communication behavior of verbal and nonverbal autistic children}.
\newblock \bibinfo{journal}{\emph{Praxis der Kinderpsychologie und Kinderpsychiatrie}} \bibinfo{volume}{49}, \bibinfo{number}{2} (\bibinfo{date}{February} \bibinfo{year}{2000}), \bibinfo{pages}{97—108}.
\newblock
\showISSN{0032-7034}
\urldef\tempurl%
\url{http://europepmc.org/abstract/MED/10721273}
\showURL{%
\tempurl}


\bibitem[Bernard-Opitz et~al\mbox{.}(1999)]%
        {Vera1999}
\bibfield{author}{\bibinfo{person}{Vera Bernard-Opitz}, \bibinfo{person}{N. Sriram}, {and} \bibinfo{person}{Sharul Sapuan}.} \bibinfo{year}{1999}\natexlab{}.
\newblock \showarticletitle{Enhancing Vocal Imitations in Children with Autism Using the IBM Speech Viewer}.
\newblock \bibinfo{journal}{\emph{Autism}} \bibinfo{volume}{3}, \bibinfo{number}{2} (\bibinfo{year}{1999}), \bibinfo{pages}{131--147}.
\newblock
\urldef\tempurl%
\url{https://doi.org/10.1177/1362361399003002004}
\showDOI{\tempurl}
\showeprint{https://doi.org/10.1177/1362361399003002004}


\bibitem[Beukelman and Light(2020)]%
        {beukelman2020}
\bibfield{author}{\bibinfo{person}{David Beukelman} {and} \bibinfo{person}{Janice Light}.} \bibinfo{year}{2020}\natexlab{}.
\newblock \showarticletitle{Augmentative and alternative communication: Supporting children and adults with complex communication needs}.
\newblock  (\bibinfo{year}{2020}).
\newblock


\bibitem[Beukelman and Mirenda(2013)]%
        {beukelman2013}
\bibfield{author}{\bibinfo{person}{D.R. Beukelman} {and} \bibinfo{person}{P. Mirenda}.} \bibinfo{year}{2013}\natexlab{}.
\newblock \bibinfo{booktitle}{\emph{Augmentative and Alternative Communication: Supporting Children and Adults with Complex Communication Needs}}.
\newblock \bibinfo{publisher}{Paul H. Brookes Pub.}
\newblock
\showISBNx{9781598571967}
\showLCCN{2012015676}
\urldef\tempurl%
\url{https://books.google.co.kr/books?id=oUQZMAEACAAJ}
\showURL{%
\tempurl}


\bibitem[Bindlish et~al\mbox{.}(2018)]%
        {bindlish2018}
\bibfield{author}{\bibinfo{person}{Nitin Bindlish}, \bibinfo{person}{Roma Kumar}, \bibinfo{person}{Manju Mehta}, {and} \bibinfo{person}{Kristine~Thompson Dubey}.} \bibinfo{year}{2018}\natexlab{}.
\newblock \showarticletitle{Effectiveness of parent-led interventions for autism and other developmental disorders.}
\newblock \bibinfo{journal}{\emph{Indian Journal of Health \& Wellbeing}} \bibinfo{volume}{9}, \bibinfo{number}{2} (\bibinfo{year}{2018}).
\newblock


\bibitem[Black et~al\mbox{.}(2010)]%
        {black2010}
\bibfield{author}{\bibinfo{person}{Rolf Black}, \bibinfo{person}{Joseph Reddington}, \bibinfo{person}{Ehud Reiter}, \bibinfo{person}{Nava Tintarev}, {and} \bibinfo{person}{Annalu Waller}.} \bibinfo{year}{2010}\natexlab{}.
\newblock \showarticletitle{Using NLG and sensors to support personal narrative for children with complex communication needs}. In \bibinfo{booktitle}{\emph{Proceedings of the NAACL HLT 2010 Workshop on Speech and Language Processing for Assistive Technologies}}. Association for Computational Linguists.
\newblock


\bibitem[Boyd(2002)]%
        {Brian2002}
\bibfield{author}{\bibinfo{person}{Brian~A. Boyd}.} \bibinfo{year}{2002}\natexlab{}.
\newblock \showarticletitle{Examining the Relationship BetWeen Stress and Lack of Social Support in Mothers of Children With Autism}.
\newblock \bibinfo{journal}{\emph{Focus on Autism and Other Developmental Disabilities}} \bibinfo{volume}{17}, \bibinfo{number}{4} (\bibinfo{year}{2002}), \bibinfo{pages}{208--215}.
\newblock
\urldef\tempurl%
\url{https://doi.org/10.1177/10883576020170040301}
\showDOI{\tempurl}
\showeprint{https://doi.org/10.1177/10883576020170040301}


\bibitem[Braun et~al\mbox{.}(2017)]%
        {Matthew2017}
\bibfield{author}{\bibinfo{person}{Matthew~J Braun}, \bibinfo{person}{Winnie Dunn}, {and} \bibinfo{person}{Scott~D Tomchek}.} \bibinfo{year}{2017}\natexlab{}.
\newblock \showarticletitle{A pilot study on professional documentation: Do we write from a strengths perspective?}
\newblock \bibinfo{journal}{\emph{American journal of speech-language pathology}} \bibinfo{volume}{26}, \bibinfo{number}{3} (\bibinfo{year}{2017}), \bibinfo{pages}{972--981}.
\newblock


\bibitem[Braun and Clarke(2006)]%
        {Braun2006ThematicAnalysis}
\bibfield{author}{\bibinfo{person}{Virginia Braun} {and} \bibinfo{person}{Victoria Clarke}.} \bibinfo{year}{2006}\natexlab{}.
\newblock \showarticletitle{Using Thematic Analysis in Psychology}.
\newblock \bibinfo{journal}{\emph{Qualitative Research in Psychology}} \bibinfo{volume}{3}, \bibinfo{number}{2} (\bibinfo{year}{2006}), \bibinfo{pages}{77--101}.
\newblock
\urldef\tempurl%
\url{https://doi.org/10.1191/1478088706qp063oa}
\showDOI{\tempurl}


\bibitem[Brock et~al\mbox{.}(2014)]%
        {Matthew2014}
\bibfield{author}{\bibinfo{person}{Matthew~E. Brock}, \bibinfo{person}{Heartley~B. Huber}, \bibinfo{person}{Erik~W. Carter}, \bibinfo{person}{A.~Pablo Juarez}, {and} \bibinfo{person}{Zachary~E. Warren}.} \bibinfo{year}{2014}\natexlab{}.
\newblock \showarticletitle{Statewide Assessment of Professional Development Needs Related to Educating Students With Autism Spectrum Disorder}.
\newblock \bibinfo{journal}{\emph{Focus on Autism and Other Developmental Disabilities}} \bibinfo{volume}{29}, \bibinfo{number}{2} (\bibinfo{year}{2014}), \bibinfo{pages}{67--79}.
\newblock
\urldef\tempurl%
\url{https://doi.org/10.1177/1088357614522290}
\showDOI{\tempurl}
\showeprint{https://doi.org/10.1177/1088357614522290}


\bibitem[Cai et~al\mbox{.}(2023)]%
        {cai2023}
\bibfield{author}{\bibinfo{person}{Shanqing Cai}, \bibinfo{person}{Subhashini Venugopalan}, \bibinfo{person}{Katie Seaver}, \bibinfo{person}{Xiang Xiao}, \bibinfo{person}{Katrin Tomanek}, \bibinfo{person}{Sri Jalasutram}, \bibinfo{person}{Meredith~Ringel Morris}, \bibinfo{person}{Shaun Kane}, \bibinfo{person}{Ajit Narayanan}, \bibinfo{person}{Robert~L. MacDonald}, \bibinfo{person}{Emily Kornman}, \bibinfo{person}{Daniel Vance}, \bibinfo{person}{Blair Casey}, \bibinfo{person}{Steve~M. Gleason}, \bibinfo{person}{Philip~Q. Nelson}, {and} \bibinfo{person}{Michael~P. Brenner}.} \bibinfo{year}{2023}\natexlab{}.
\newblock \bibinfo{title}{Using Large Language Models to Accelerate Communication for Users with Severe Motor Impairments}.
\newblock
\newblock
\showeprint[arxiv]{2312.01532}~[cs.HC]
\urldef\tempurl%
\url{https://arxiv.org/abs/2312.01532}
\showURL{%
\tempurl}


\bibitem[Chen(2022)]%
        {Rachel2022}
\bibfield{author}{\bibinfo{person}{Rachel S.~Y. Chen}.} \bibinfo{year}{2022}\natexlab{}.
\newblock \showarticletitle{Improvisations in the embodied interactions of a non-speaking autistic child and his mother: practices for creating intersubjective understanding}.
\newblock \bibinfo{journal}{\emph{Cognitive Linguistics}} \bibinfo{volume}{33}, \bibinfo{number}{1} (\bibinfo{year}{2022}), \bibinfo{pages}{155--191}.
\newblock
\urldef\tempurl%
\url{https://doi.org/doi:10.1515/cog-2021-0047}
\showDOI{\tempurl}


\bibitem[Chen et~al\mbox{.}(2024)]%
        {Chen2024}
\bibfield{author}{\bibinfo{person}{Yanru Chen}, \bibinfo{person}{Brynn Siles}, {and} \bibinfo{person}{Helen Tager-Flusberg}.} \bibinfo{year}{2024}\natexlab{}.
\newblock \showarticletitle{Receptive language and receptive-expressive discrepancy in minimally verbal autistic children and adolescents}.
\newblock \bibinfo{journal}{\emph{Autism Research}} \bibinfo{volume}{17}, \bibinfo{number}{2} (\bibinfo{year}{2024}), \bibinfo{pages}{381--394}.
\newblock
\urldef\tempurl%
\url{https://doi.org/10.1002/aur.3079}
\showDOI{\tempurl}
\showeprint{https://onlinelibrary.wiley.com/doi/pdf/10.1002/aur.3079}


\bibitem[Chien et~al\mbox{.}(2015)]%
        {CHIEN201579}
\bibfield{author}{\bibinfo{person}{Miao-En Chien}, \bibinfo{person}{Cyun-Meng Jheng}, \bibinfo{person}{Ni-Miao Lin}, \bibinfo{person}{Hsien-Hui Tang}, \bibinfo{person}{Paul Taele}, \bibinfo{person}{Wen-Sheng Tseng}, {and} \bibinfo{person}{Mike~Y. Chen}.} \bibinfo{year}{2015}\natexlab{}.
\newblock \showarticletitle{iCAN: A tablet-based pedagogical system for improving communication skills of children with autism}.
\newblock \bibinfo{journal}{\emph{International Journal of Human-Computer Studies}}  \bibinfo{volume}{73} (\bibinfo{year}{2015}), \bibinfo{pages}{79--90}.
\newblock
\showISSN{1071-5819}
\urldef\tempurl%
\url{https://doi.org/10.1016/j.ijhcs.2014.06.001}
\showDOI{\tempurl}


\bibitem[Choi et~al\mbox{.}(2020)]%
        {Choi2020}
\bibfield{author}{\bibinfo{person}{Boin Choi}, \bibinfo{person}{Charles~A. Nelson}, \bibinfo{person}{Meredith~L. Rowe}, {and} \bibinfo{person}{Helen Tager-Flusberg}.} \bibinfo{year}{2020}\natexlab{}.
\newblock \showarticletitle{Reciprocal Influences Between Parent Input and Child Language Skills in Dyads Involving High- and Low-Risk Infants for Autism Spectrum Disorder}.
\newblock \bibinfo{journal}{\emph{Autism Research}} \bibinfo{volume}{13}, \bibinfo{number}{7} (\bibinfo{year}{2020}), \bibinfo{pages}{1168--1183}.
\newblock
\urldef\tempurl%
\url{https://doi.org/10.1002/aur.2270}
\showDOI{\tempurl}
\showeprint{https://onlinelibrary.wiley.com/doi/pdf/10.1002/aur.2270}


\bibitem[Choi et~al\mbox{.}(2024)]%
        {Choi2024}
\bibfield{author}{\bibinfo{person}{Dasom Choi}, \bibinfo{person}{Sunok Lee}, \bibinfo{person}{Sung-In Kim}, \bibinfo{person}{Kyungah Lee}, \bibinfo{person}{Hee~Jeong Yoo}, \bibinfo{person}{Sangsu Lee}, {and} \bibinfo{person}{Hwajung Hong}.} \bibinfo{year}{2024}\natexlab{}.
\newblock \showarticletitle{Unlock Life with a Chat(GPT): Integrating Conversational AI with Large Language Models into Everyday Lives of Autistic Individuals}. In \bibinfo{booktitle}{\emph{Proceedings of the CHI Conference on Human Factors in Computing Systems}} (Honolulu, HI, USA) \emph{(\bibinfo{series}{CHI '24})}. \bibinfo{publisher}{Association for Computing Machinery}, \bibinfo{address}{New York, NY, USA}, Article \bibinfo{articleno}{72}, \bibinfo{numpages}{17}~pages.
\newblock
\showISBNx{9798400703300}
\urldef\tempurl%
\url{https://doi.org/10.1145/3613904.3641989}
\showDOI{\tempurl}


\bibitem[Chung and Douglas(2015)]%
        {chung2015}
\bibfield{author}{\bibinfo{person}{Yun-Ching Chung} {and} \bibinfo{person}{Karen~H Douglas}.} \bibinfo{year}{2015}\natexlab{}.
\newblock \showarticletitle{A peer interaction package for students with autism spectrum disorders who use speech-generating devices}.
\newblock \bibinfo{journal}{\emph{Journal of Developmental and Physical Disabilities}}  \bibinfo{volume}{27} (\bibinfo{year}{2015}), \bibinfo{pages}{831--849}.
\newblock


\bibitem[Cloud(2024)]%
        {yoo2024hyperclovaxtechnicalreport}
\bibfield{author}{\bibinfo{person}{NAVER Cloud}.} \bibinfo{year}{2024}\natexlab{}.
\newblock \bibinfo{title}{HyperCLOVA X Technical Report}.
\newblock
\newblock
\showeprint[arxiv]{2404.01954}~[cs.CL]
\urldef\tempurl%
\url{https://arxiv.org/abs/2404.01954}
\showURL{%
\tempurl}


\bibitem[Cynthia~Donato and Arthur-Kelly(2018)]%
        {Cynthia2018}
\bibfield{author}{\bibinfo{person}{Elizabeth~Spencer Cynthia~Donato} {and} \bibinfo{person}{Michael Arthur-Kelly}.} \bibinfo{year}{2018}\natexlab{}.
\newblock \showarticletitle{A critical synthesis of barriers and facilitators to the use of AAC by children with autism spectrum disorder and their communication partners}.
\newblock \bibinfo{journal}{\emph{Augmentative and Alternative Communication}} \bibinfo{volume}{34}, \bibinfo{number}{3} (\bibinfo{year}{2018}), \bibinfo{pages}{242--253}.
\newblock
\urldef\tempurl%
\url{https://doi.org/10.1080/07434618.2018.1493141}
\showDOI{\tempurl}
\showeprint{https://doi.org/10.1080/07434618.2018.1493141}
\newblock
\shownote{PMID: 30231643}.


\bibitem[Dai et~al\mbox{.}(2018)]%
        {DAI201836}
\bibfield{author}{\bibinfo{person}{Yael~G. Dai}, \bibinfo{person}{Lynn Brennan}, \bibinfo{person}{Ariel Como}, \bibinfo{person}{Jamie Hughes-Lika}, \bibinfo{person}{Thyde Dumont-Mathieu}, \bibinfo{person}{Iris Carcani-Rathwell}, \bibinfo{person}{Ola Minxhozi}, \bibinfo{person}{Blerina Aliaj}, {and} \bibinfo{person}{Deborah~A. Fein}.} \bibinfo{year}{2018}\natexlab{}.
\newblock \showarticletitle{A video parent-training program for families of children with autism spectrum disorder in Albania}.
\newblock \bibinfo{journal}{\emph{Research in Autism Spectrum Disorders}}  \bibinfo{volume}{56} (\bibinfo{year}{2018}), \bibinfo{pages}{36--49}.
\newblock
\showISSN{1750-9467}
\urldef\tempurl%
\url{https://doi.org/10.1016/j.rasd.2018.08.008}
\showDOI{\tempurl}


\bibitem[Damiao et~al\mbox{.}(2023)]%
        {Damiao2023}
\bibfield{author}{\bibinfo{person}{John Damiao}, \bibinfo{person}{Galilee Damiao}, \bibinfo{person}{Catherine Cavaliere}, \bibinfo{person}{Susanna Dunscomb}, \bibinfo{person}{Kirsten Ekelund}, \bibinfo{person}{Renee Lago}, {and} \bibinfo{person}{Ashley Volpe}.} \bibinfo{year}{2023}\natexlab{}.
\newblock \showarticletitle{{Parent Perspectives on Assisted Communication and Autism Spectrum Disorder}}.
\newblock \bibinfo{journal}{\emph{The American Journal of Occupational Therapy}} \bibinfo{volume}{78}, \bibinfo{number}{1} (\bibinfo{date}{12} \bibinfo{year}{2023}), \bibinfo{pages}{7801205070}.
\newblock
\showISSN{0272-9490}
\urldef\tempurl%
\url{https://doi.org/10.5014/ajot.2024.050343}
\showDOI{\tempurl}
\showeprint{https://research.aota.org/ajot/article-pdf/78/1/7801205070/84454/7801205070.pdf}


\bibitem[DeepL(2024)]%
        {DeepL}
\bibfield{author}{\bibinfo{person}{DeepL}.} \bibinfo{year}{2024}\natexlab{}.
\newblock \bibinfo{title}{{DeepL - Build multilingual experiences with the DeepL API}}.
\newblock
\newblock
\urldef\tempurl%
\url{https://www.deepl.com/en/pro-api}
\showURL{%
Retrieved Sep 10, 2024 from \tempurl}


\bibitem[Del~Bianco et~al\mbox{.}(2018)]%
        {del2018thorn}
\bibfield{author}{\bibinfo{person}{Teresa Del~Bianco}, \bibinfo{person}{Yagmur Ozturk}, \bibinfo{person}{Ilaria Basadonne}, \bibinfo{person}{Noemi Mazzoni}, {and} \bibinfo{person}{Paola Venuti}.} \bibinfo{year}{2018}\natexlab{}.
\newblock \showarticletitle{The thorn in the dyad: A vision on parent-child relationship in autism spectrum disorder}.
\newblock \bibinfo{journal}{\emph{Europe's journal of psychology}} \bibinfo{volume}{14}, \bibinfo{number}{3} (\bibinfo{year}{2018}), \bibinfo{pages}{695}.
\newblock


\bibitem[Demmans~Epp et~al\mbox{.}(2012)]%
        {Epp2012}
\bibfield{author}{\bibinfo{person}{Carrie Demmans~Epp}, \bibinfo{person}{Justin Djordjevic}, \bibinfo{person}{Shimu Wu}, \bibinfo{person}{Karyn Moffatt}, {and} \bibinfo{person}{Ronald~M. Baecker}.} \bibinfo{year}{2012}\natexlab{}.
\newblock \showarticletitle{Towards providing just-in-time vocabulary support for assistive and augmentative communication}. In \bibinfo{booktitle}{\emph{Proceedings of the 2012 ACM International Conference on Intelligent User Interfaces}} (Lisbon, Portugal) \emph{(\bibinfo{series}{IUI '12})}. \bibinfo{publisher}{Association for Computing Machinery}, \bibinfo{address}{New York, NY, USA}, \bibinfo{pages}{33–36}.
\newblock
\showISBNx{9781450310482}
\urldef\tempurl%
\url{https://doi.org/10.1145/2166966.2166973}
\showDOI{\tempurl}


\bibitem[Erinn H.~Finke and Drager(2009)]%
        {Erinn2009}
\bibfield{author}{\bibinfo{person}{David B.~McNaughton Erinn H.~Finke, Erinn H.~Finke} {and} \bibinfo{person}{Kathryn D.~R. Drager}.} \bibinfo{year}{2009}\natexlab{}.
\newblock \showarticletitle{“All Children Can and Should Have the Opportunity to Learn”: General Education Teachers' Perspectives on Including Children with Autism Spectrum Disorder who Require AAC}.
\newblock \bibinfo{journal}{\emph{Augmentative and Alternative Communication}} \bibinfo{volume}{25}, \bibinfo{number}{2} (\bibinfo{year}{2009}), \bibinfo{pages}{110--122}.
\newblock
\urldef\tempurl%
\url{https://doi.org/10.1080/07434610902886206}
\showDOI{\tempurl}
\showeprint{https://doi.org/10.1080/07434610902886206}
\newblock
\shownote{PMID: 19444682}.


\bibitem[et~al.(2024)]%
        {google2024gemini}
\bibfield{author}{\bibinfo{person}{Gemini~Team et al.}} \bibinfo{year}{2024}\natexlab{}.
\newblock \bibinfo{title}{Gemini: A Family of Highly Capable Multimodal Models}.
\newblock
\newblock
\showeprint[arxiv]{2312.11805}~[cs.CL]
\urldef\tempurl%
\url{https://arxiv.org/abs/2312.11805}
\showURL{%
\tempurl}


\bibitem[Eyberg and Funderburk(2011)]%
        {eyberg2011}
\bibfield{author}{\bibinfo{person}{SM Eyberg} {and} \bibinfo{person}{B Funderburk}.} \bibinfo{year}{2011}\natexlab{}.
\newblock \showarticletitle{Parent-child interaction therapy protocol}.
\newblock \bibinfo{journal}{\emph{Gainesville, FL: PCIT International}} (\bibinfo{year}{2011}).
\newblock


\bibitem[Fallon et~al\mbox{.}(2001)]%
        {Karen2001}
\bibfield{author}{\bibinfo{person}{Karen~A. Fallon}, \bibinfo{person}{Janice~C. Light}, {and} \bibinfo{person}{Tara~Kramer Paige}.} \bibinfo{year}{2001}\natexlab{}.
\newblock \showarticletitle{Enhancing Vocabulary Selection for Preschoolers Who Require Augmentative and Alternative Communication (AAC)}.
\newblock \bibinfo{journal}{\emph{American Journal of Speech-Language Pathology}} \bibinfo{volume}{10}, \bibinfo{number}{1} (\bibinfo{year}{2001}), \bibinfo{pages}{81--94}.
\newblock
\urldef\tempurl%
\url{https://doi.org/10.1044/1058-0360(2001/010)}
\showDOI{\tempurl}
\showeprint{https://pubs.asha.org/doi/pdf/10.1044/1058-0360\%282001/010\%29}


\bibitem[FastAPI(2024)]%
        {FastAPI}
\bibfield{author}{\bibinfo{person}{FastAPI}.} \bibinfo{year}{2024}\natexlab{}.
\newblock \bibinfo{title}{{FastAPI framework, high performance, easy to learn, fast to code, ready for production}}.
\newblock
\newblock
\urldef\tempurl%
\url{https://fastapi.tiangolo.com/}
\showURL{%
Retrieved Sep 10, 2024 from \tempurl}


\bibitem[Fenson et~al\mbox{.}(2007)]%
        {fenson2007}
\bibfield{author}{\bibinfo{person}{Larry Fenson} {et~al\mbox{.}}} \bibinfo{year}{2007}\natexlab{}.
\newblock \bibinfo{booktitle}{\emph{MacArthur-Bates communicative development inventories}}.
\newblock \bibinfo{publisher}{Paul H. Brookes Publishing Company Baltimore, MD}.
\newblock


\bibitem[Fontana~de Vargas et~al\mbox{.}(2022)]%
        {Fontana2022}
\bibfield{author}{\bibinfo{person}{Mauricio Fontana~de Vargas}, \bibinfo{person}{Jiamin Dai}, {and} \bibinfo{person}{Karyn Moffatt}.} \bibinfo{year}{2022}\natexlab{}.
\newblock \showarticletitle{AAC with Automated Vocabulary from Photographs: Insights from School and Speech-Language Therapy Settings}. In \bibinfo{booktitle}{\emph{Proceedings of the 24th International ACM SIGACCESS Conference on Computers and Accessibility}} (Athens, Greece) \emph{(\bibinfo{series}{ASSETS '22})}. \bibinfo{publisher}{Association for Computing Machinery}, \bibinfo{address}{New York, NY, USA}, Article \bibinfo{articleno}{23}, \bibinfo{numpages}{18}~pages.
\newblock
\showISBNx{9781450392587}
\urldef\tempurl%
\url{https://doi.org/10.1145/3517428.3544805}
\showDOI{\tempurl}


\bibitem[Fontana De~Vargas et~al\mbox{.}(2024a)]%
        {Vargas2024}
\bibfield{author}{\bibinfo{person}{Mauricio Fontana De~Vargas}, \bibinfo{person}{Christina Yu}, \bibinfo{person}{Howard~C. Shane}, {and} \bibinfo{person}{Karyn Moffatt}.} \bibinfo{year}{2024}\natexlab{a}.
\newblock \showarticletitle{Co-Designing QuickPic: Automated Topic-Specific Communication Boards from Photographs for AAC-Based Language Instruction}. In \bibinfo{booktitle}{\emph{Proceedings of the CHI Conference on Human Factors in Computing Systems}} (Honolulu, HI, USA) \emph{(\bibinfo{series}{CHI '24})}. \bibinfo{publisher}{Association for Computing Machinery}, \bibinfo{address}{New York, NY, USA}, Article \bibinfo{articleno}{910}, \bibinfo{numpages}{16}~pages.
\newblock
\showISBNx{9798400703300}
\urldef\tempurl%
\url{https://doi.org/10.1145/3613904.3642080}
\showDOI{\tempurl}


\bibitem[Fontana De~Vargas et~al\mbox{.}(2024b)]%
        {Fontana2024}
\bibfield{author}{\bibinfo{person}{Mauricio Fontana De~Vargas}, \bibinfo{person}{Christina Yu}, \bibinfo{person}{Howard~C. Shane}, {and} \bibinfo{person}{Karyn Moffatt}.} \bibinfo{year}{2024}\natexlab{b}.
\newblock \showarticletitle{Co-Designing QuickPic: Automated Topic-Specific Communication Boards from Photographs for AAC-Based Language Instruction}. In \bibinfo{booktitle}{\emph{Proceedings of the CHI Conference on Human Factors in Computing Systems}} (Honolulu, HI, USA) \emph{(\bibinfo{series}{CHI '24})}. \bibinfo{publisher}{Association for Computing Machinery}, \bibinfo{address}{New York, NY, USA}, Article \bibinfo{articleno}{910}, \bibinfo{numpages}{16}~pages.
\newblock
\showISBNx{9798400703300}
\urldef\tempurl%
\url{https://doi.org/10.1145/3613904.3642080}
\showDOI{\tempurl}


\bibitem[Gadberry(2011)]%
        {Gadberry2011}
\bibfield{author}{\bibinfo{person}{Anita~L. Gadberry}.} \bibinfo{year}{2011}\natexlab{}.
\newblock \showarticletitle{{A Survey of the Use of Aided Augmentative and Alternative Communication during Music Therapy Sessions with Persons with Autism Spectrum Disorders}}.
\newblock \bibinfo{journal}{\emph{Journal of Music Therapy}} \bibinfo{volume}{48}, \bibinfo{number}{1} (\bibinfo{date}{03} \bibinfo{year}{2011}), \bibinfo{pages}{74--89}.
\newblock
\showISSN{0022-2917}
\urldef\tempurl%
\url{https://doi.org/10.1093/jmt/48.1.74}
\showDOI{\tempurl}
\showeprint{https://academic.oup.com/jmt/article-pdf/48/1/74/5437339/48-1-74.pdf}


\bibitem[Gadiraju et~al\mbox{.}(2023)]%
        {Gadiraju2023}
\bibfield{author}{\bibinfo{person}{Vinitha Gadiraju}, \bibinfo{person}{Shaun Kane}, \bibinfo{person}{Sunipa Dev}, \bibinfo{person}{Alex Taylor}, \bibinfo{person}{Ding Wang}, \bibinfo{person}{Emily Denton}, {and} \bibinfo{person}{Robin Brewer}.} \bibinfo{year}{2023}\natexlab{}.
\newblock \showarticletitle{"I wouldn’t say offensive but...": Disability-Centered Perspectives on Large Language Models}. In \bibinfo{booktitle}{\emph{Proceedings of the 2023 ACM Conference on Fairness, Accountability, and Transparency}} (Chicago, IL, USA) \emph{(\bibinfo{series}{FAccT '23})}. \bibinfo{publisher}{Association for Computing Machinery}, \bibinfo{address}{New York, NY, USA}, \bibinfo{pages}{205–216}.
\newblock
\showISBNx{9798400701924}
\urldef\tempurl%
\url{https://doi.org/10.1145/3593013.3593989}
\showDOI{\tempurl}


\bibitem[Ganz et~al\mbox{.}(2013)]%
        {Jennifer2013}
\bibfield{author}{\bibinfo{person}{Jennifer~B. Ganz}, \bibinfo{person}{Fara~D. Goodwyn}, \bibinfo{person}{Margot~M. Boles}, \bibinfo{person}{Ee~Rea Hong}, \bibinfo{person}{Mandy~J. Rispoli}, \bibinfo{person}{Emily~M. Lund}, {and} \bibinfo{person}{Elizabeth Kite}.} \bibinfo{year}{2013}\natexlab{}.
\newblock \showarticletitle{Impacts of a PECS Instructional Coaching Intervention on Practitioners and Children with Autism}.
\newblock \bibinfo{journal}{\emph{Augmentative and Alternative Communication}} \bibinfo{volume}{29}, \bibinfo{number}{3} (\bibinfo{year}{2013}), \bibinfo{pages}{210--221}.
\newblock
\urldef\tempurl%
\url{https://doi.org/10.3109/07434618.2013.818058}
\showDOI{\tempurl}
\showeprint{https://doi.org/10.3109/07434618.2013.818058}
\newblock
\shownote{PMID: 23952566}.


\bibitem[Gibaud-Wallston and Wandersmann(1978)]%
        {gibaud1978}
\bibfield{author}{\bibinfo{person}{Jonatha Gibaud-Wallston} {and} \bibinfo{person}{Lois~Pall Wandersmann}.} \bibinfo{year}{1978}\natexlab{}.
\newblock \bibinfo{booktitle}{\emph{Development and utility of the Parenting Sense of Competence Scale}}.
\newblock \bibinfo{publisher}{John F. Kennedy center for research on education and human development}.
\newblock


\bibitem[Gunilla~Thunberg and Sandberg(2007)]%
        {Gunilla2007}
\bibfield{author}{\bibinfo{person}{Elisabeth~Ahlsén Gunilla~Thunberg} {and} \bibinfo{person}{Annika~Dahlgren Sandberg}.} \bibinfo{year}{2007}\natexlab{}.
\newblock \showarticletitle{Children with autistic spectrum disorders and speech‐generating devices: Communication in different activities at home}.
\newblock \bibinfo{journal}{\emph{Clinical Linguistics \& Phonetics}} \bibinfo{volume}{21}, \bibinfo{number}{6} (\bibinfo{year}{2007}), \bibinfo{pages}{457--479}.
\newblock
\urldef\tempurl%
\url{https://doi.org/10.1080/02699200701314963}
\showDOI{\tempurl}
\showeprint{https://doi.org/10.1080/02699200701314963}
\newblock
\shownote{PMID: 17516231}.


\bibitem[Hailpern et~al\mbox{.}(2012)]%
        {Hailpern2012}
\bibfield{author}{\bibinfo{person}{Joshua Hailpern}, \bibinfo{person}{Andrew Harris}, \bibinfo{person}{Reed La~Botz}, \bibinfo{person}{Brianna Birman}, {and} \bibinfo{person}{Karrie Karahalios}.} \bibinfo{year}{2012}\natexlab{}.
\newblock \showarticletitle{Designing visualizations to facilitate multisyllabic speech with children with autism and speech delays}. In \bibinfo{booktitle}{\emph{Proceedings of the Designing Interactive Systems Conference}} (Newcastle Upon Tyne, United Kingdom) \emph{(\bibinfo{series}{DIS '12})}. \bibinfo{publisher}{Association for Computing Machinery}, \bibinfo{address}{New York, NY, USA}, \bibinfo{pages}{126–135}.
\newblock
\showISBNx{9781450312103}
\urldef\tempurl%
\url{https://doi.org/10.1145/2317956.2317977}
\showDOI{\tempurl}


\bibitem[Hailpern et~al\mbox{.}(2010)]%
        {Hailpern2010}
\bibfield{author}{\bibinfo{person}{Joshua Hailpern}, \bibinfo{person}{Karrie Karahalios}, \bibinfo{person}{Laura DeThorne}, {and} \bibinfo{person}{Jim Halle}.} \bibinfo{year}{2010}\natexlab{}.
\newblock \showarticletitle{Vocsyl: visualizing syllable production for children with ASD and speech delays}. In \bibinfo{booktitle}{\emph{Proceedings of the 12th International ACM SIGACCESS Conference on Computers and Accessibility}} (Orlando, Florida, USA) \emph{(\bibinfo{series}{ASSETS '10})}. \bibinfo{publisher}{Association for Computing Machinery}, \bibinfo{address}{New York, NY, USA}, \bibinfo{pages}{297–298}.
\newblock
\showISBNx{9781605588810}
\urldef\tempurl%
\url{https://doi.org/10.1145/1878803.1878879}
\showDOI{\tempurl}


\bibitem[Hayes et~al\mbox{.}(2010)]%
        {hayes2010}
\bibfield{author}{\bibinfo{person}{Gillian~R Hayes}, \bibinfo{person}{Sen Hirano}, \bibinfo{person}{Gabriela Marcu}, \bibinfo{person}{Mohamad Monibi}, \bibinfo{person}{David~H Nguyen}, {and} \bibinfo{person}{Michael Yeganyan}.} \bibinfo{year}{2010}\natexlab{}.
\newblock \showarticletitle{Interactive visual supports for children with autism}.
\newblock \bibinfo{journal}{\emph{Personal and ubiquitous computing}}  \bibinfo{volume}{14} (\bibinfo{year}{2010}), \bibinfo{pages}{663--680}.
\newblock


\bibitem[Hong et~al\mbox{.}(2018)]%
        {hong2018}
\bibfield{author}{\bibinfo{person}{Ee~Rea Hong}, \bibinfo{person}{Liyuan Gong}, \bibinfo{person}{Jennifer~B Ganz}, {and} \bibinfo{person}{Leslie Neely}.} \bibinfo{year}{2018}\natexlab{}.
\newblock \showarticletitle{Self-paced and video-based learning: parent training and language skills in Japanese children with ASD}.
\newblock \bibinfo{journal}{\emph{Exceptionality Education International}} \bibinfo{volume}{28}, \bibinfo{number}{2} (\bibinfo{year}{2018}).
\newblock


\bibitem[Hossain et~al\mbox{.}(2022)]%
        {Hossain2022}
\bibfield{author}{\bibinfo{person}{Ekram Hossain}, \bibinfo{person}{Merritt~Lee Cahoon}, \bibinfo{person}{Yao Liu}, \bibinfo{person}{Chigusa Kurumada}, {and} \bibinfo{person}{Zhen Bai}.} \bibinfo{year}{2022}\natexlab{}.
\newblock \showarticletitle{Context-responsive ASL Recommendation for Parent-Child Interaction}. In \bibinfo{booktitle}{\emph{Proceedings of the 24th International ACM SIGACCESS Conference on Computers and Accessibility}} (Athens, Greece) \emph{(\bibinfo{series}{ASSETS '22})}. \bibinfo{publisher}{Association for Computing Machinery}, \bibinfo{address}{New York, NY, USA}, Article \bibinfo{articleno}{76}, \bibinfo{numpages}{5}~pages.
\newblock
\showISBNx{9781450392587}
\urldef\tempurl%
\url{https://doi.org/10.1145/3517428.3550366}
\showDOI{\tempurl}


\bibitem[Huber et~al\mbox{.}(2019)]%
        {Huber2019}
\bibfield{author}{\bibinfo{person}{Bernd Huber}, \bibinfo{person}{Richard~F. Davis}, \bibinfo{person}{Allison Cotter}, \bibinfo{person}{Emily Junkin}, \bibinfo{person}{Mindy Yard}, \bibinfo{person}{Stuart Shieber}, \bibinfo{person}{Elizabeth Brestan-Knight}, {and} \bibinfo{person}{Krzysztof~Z. Gajos}.} \bibinfo{year}{2019}\natexlab{}.
\newblock \showarticletitle{SpecialTime: Automatically Detecting Dialogue Acts from Speech to Support Parent-Child Interaction Therapy}. In \bibinfo{booktitle}{\emph{Proceedings of the 13th EAI International Conference on Pervasive Computing Technologies for Healthcare}} (Trento, Italy) \emph{(\bibinfo{series}{PervasiveHealth'19})}. \bibinfo{publisher}{Association for Computing Machinery}, \bibinfo{address}{New York, NY, USA}, \bibinfo{pages}{139–148}.
\newblock
\showISBNx{9781450361262}
\urldef\tempurl%
\url{https://doi.org/10.1145/3329189.3329203}
\showDOI{\tempurl}


\bibitem[Hwang et~al\mbox{.}(2014)]%
        {Hwang2014}
\bibfield{author}{\bibinfo{person}{Inseok Hwang}, \bibinfo{person}{Chungkuk Yoo}, \bibinfo{person}{Chanyou Hwang}, \bibinfo{person}{Dongsun Yim}, \bibinfo{person}{Youngki Lee}, \bibinfo{person}{Chulhong Min}, \bibinfo{person}{John Kim}, {and} \bibinfo{person}{Junehwa Song}.} \bibinfo{year}{2014}\natexlab{}.
\newblock \showarticletitle{TalkBetter: family-driven mobile intervention care for children with language delay}. In \bibinfo{booktitle}{\emph{Proceedings of the 17th ACM Conference on Computer Supported Cooperative Work \& Social Computing}} (Baltimore, Maryland, USA) \emph{(\bibinfo{series}{CSCW '14})}. \bibinfo{publisher}{Association for Computing Machinery}, \bibinfo{address}{New York, NY, USA}, \bibinfo{pages}{1283–1296}.
\newblock
\showISBNx{9781450325400}
\urldef\tempurl%
\url{https://doi.org/10.1145/2531602.2531668}
\showDOI{\tempurl}


\bibitem[Ingersoll and Gergans(2007)]%
        {INGERSOLL2007163}
\bibfield{author}{\bibinfo{person}{Brooke Ingersoll} {and} \bibinfo{person}{Samantha Gergans}.} \bibinfo{year}{2007}\natexlab{}.
\newblock \showarticletitle{The effect of a parent-implemented imitation intervention on spontaneous imitation skills in young children with autism}.
\newblock \bibinfo{journal}{\emph{Research in Developmental Disabilities}} \bibinfo{volume}{28}, \bibinfo{number}{2} (\bibinfo{year}{2007}), \bibinfo{pages}{163--175}.
\newblock
\showISSN{0891-4222}
\urldef\tempurl%
\url{https://doi.org/10.1016/j.ridd.2006.02.004}
\showDOI{\tempurl}


\bibitem[Johnson et~al\mbox{.}(2023)]%
        {johnson2023}
\bibfield{author}{\bibinfo{person}{Kristina~T Johnson}, \bibinfo{person}{Jaya Narain}, \bibinfo{person}{Thomas Quatieri}, \bibinfo{person}{Pattie Maes}, {and} \bibinfo{person}{Rosalind~W Picard}.} \bibinfo{year}{2023}\natexlab{}.
\newblock \showarticletitle{ReCANVo: A database of real-world communicative and affective nonverbal vocalizations}.
\newblock \bibinfo{journal}{\emph{Scientific Data}} \bibinfo{volume}{10}, \bibinfo{number}{1} (\bibinfo{year}{2023}), \bibinfo{pages}{523}.
\newblock


\bibitem[Kasari et~al\mbox{.}(2013)]%
        {Kasari2013}
\bibfield{author}{\bibinfo{person}{Connie Kasari}, \bibinfo{person}{Nancy Brady}, \bibinfo{person}{Catherine Lord}, {and} \bibinfo{person}{Helen Tager-Flusberg}.} \bibinfo{year}{2013}\natexlab{}.
\newblock \showarticletitle{Assessing the Minimally Verbal School-Aged Child With Autism Spectrum Disorder}.
\newblock \bibinfo{journal}{\emph{Autism Research}} \bibinfo{volume}{6}, \bibinfo{number}{6} (\bibinfo{year}{2013}), \bibinfo{pages}{479--493}.
\newblock
\urldef\tempurl%
\url{https://doi.org/10.1002/aur.1334}
\showDOI{\tempurl}
\showeprint{https://onlinelibrary.wiley.com/doi/pdf/10.1002/aur.1334}


\bibitem[Kasari et~al\mbox{.}(2014)]%
        {KASARI2014635}
\bibfield{author}{\bibinfo{person}{Connie Kasari}, \bibinfo{person}{Ann Kaiser}, \bibinfo{person}{Kelly Goods}, \bibinfo{person}{Jennifer Nietfeld}, \bibinfo{person}{Pamela Mathy}, \bibinfo{person}{Rebecca Landa}, \bibinfo{person}{Susan Murphy}, {and} \bibinfo{person}{Daniel Almirall}.} \bibinfo{year}{2014}\natexlab{}.
\newblock \showarticletitle{Communication Interventions for Minimally Verbal Children With Autism: A Sequential Multiple Assignment Randomized Trial}.
\newblock \bibinfo{journal}{\emph{Journal of the American Academy of Child \& Adolescent Psychiatry}} \bibinfo{volume}{53}, \bibinfo{number}{6} (\bibinfo{year}{2014}), \bibinfo{pages}{635--646}.
\newblock
\showISSN{0890-8567}
\urldef\tempurl%
\url{https://doi.org/10.1016/j.jaac.2014.01.019}
\showDOI{\tempurl}


\bibitem[Kenny et~al\mbox{.}(2016)]%
        {Lorcan2016}
\bibfield{author}{\bibinfo{person}{Lorcan Kenny}, \bibinfo{person}{Caroline Hattersley}, \bibinfo{person}{Bonnie Molins}, \bibinfo{person}{Carole Buckley}, \bibinfo{person}{Carol Povey}, {and} \bibinfo{person}{Elizabeth Pellicano}.} \bibinfo{year}{2016}\natexlab{}.
\newblock \showarticletitle{Which terms should be used to describe autism? Perspectives from the UK autism community}.
\newblock \bibinfo{journal}{\emph{Autism}} \bibinfo{volume}{20}, \bibinfo{number}{4} (\bibinfo{year}{2016}), \bibinfo{pages}{442--462}.
\newblock
\urldef\tempurl%
\url{https://doi.org/10.1177/1362361315588200}
\showDOI{\tempurl}
\showeprint{https://doi.org/10.1177/1362361315588200}
\newblock
\shownote{PMID: 26134030}.


\bibitem[Kientz et~al\mbox{.}(2014)]%
        {kientz2014}
\bibfield{author}{\bibinfo{person}{JA Kientz}, \bibinfo{person}{MS Goodwin}, \bibinfo{person}{GR Hayes}, {and} \bibinfo{person}{GD Abowd}.} \bibinfo{year}{2014}\natexlab{}.
\newblock \showarticletitle{Interactive technologies for autism. Synthesis lectures on assistive, rehabilitative, and healthpreserving technologies}.
\newblock \bibinfo{journal}{\emph{San Rafael, CA: Morgan \& Claypool Publishers}} (\bibinfo{year}{2014}).
\newblock


\bibitem[Kim et~al\mbox{.}(2024)]%
        {kim2024mindfuldiary}
\bibfield{author}{\bibinfo{person}{Taewan Kim}, \bibinfo{person}{Seolyeong Bae}, \bibinfo{person}{Hyun~Ah Kim}, \bibinfo{person}{Su-Woo Lee}, \bibinfo{person}{Hwajung Hong}, \bibinfo{person}{Chanmo Yang}, {and} \bibinfo{person}{Young-Ho Kim}.} \bibinfo{year}{2024}\natexlab{}.
\newblock \showarticletitle{MindfulDiary: Harnessing Large Language Model to Support Psychiatric Patients' Journaling}. In \bibinfo{booktitle}{\emph{Proceedings of the CHI Conference on Human Factors in Computing Systems}} (Honolulu, HI, USA) \emph{(\bibinfo{series}{CHI '24})}. \bibinfo{publisher}{Association for Computing Machinery}, \bibinfo{address}{New York, NY, USA}, Article \bibinfo{articleno}{701}, \bibinfo{numpages}{20}~pages.
\newblock
\showISBNx{9798400703300}
\urldef\tempurl%
\url{https://doi.org/10.1145/3613904.3642937}
\showDOI{\tempurl}


\bibitem[Kim et~al\mbox{.}(2020)]%
        {Wonjung2020}
\bibfield{author}{\bibinfo{person}{Wonjung Kim}, \bibinfo{person}{Seungchul Lee}, \bibinfo{person}{Seonghoon Kim}, \bibinfo{person}{Sungbin Jo}, \bibinfo{person}{Chungkuk Yoo}, \bibinfo{person}{Inseok Hwang}, \bibinfo{person}{Seungwoo Kang}, {and} \bibinfo{person}{Junehwa Song}.} \bibinfo{year}{2020}\natexlab{}.
\newblock \showarticletitle{Dyadic Mirror: Everyday Second-person Live-view for Empathetic Reflection upon Parent-child Interaction}.
\newblock \bibinfo{journal}{\emph{Proc. ACM Interact. Mob. Wearable Ubiquitous Technol.}} \bibinfo{volume}{4}, \bibinfo{number}{3}, Article \bibinfo{articleno}{86} (\bibinfo{date}{sep} \bibinfo{year}{2020}), \bibinfo{numpages}{29}~pages.
\newblock
\urldef\tempurl%
\url{https://doi.org/10.1145/3411815}
\showDOI{\tempurl}


\bibitem[Kluth and Darmody-Latham(2003)]%
        {Paula2003}
\bibfield{author}{\bibinfo{person}{Paula Kluth} {and} \bibinfo{person}{Julie Darmody-Latham}.} \bibinfo{year}{2003}\natexlab{}.
\newblock \showarticletitle{Beyond Sight Words: Literacy Opportunities for Students with Autism}.
\newblock \bibinfo{journal}{\emph{The Reading Teacher}} \bibinfo{volume}{56}, \bibinfo{number}{6} (\bibinfo{year}{2003}), \bibinfo{pages}{532--535}.
\newblock
\showISSN{00340561}
\urldef\tempurl%
\url{http://www.jstor.org/stable/20205240}
\showURL{%
\tempurl}


\bibitem[Knapp(1978)]%
        {knapp1978}
\bibfield{author}{\bibinfo{person}{Mark~L Knapp}.} \bibinfo{year}{1978}\natexlab{}.
\newblock \showarticletitle{Nonverbal communication in human interaction}.
\newblock \bibinfo{journal}{\emph{Rinchart \& Winston}} (\bibinfo{year}{1978}).
\newblock


\bibitem[Kory-Westlund and Breazeal(2019)]%
        {Kory2019}
\bibfield{author}{\bibinfo{person}{Jacqueline~M. Kory-Westlund} {and} \bibinfo{person}{Cynthia Breazeal}.} \bibinfo{year}{2019}\natexlab{}.
\newblock \showarticletitle{Assessing Children's Perceptions and Acceptance of a Social Robot}. In \bibinfo{booktitle}{\emph{Proceedings of the 18th ACM International Conference on Interaction Design and Children}} (Boise, ID, USA) \emph{(\bibinfo{series}{IDC '19})}. \bibinfo{publisher}{Association for Computing Machinery}, \bibinfo{address}{New York, NY, USA}, \bibinfo{pages}{38–50}.
\newblock
\showISBNx{9781450366908}
\urldef\tempurl%
\url{https://doi.org/10.1145/3311927.3323143}
\showDOI{\tempurl}


\bibitem[Kuttler et~al\mbox{.}(1998)]%
        {Shari1998}
\bibfield{author}{\bibinfo{person}{Shari Kuttler}, \bibinfo{person}{Brenda~Smith Myles}, {and} \bibinfo{person}{Judith~K. Carlson}.} \bibinfo{year}{1998}\natexlab{}.
\newblock \showarticletitle{The Use of Social Stories to Reduce Precursors to Tantrum Behavior in a Student with Autism}.
\newblock \bibinfo{journal}{\emph{Focus on Autism and Other Developmental Disabilities}} \bibinfo{volume}{13}, \bibinfo{number}{3} (\bibinfo{year}{1998}), \bibinfo{pages}{176--182}.
\newblock
\urldef\tempurl%
\url{https://doi.org/10.1177/108835769801300306}
\showDOI{\tempurl}
\showeprint{https://doi.org/10.1177/108835769801300306}


\bibitem[Kwon et~al\mbox{.}(2022)]%
        {Kwon2022}
\bibfield{author}{\bibinfo{person}{Taeahn Kwon}, \bibinfo{person}{Minkyung Jeong}, \bibinfo{person}{Eon-Suk Ko}, {and} \bibinfo{person}{Youngki Lee}.} \bibinfo{year}{2022}\natexlab{}.
\newblock \showarticletitle{Captivate! Contextual Language Guidance for Parent–Child Interaction}. In \bibinfo{booktitle}{\emph{Proceedings of the 2022 CHI Conference on Human Factors in Computing Systems}} (New Orleans, LA, USA) \emph{(\bibinfo{series}{CHI '22})}. \bibinfo{publisher}{Association for Computing Machinery}, \bibinfo{address}{New York, NY, USA}, Article \bibinfo{articleno}{219}, \bibinfo{numpages}{17}~pages.
\newblock
\showISBNx{9781450391573}
\urldef\tempurl%
\url{https://doi.org/10.1145/3491102.3501865}
\showDOI{\tempurl}


\bibitem[Laubscher and Light(2020)]%
        {Emily2020}
\bibfield{author}{\bibinfo{person}{Emily Laubscher} {and} \bibinfo{person}{Janice Light}.} \bibinfo{year}{2020}\natexlab{}.
\newblock \showarticletitle{Core vocabulary lists for young children and considerations for early language development: a narrative review}.
\newblock \bibinfo{journal}{\emph{Augmentative and Alternative Communication}} \bibinfo{volume}{36}, \bibinfo{number}{1} (\bibinfo{year}{2020}), \bibinfo{pages}{43--53}.
\newblock
\urldef\tempurl%
\url{https://doi.org/10.1080/07434618.2020.1737964}
\showDOI{\tempurl}
\showeprint{https://doi.org/10.1080/07434618.2020.1737964}
\newblock
\shownote{PMID: 32172598}.


\bibitem[Light and McNaughton(2012)]%
        {Janice2012}
\bibfield{author}{\bibinfo{person}{Janice Light} {and} \bibinfo{person}{David McNaughton}.} \bibinfo{year}{2012}\natexlab{}.
\newblock \showarticletitle{Supporting the Communication, Language, and Literacy Development of Children with Complex Communication Needs: State of the Science and Future Research Priorities}.
\newblock \bibinfo{journal}{\emph{Assistive Technology}} \bibinfo{volume}{24}, \bibinfo{number}{1} (\bibinfo{year}{2012}), \bibinfo{pages}{34--44}.
\newblock
\urldef\tempurl%
\url{https://doi.org/10.1080/10400435.2011.648717}
\showDOI{\tempurl}
\showeprint{https://doi.org/10.1080/10400435.2011.648717}


\bibitem[Lin and Faldowski(2023)]%
        {lin2023}
\bibfield{author}{\bibinfo{person}{Mei-Ling Lin} {and} \bibinfo{person}{Richard~A Faldowski}.} \bibinfo{year}{2023}\natexlab{}.
\newblock \showarticletitle{The relationship of parent support and child emotional regulation to school readiness}.
\newblock \bibinfo{journal}{\emph{International Journal of Environmental Research and Public Health}} \bibinfo{volume}{20}, \bibinfo{number}{6} (\bibinfo{year}{2023}), \bibinfo{pages}{4867}.
\newblock


\bibitem[Liu et~al\mbox{.}(2024)]%
        {liu2024corpus}
\bibfield{author}{\bibinfo{person}{Cuilin Liu}, \bibinfo{person}{Se-Eun Jhang}, \bibinfo{person}{Homin Park}, {and} \bibinfo{person}{Hyunjong Hahm}.} \bibinfo{year}{2024}\natexlab{}.
\newblock \showarticletitle{A Corpus-based Multilingual Comparison of AI-based Machine Translations}.
\newblock \bibinfo{journal}{\emph{Korea Journal of English Language and Linguistics}}  \bibinfo{volume}{24} (\bibinfo{year}{2024}), \bibinfo{pages}{257--276}.
\newblock


\bibitem[Mahoney and Perales(2003)]%
        {Gerald2003}
\bibfield{author}{\bibinfo{person}{Gerald Mahoney} {and} \bibinfo{person}{Frida Perales}.} \bibinfo{year}{2003}\natexlab{}.
\newblock \showarticletitle{Using Relationship-Focused Intervention to Enhance the Social—Emotional Functioning of Young Children with Autism Spectrum Disorders}.
\newblock \bibinfo{journal}{\emph{Topics in Early Childhood Special Education}} \bibinfo{volume}{23}, \bibinfo{number}{2} (\bibinfo{year}{2003}), \bibinfo{pages}{74--86}.
\newblock
\urldef\tempurl%
\url{https://doi.org/10.1177/02711214030230020301}
\showDOI{\tempurl}
\showeprint{https://doi.org/10.1177/02711214030230020301}


\bibitem[Mankoff et~al\mbox{.}(2010)]%
        {Mankoff2010}
\bibfield{author}{\bibinfo{person}{Jennifer Mankoff}, \bibinfo{person}{Gillian~R. Hayes}, {and} \bibinfo{person}{Devva Kasnitz}.} \bibinfo{year}{2010}\natexlab{}.
\newblock \showarticletitle{Disability studies as a source of critical inquiry for the field of assistive technology}. In \bibinfo{booktitle}{\emph{Proceedings of the 12th International ACM SIGACCESS Conference on Computers and Accessibility}} (Orlando, Florida, USA) \emph{(\bibinfo{series}{ASSETS '10})}. \bibinfo{publisher}{Association for Computing Machinery}, \bibinfo{address}{New York, NY, USA}, \bibinfo{pages}{3–10}.
\newblock
\showISBNx{9781605588810}
\urldef\tempurl%
\url{https://doi.org/10.1145/1878803.1878807}
\showDOI{\tempurl}


\bibitem[Massaro and Bosseler(2006)]%
        {Dominic2006}
\bibfield{author}{\bibinfo{person}{Dominic~W. Massaro} {and} \bibinfo{person}{Alexis Bosseler}.} \bibinfo{year}{2006}\natexlab{}.
\newblock \showarticletitle{Read my lips: The importance of the face in a computer-animated tutor for vocabulary learning by children with autism}.
\newblock \bibinfo{journal}{\emph{Autism}} \bibinfo{volume}{10}, \bibinfo{number}{5} (\bibinfo{year}{2006}), \bibinfo{pages}{495--510}.
\newblock
\urldef\tempurl%
\url{https://doi.org/10.1177/1362361306066599}
\showDOI{\tempurl}
\showeprint{https://doi.org/10.1177/1362361306066599}
\newblock
\shownote{PMID: 16940315}.


\bibitem[McCauley and Solomon(2022)]%
        {McCauley2022}
\bibfield{author}{\bibinfo{person}{James~B. McCauley} {and} \bibinfo{person}{Marjorie Solomon}.} \bibinfo{year}{2022}\natexlab{}.
\newblock \showarticletitle{Characterizing Parent–Child Interactions in Families of Autistic Children in Late Childhood}.
\newblock \bibinfo{journal}{\emph{Social Sciences}} \bibinfo{volume}{11}, \bibinfo{number}{3} (\bibinfo{year}{2022}).
\newblock
\showISSN{2076-0760}
\urldef\tempurl%
\url{https://doi.org/10.3390/socsci11030100}
\showDOI{\tempurl}


\bibitem[McEwen(2014)]%
        {Rhonda2014}
\bibfield{author}{\bibinfo{person}{Rhonda McEwen}.} \bibinfo{year}{2014}\natexlab{}.
\newblock \showarticletitle{Mediating sociality: the use of iPod Touch™ devices in the classrooms of students with autism in Canada}.
\newblock \bibinfo{journal}{\emph{Information, Communication \& Society}} \bibinfo{volume}{17}, \bibinfo{number}{10} (\bibinfo{year}{2014}), \bibinfo{pages}{1264--1279}.
\newblock
\urldef\tempurl%
\url{https://doi.org/10.1080/1369118X.2014.920041}
\showDOI{\tempurl}
\showeprint{https://doi.org/10.1080/1369118X.2014.920041}


\bibitem[Meta(2024)]%
        {ReactNative}
\bibfield{author}{\bibinfo{person}{Meta}.} \bibinfo{year}{2024}\natexlab{}.
\newblock \bibinfo{title}{{React Native - Learn Once, Write Everywhere}}.
\newblock
\newblock
\urldef\tempurl%
\url{https://reactnative.dev/}
\showURL{%
Retrieved Sep 10, 2024 from \tempurl}


\bibitem[Microsoft(2024)]%
        {TypeScript}
\bibfield{author}{\bibinfo{person}{Microsoft}.} \bibinfo{year}{2024}\natexlab{}.
\newblock \bibinfo{title}{{TypeScript}}.
\newblock
\newblock
\urldef\tempurl%
\url{https://www.typescriptlang.org}
\showURL{%
Retrieved Sep 10, 2024 from \tempurl}


\bibitem[Mucchetti(2013)]%
        {Charlotte2013}
\bibfield{author}{\bibinfo{person}{Charlotte~A Mucchetti}.} \bibinfo{year}{2013}\natexlab{}.
\newblock \showarticletitle{Adapted shared reading at school for minimally verbal students with autism}.
\newblock \bibinfo{journal}{\emph{Autism}} \bibinfo{volume}{17}, \bibinfo{number}{3} (\bibinfo{year}{2013}), \bibinfo{pages}{358--372}.
\newblock
\urldef\tempurl%
\url{https://doi.org/10.1177/1362361312470495}
\showDOI{\tempurl}
\showeprint{https://doi.org/10.1177/1362361312470495}
\newblock
\shownote{PMID: 23592847}.


\bibitem[Naous et~al\mbox{.}(2024)]%
        {naous2024CulturalBiasLLM}
\bibfield{author}{\bibinfo{person}{Tarek Naous}, \bibinfo{person}{Michael Ryan}, \bibinfo{person}{Alan Ritter}, {and} \bibinfo{person}{Wei Xu}.} \bibinfo{year}{2024}\natexlab{}.
\newblock \showarticletitle{Having Beer after Prayer? Measuring Cultural Bias in Large Language Models}. In \bibinfo{booktitle}{\emph{Proceedings of the 62nd Annual Meeting of the Association for Computational Linguistics (Volume 1: Long Papers)}}, \bibfield{editor}{\bibinfo{person}{Lun-Wei Ku}, \bibinfo{person}{Andre Martins}, {and} \bibinfo{person}{Vivek Srikumar}} (Eds.). \bibinfo{publisher}{Association for Computational Linguistics}, \bibinfo{address}{Bangkok, Thailand}, \bibinfo{pages}{16366--16393}.
\newblock
\urldef\tempurl%
\url{https://aclanthology.org/2024.acl-long.862}
\showURL{%
\tempurl}


\bibitem[Naples et~al\mbox{.}(2023)]%
        {Adam2023}
\bibfield{author}{\bibinfo{person}{Adam Naples}, \bibinfo{person}{Elena~J Tenenbaum}, \bibinfo{person}{Richard~N Jones}, \bibinfo{person}{Giulia Righi}, \bibinfo{person}{Stephen~J Sheinkopf}, {and} \bibinfo{person}{Inge-Marie Eigsti}.} \bibinfo{year}{2023}\natexlab{}.
\newblock \showarticletitle{Exploring communicative competence in autistic children who are minimally verbal: The Low Verbal Investigatory Survey for Autism (LVIS)}.
\newblock \bibinfo{journal}{\emph{Autism}} \bibinfo{volume}{27}, \bibinfo{number}{5} (\bibinfo{year}{2023}), \bibinfo{pages}{1391--1406}.
\newblock
\urldef\tempurl%
\url{https://doi.org/10.1177/13623613221136657}
\showDOI{\tempurl}
\showeprint{https://doi.org/10.1177/13623613221136657}
\newblock
\shownote{PMID: 36373838}.


\bibitem[Nazari et~al\mbox{.}(2024)]%
        {Nazari2024}
\bibfield{author}{\bibinfo{person}{Ahmadreza Nazari}, \bibinfo{person}{Lorans Alabood}, \bibinfo{person}{Kaylyn~B Feeley}, \bibinfo{person}{Vikram~K. Jaswal}, {and} \bibinfo{person}{Diwakar Krishnamurthy}.} \bibinfo{year}{2024}\natexlab{}.
\newblock \showarticletitle{Personalizing an AR-based Communication System for Nonspeaking Autistic Users}. In \bibinfo{booktitle}{\emph{Proceedings of the 29th International Conference on Intelligent User Interfaces}} (Greenville, SC, USA) \emph{(\bibinfo{series}{IUI '24})}. \bibinfo{publisher}{Association for Computing Machinery}, \bibinfo{address}{New York, NY, USA}, \bibinfo{pages}{731–741}.
\newblock
\showISBNx{9798400705083}
\urldef\tempurl%
\url{https://doi.org/10.1145/3640543.3645153}
\showDOI{\tempurl}


\bibitem[Neamtu et~al\mbox{.}(2019)]%
        {neamtu2019}
\bibfield{author}{\bibinfo{person}{Rodica Neamtu}, \bibinfo{person}{Andr{\'e} Camara}, \bibinfo{person}{Carlos Pereira}, {and} \bibinfo{person}{Rafael Ferreira}.} \bibinfo{year}{2019}\natexlab{}.
\newblock \showarticletitle{Using artificial intelligence for augmentative alternative communication for children with disabilities}. In \bibinfo{booktitle}{\emph{Human-Computer Interaction--INTERACT 2019: 17th IFIP TC 13 International Conference, Paphos, Cyprus, September 2--6, 2019, Proceedings, Part I 17}}. Springer, \bibinfo{pages}{234--243}.
\newblock


\bibitem[Neimy and Fossett(2022)]%
        {Neimy2022}
\bibfield{author}{\bibinfo{person}{Hayley Neimy} {and} \bibinfo{person}{Brenda Fossett}.} \bibinfo{year}{2022}\natexlab{}.
\newblock \bibinfo{booktitle}{\emph{Augmentative and Alternative Communication (AAC) Systems}}.
\newblock \bibinfo{publisher}{Springer International Publishing}, \bibinfo{address}{Cham}, \bibinfo{pages}{375--401}.
\newblock
\showISBNx{978-3-030-96478-8}
\urldef\tempurl%
\url{https://doi.org/10.1007/978-3-030-96478-8_20}
\showDOI{\tempurl}


\bibitem[Nunes and Hanline(2007)]%
        {Debora2007}
\bibfield{author}{\bibinfo{person}{Debora Nunes} {and} \bibinfo{person}{Mary~Frances Hanline}.} \bibinfo{year}{2007}\natexlab{}.
\newblock \showarticletitle{Enhancing the Alternative and Augmentative Communication Use of a Child with Autism through a Parent‐implemented Naturalistic Intervention}.
\newblock \bibinfo{journal}{\emph{International Journal of Disability, Development and Education}} \bibinfo{volume}{54}, \bibinfo{number}{2} (\bibinfo{year}{2007}), \bibinfo{pages}{177--197}.
\newblock
\urldef\tempurl%
\url{https://doi.org/10.1080/10349120701330495}
\showDOI{\tempurl}
\showeprint{https://doi.org/10.1080/10349120701330495}


\bibitem[Odermatt et~al\mbox{.}(2022)]%
        {Odermatt2022}
\bibfield{author}{\bibinfo{person}{Salome~D. Odermatt}, \bibinfo{person}{Wenke Möhring}, \bibinfo{person}{Silvia Grieder}, {and} \bibinfo{person}{Alexander Grob}.} \bibinfo{year}{2022}\natexlab{}.
\newblock \showarticletitle{Cognitive and Developmental Functions in Autistic and Non-Autistic Children and Adolescents: Evidence from the Intelligence and Development Scales–2}.
\newblock \bibinfo{journal}{\emph{Journal of Intelligence}} \bibinfo{volume}{10}, \bibinfo{number}{4} (\bibinfo{year}{2022}).
\newblock
\showISSN{2079-3200}
\urldef\tempurl%
\url{https://doi.org/10.3390/jintelligence10040112}
\showDOI{\tempurl}


\bibitem[Oono et~al\mbox{.}(2013)]%
        {Oono2013}
\bibfield{author}{\bibinfo{person}{Inalegwu~P Oono}, \bibinfo{person}{Emma~J Honey}, {and} \bibinfo{person}{Helen McConachie}.} \bibinfo{year}{2013}\natexlab{}.
\newblock \showarticletitle{Parent-mediated early intervention for young children with autism spectrum disorders (ASD)}.
\newblock \bibinfo{journal}{\emph{Evidence-Based Child Health: A Cochrane Review Journal}} \bibinfo{volume}{8}, \bibinfo{number}{6} (\bibinfo{year}{2013}), \bibinfo{pages}{2380--2479}.
\newblock
\urldef\tempurl%
\url{https://doi.org/10.1002/ebch.1952}
\showDOI{\tempurl}
\showeprint{https://onlinelibrary.wiley.com/doi/pdf/10.1002/ebch.1952}


\bibitem[OpenAI(2023)]%
        {openai2023gpt4}
\bibfield{author}{\bibinfo{person}{OpenAI}.} \bibinfo{year}{2023}\natexlab{}.
\newblock \bibinfo{title}{GPT-4 Technical Report}.
\newblock
\newblock
\showeprint[arxiv]{2303.08774}~[cs.CL]


\bibitem[OpenAI(2024)]%
        {openai}
\bibfield{author}{\bibinfo{person}{OpenAI}.} \bibinfo{year}{2024}\natexlab{}.
\newblock \bibinfo{title}{{OpenAI API}}.
\newblock
\newblock
\urldef\tempurl%
\url{https://openai.com/api/}
\showURL{%
Retrieved Sep 10, 2024 from \tempurl}


\bibitem[Park et~al\mbox{.}(2016)]%
        {Parl2016KAAC}
\bibfield{author}{\bibinfo{person}{Eun-Hye Park}, \bibinfo{person}{Young~Tae Kim}, \bibinfo{person}{Ki-Hyung Hong}, \bibinfo{person}{Seokjeong Yeon}, \bibinfo{person}{Kyung~Yang Kim}, {and} \bibinfo{person}{Janghyun Lim}.} \bibinfo{year}{2016}\natexlab{}.
\newblock \showarticletitle{Development of Korean Ewha-AAC Symbols: Validity of Vocabulary and Graphic Symbols}.
\newblock \bibinfo{journal}{\emph{AAC Research \& Practice}} \bibinfo{volume}{4}, \bibinfo{number}{2} (\bibinfo{year}{2016}), \bibinfo{pages}{19--40}.
\newblock
\showISSN{2288-7180}


\bibitem[Patel and Radhakrishnan(2007)]%
        {patel2007}
\bibfield{author}{\bibinfo{person}{Rupal Patel} {and} \bibinfo{person}{Rajiv Radhakrishnan}.} \bibinfo{year}{2007}\natexlab{}.
\newblock \showarticletitle{Enhancing Access to Situational Vocabulary by Leveraging Geographic Context.}
\newblock \bibinfo{journal}{\emph{Assistive Technology Outcomes and Benefits}} \bibinfo{volume}{4}, \bibinfo{number}{1} (\bibinfo{year}{2007}), \bibinfo{pages}{99--114}.
\newblock


\bibitem[Paynter et~al\mbox{.}(2018)]%
        {Paynter2018}
\bibfield{author}{\bibinfo{person}{J. Paynter}, \bibinfo{person}{D. Trembath}, {and} \bibinfo{person}{A. Lane}.} \bibinfo{year}{2018}\natexlab{}.
\newblock \showarticletitle{Differential outcome subgroups in children with autism spectrum disorder attending early intervention}.
\newblock \bibinfo{journal}{\emph{Journal of Intellectual Disability Research}} \bibinfo{volume}{62}, \bibinfo{number}{7} (\bibinfo{year}{2018}), \bibinfo{pages}{650--659}.
\newblock
\urldef\tempurl%
\url{https://doi.org/10.1111/jir.12504}
\showDOI{\tempurl}
\showeprint{https://onlinelibrary.wiley.com/doi/pdf/10.1111/jir.12504}


\bibitem[Pinheiro and Bates(2000)]%
        {Pinheiro2000MixedEffects}
\bibfield{author}{\bibinfo{person}{Jos{\'{e}} Pinheiro} {and} \bibinfo{person}{Douglas Bates}.} \bibinfo{year}{2000}\natexlab{}.
\newblock \bibinfo{booktitle}{\emph{{Mixed-Effects Models in S and S-PLUS}} (\bibinfo{edition}{1} ed.)}.
\newblock \bibinfo{publisher}{Springer-Verlag}, \bibinfo{address}{New York}. 528 pages.
\newblock
\showISBNx{0-387-98957-9}
\urldef\tempurl%
\url{https://doi.org/10.1007/b98882}
\showDOI{\tempurl}


\bibitem[Plutchik(1980)]%
        {plutchik1980general}
\bibfield{author}{\bibinfo{person}{Robert Plutchik}.} \bibinfo{year}{1980}\natexlab{}.
\newblock \showarticletitle{A general psychoevolutionary theory of emotion}.
\newblock In \bibinfo{booktitle}{\emph{Theories of emotion}}. \bibinfo{publisher}{Elsevier}, \bibinfo{pages}{3--33}.
\newblock


\bibitem[R.~Michael~Barker and Thiemann-Bourque(2013)]%
        {Michael2013}
\bibfield{author}{\bibinfo{person}{Nancy C.~Brady R.~Michael~Barker, Sanae~Akaba} {and} \bibinfo{person}{Kathy Thiemann-Bourque}.} \bibinfo{year}{2013}\natexlab{}.
\newblock \showarticletitle{Support for AAC Use in Preschool, and Growth in Language Skills, for Young Children with Developmental Disabilities}.
\newblock \bibinfo{journal}{\emph{Augmentative and Alternative Communication}} \bibinfo{volume}{29}, \bibinfo{number}{4} (\bibinfo{year}{2013}), \bibinfo{pages}{334--346}.
\newblock
\urldef\tempurl%
\url{https://doi.org/10.3109/07434618.2013.848933}
\showDOI{\tempurl}
\showeprint{https://doi.org/10.3109/07434618.2013.848933}
\newblock
\shownote{PMID: 24229337}.


\bibitem[Roberts and Kaiser(2011)]%
        {Megan2011}
\bibfield{author}{\bibinfo{person}{Megan~Y. Roberts} {and} \bibinfo{person}{Ann~P. Kaiser}.} \bibinfo{year}{2011}\natexlab{}.
\newblock \showarticletitle{The Effectiveness of Parent-Implemented Language Interventions: A Meta-Analysis}.
\newblock \bibinfo{journal}{\emph{American Journal of Speech-Language Pathology}} \bibinfo{volume}{20}, \bibinfo{number}{3} (\bibinfo{year}{2011}), \bibinfo{pages}{180--199}.
\newblock
\urldef\tempurl%
\url{https://doi.org/10.1044/1058-0360(2011/10-0055)}
\showDOI{\tempurl}
\showeprint{https://pubs.asha.org/doi/pdf/10.1044/1058-0360\%282011/10-0055\%29}


\bibitem[Sacks et~al\mbox{.}(1974)]%
        {sacks1974}
\bibfield{author}{\bibinfo{person}{Harvey Sacks}, \bibinfo{person}{Emanuel~A Schegloff}, {and} \bibinfo{person}{Gail Jefferson}.} \bibinfo{year}{1974}\natexlab{}.
\newblock \showarticletitle{A simplest systematics for the organization of turn-taking for conversation}.
\newblock \bibinfo{journal}{\emph{language}} \bibinfo{volume}{50}, \bibinfo{number}{4} (\bibinfo{year}{1974}), \bibinfo{pages}{696--735}.
\newblock


\bibitem[Schlosser and Wendt(2008)]%
        {Ralf2008}
\bibfield{author}{\bibinfo{person}{Ralf~W. Schlosser} {and} \bibinfo{person}{Oliver Wendt}.} \bibinfo{year}{2008}\natexlab{}.
\newblock \showarticletitle{Effects of Augmentative and Alternative Communication Intervention on Speech Production in Children With Autism: A Systematic Review}.
\newblock \bibinfo{journal}{\emph{American Journal of Speech-Language Pathology}} \bibinfo{volume}{17}, \bibinfo{number}{3} (\bibinfo{year}{2008}), \bibinfo{pages}{212--230}.
\newblock
\urldef\tempurl%
\url{https://doi.org/10.1044/1058-0360(2008/021)}
\showDOI{\tempurl}
\showeprint{https://pubs.asha.org/doi/pdf/10.1044/1058-0360\%282008/021\%29}


\bibitem[Schuler et~al\mbox{.}(1997)]%
        {schuler1997}
\bibfield{author}{\bibinfo{person}{Adriana~L Schuler}, \bibinfo{person}{Barry~M Prizant}, {and} \bibinfo{person}{Amy~M Wetherby}.} \bibinfo{year}{1997}\natexlab{}.
\newblock \showarticletitle{Enhancing language and communication development: Prelinguistic approaches}.
\newblock \bibinfo{journal}{\emph{Handbook of autism and pervasive developmental disorders}}  \bibinfo{volume}{2} (\bibinfo{year}{1997}).
\newblock


\bibitem[Scott-Van~Zeeland et~al\mbox{.}(2010)]%
        {Scott2010}
\bibfield{author}{\bibinfo{person}{Ashley~A. Scott-Van~Zeeland}, \bibinfo{person}{Mirella Dapretto}, \bibinfo{person}{Dara~G. Ghahremani}, \bibinfo{person}{Russell~A. Poldrack}, {and} \bibinfo{person}{Susan~Y. Bookheimer}.} \bibinfo{year}{2010}\natexlab{}.
\newblock \showarticletitle{Reward processing in autism}.
\newblock \bibinfo{journal}{\emph{Autism Research}} \bibinfo{volume}{3}, \bibinfo{number}{2} (\bibinfo{year}{2010}), \bibinfo{pages}{53--67}.
\newblock
\urldef\tempurl%
\url{https://doi.org/10.1002/aur.122}
\showDOI{\tempurl}
\showeprint{https://onlinelibrary.wiley.com/doi/pdf/10.1002/aur.122}


\bibitem[Seo et~al\mbox{.}(2024)]%
        {seo2024chacha}
\bibfield{author}{\bibinfo{person}{Woosuk Seo}, \bibinfo{person}{Chanmo Yang}, {and} \bibinfo{person}{Young-Ho Kim}.} \bibinfo{year}{2024}\natexlab{}.
\newblock \showarticletitle{ChaCha: Leveraging Large Language Models to Prompt Children to Share Their Emotions about Personal Events}. In \bibinfo{booktitle}{\emph{Proceedings of the CHI Conference on Human Factors in Computing Systems}} (Honolulu, HI, USA) \emph{(\bibinfo{series}{CHI '24})}. \bibinfo{publisher}{Association for Computing Machinery}, \bibinfo{address}{New York, NY, USA}, Article \bibinfo{articleno}{903}, \bibinfo{numpages}{20}~pages.
\newblock
\showISBNx{9798400703300}
\urldef\tempurl%
\url{https://doi.org/10.1145/3613904.3642152}
\showDOI{\tempurl}


\bibitem[Shin et~al\mbox{.}(2023)]%
        {shin2023planfitting}
\bibfield{author}{\bibinfo{person}{Donghoon Shin}, \bibinfo{person}{Gary Hsieh}, {and} \bibinfo{person}{Young-Ho Kim}.} \bibinfo{year}{2023}\natexlab{}.
\newblock \bibinfo{title}{PlanFitting: Tailoring Personalized Exercise Plans with Large Language Models}.
\newblock
\newblock
\showeprint[arxiv]{2309.12555}~[cs.HC]
\urldef\tempurl%
\url{https://arxiv.org/abs/2309.12555}
\showURL{%
\tempurl}


\bibitem[Sigafoos and Reichle(1993)]%
        {sigafoos1993}
\bibfield{author}{\bibinfo{person}{J Sigafoos} {and} \bibinfo{person}{J Reichle}.} \bibinfo{year}{1993}\natexlab{}.
\newblock \showarticletitle{Establishing spontaneous verbal behavior}.
\newblock \bibinfo{journal}{\emph{Strategies for teaching students with mild to severe mental retardation}}  \bibinfo{volume}{5} (\bibinfo{year}{1993}), \bibinfo{pages}{191--230}.
\newblock


\bibitem[Siller and Sigman(2002)]%
        {siller2002}
\bibfield{author}{\bibinfo{person}{Michael Siller} {and} \bibinfo{person}{Marian Sigman}.} \bibinfo{year}{2002}\natexlab{}.
\newblock \showarticletitle{The behaviors of parents of children with autism predict the subsequent development of their children's communication}.
\newblock \bibinfo{journal}{\emph{Journal of autism and developmental disorders}}  \bibinfo{volume}{32} (\bibinfo{year}{2002}), \bibinfo{pages}{77--89}.
\newblock


\bibitem[Singh et~al\mbox{.}(1995)]%
        {Nirbhay1995}
\bibfield{author}{\bibinfo{person}{Nirbhay~N. Singh}, \bibinfo{person}{W.~John Curtis}, \bibinfo{person}{Cynthia~R. Ellis}, \bibinfo{person}{Mary~W. Nicholson}, \bibinfo{person}{Terri~M. Villani}, {and} \bibinfo{person}{Hollis~A. Wechsler}.} \bibinfo{year}{1995}\natexlab{}.
\newblock \showarticletitle{Psychometric Analysis of the Family Empowerment Scale}.
\newblock \bibinfo{journal}{\emph{Journal of Emotional and Behavioral Disorders}} \bibinfo{volume}{3}, \bibinfo{number}{2} (\bibinfo{year}{1995}), \bibinfo{pages}{85--91}.
\newblock
\urldef\tempurl%
\url{https://doi.org/10.1177/106342669500300203}
\showDOI{\tempurl}
\showeprint{https://doi.org/10.1177/106342669500300203}


\bibitem[Singh and Wilson(2024)]%
        {Singh2024}
\bibfield{author}{\bibinfo{person}{Sanyukta Singh} {and} \bibinfo{person}{Cara Wilson}.} \bibinfo{year}{2024}\natexlab{}.
\newblock \showarticletitle{Autistic Expression Beyond the Verbal - Studying Minimally-Verbal Autistic Indian Children's Embodied Interactions with Screen-Based Technology}. In \bibinfo{booktitle}{\emph{Proceedings of the 23rd Annual ACM Interaction Design and Children Conference}} (Delft, Netherlands) \emph{(\bibinfo{series}{IDC '24})}. \bibinfo{publisher}{Association for Computing Machinery}, \bibinfo{address}{New York, NY, USA}, \bibinfo{pages}{612–624}.
\newblock
\showISBNx{9798400704420}
\urldef\tempurl%
\url{https://doi.org/10.1145/3628516.3655792}
\showDOI{\tempurl}


\bibitem[Skwerer et~al\mbox{.}(2016)]%
        {Daniela2016}
\bibfield{author}{\bibinfo{person}{Daniela~Plesa Skwerer}, \bibinfo{person}{Samantha~E Jordan}, \bibinfo{person}{Briana~H Brukilacchio}, {and} \bibinfo{person}{Helen Tager-Flusberg}.} \bibinfo{year}{2016}\natexlab{}.
\newblock \showarticletitle{Comparing methods for assessing receptive language skills in minimally verbal children and adolescents with autism spectrum disorders}.
\newblock \bibinfo{journal}{\emph{Autism}} \bibinfo{volume}{20}, \bibinfo{number}{5} (\bibinfo{year}{2016}), \bibinfo{pages}{591--604}.
\newblock
\urldef\tempurl%
\url{https://doi.org/10.1177/1362361315600146}
\showDOI{\tempurl}
\showeprint{https://doi.org/10.1177/1362361315600146}
\newblock
\shownote{PMID: 26408635}.


\bibitem[Song et~al\mbox{.}(2016)]%
        {Song2016}
\bibfield{author}{\bibinfo{person}{Seokwoo Song}, \bibinfo{person}{Seungho Kim}, \bibinfo{person}{John Kim}, \bibinfo{person}{Wonjeong Park}, {and} \bibinfo{person}{Dongsun Yim}.} \bibinfo{year}{2016}\natexlab{}.
\newblock \showarticletitle{TalkLIME: mobile system intervention to improve parent-child interaction for children with language delay}. In \bibinfo{booktitle}{\emph{Proceedings of the 2016 ACM International Joint Conference on Pervasive and Ubiquitous Computing}} (Heidelberg, Germany) \emph{(\bibinfo{series}{UbiComp '16})}. \bibinfo{publisher}{Association for Computing Machinery}, \bibinfo{address}{New York, NY, USA}, \bibinfo{pages}{304–315}.
\newblock
\showISBNx{9781450344616}
\urldef\tempurl%
\url{https://doi.org/10.1145/2971648.2971650}
\showDOI{\tempurl}


\bibitem[Spiel et~al\mbox{.}(2019)]%
        {Spiel2019}
\bibfield{author}{\bibinfo{person}{Katta Spiel}, \bibinfo{person}{Christopher Frauenberger}, \bibinfo{person}{Os Keyes}, {and} \bibinfo{person}{Geraldine Fitzpatrick}.} \bibinfo{year}{2019}\natexlab{}.
\newblock \showarticletitle{Agency of Autistic Children in Technology Research—A Critical Literature Review}.
\newblock \bibinfo{journal}{\emph{ACM Trans. Comput.-Hum. Interact.}} \bibinfo{volume}{26}, \bibinfo{number}{6}, Article \bibinfo{articleno}{38} (\bibinfo{date}{nov} \bibinfo{year}{2019}), \bibinfo{numpages}{40}~pages.
\newblock
\showISSN{1073-0516}
\urldef\tempurl%
\url{https://doi.org/10.1145/3344919}
\showDOI{\tempurl}


\bibitem[Subakan et~al\mbox{.}(2021)]%
        {Subakan2021}
\bibfield{author}{\bibinfo{person}{Cem Subakan}, \bibinfo{person}{Mirco Ravanelli}, \bibinfo{person}{Samuele Cornell}, \bibinfo{person}{Mirko Bronzi}, {and} \bibinfo{person}{Jianyuan Zhong}.} \bibinfo{year}{2021}\natexlab{}.
\newblock \showarticletitle{Attention Is All You Need In Speech Separation}. In \bibinfo{booktitle}{\emph{ICASSP 2021 - 2021 IEEE International Conference on Acoustics, Speech and Signal Processing (ICASSP)}}. \bibinfo{pages}{21--25}.
\newblock
\urldef\tempurl%
\url{https://doi.org/10.1109/ICASSP39728.2021.9413901}
\showDOI{\tempurl}


\bibitem[Suess et~al\mbox{.}(2014)]%
        {suess2014}
\bibfield{author}{\bibinfo{person}{Alyssa~N Suess}, \bibinfo{person}{Patrick~W Romani}, \bibinfo{person}{David~P Wacker}, \bibinfo{person}{Shannon~M Dyson}, \bibinfo{person}{Jennifer~L Kuhle}, \bibinfo{person}{John~F Lee}, \bibinfo{person}{Scott~D Lindgren}, \bibinfo{person}{Todd~G Kopelman}, \bibinfo{person}{Kelly~E Pelzel}, {and} \bibinfo{person}{Debra~B Waldron}.} \bibinfo{year}{2014}\natexlab{}.
\newblock \showarticletitle{Evaluating the treatment fidelity of parents who conduct in-home functional communication training with coaching via telehealth}.
\newblock \bibinfo{journal}{\emph{Journal of Behavioral Education}}  \bibinfo{volume}{23} (\bibinfo{year}{2014}), \bibinfo{pages}{34--59}.
\newblock


\bibitem[Teresa~Iacono and Erickson(2016)]%
        {Teresa2016}
\bibfield{author}{\bibinfo{person}{David~Trembath Teresa~Iacono} {and} \bibinfo{person}{Shane Erickson}.} \bibinfo{year}{2016}\natexlab{}.
\newblock \showarticletitle{The role of augmentative and alternative communication for children with autism: current status and future trends}.
\newblock \bibinfo{journal}{\emph{Neuropsychiatric Disease and Treatment}}  \bibinfo{volume}{12} (\bibinfo{year}{2016}), \bibinfo{pages}{2349--2361}.
\newblock
\urldef\tempurl%
\url{https://doi.org/10.2147/NDT.S95967}
\showDOI{\tempurl}
\showeprint{https://www.tandfonline.com/doi/pdf/10.2147/NDT.S95967}
\newblock
\shownote{PMID: 27703354}.


\bibitem[Thunberg et~al\mbox{.}(2011)]%
        {thunberg2011}
\bibfield{author}{\bibinfo{person}{Gunilla Thunberg}, \bibinfo{person}{Elisabeth Ahls{\'e}n}, {and} \bibinfo{person}{Annika~Dahlgren Sandberg}.} \bibinfo{year}{2011}\natexlab{}.
\newblock \showarticletitle{Autism, communication and use of a speech-generating device in different environments--a case study}.
\newblock \bibinfo{journal}{\emph{Journal of Assistive Technologies}} \bibinfo{volume}{5}, \bibinfo{number}{4} (\bibinfo{year}{2011}), \bibinfo{pages}{181--198}.
\newblock


\bibitem[Tintarev et~al\mbox{.}(2016)]%
        {TINTAREV20161}
\bibfield{author}{\bibinfo{person}{Nava Tintarev}, \bibinfo{person}{Ehud Reiter}, \bibinfo{person}{Rolf Black}, \bibinfo{person}{Annalu Waller}, {and} \bibinfo{person}{Joe Reddington}.} \bibinfo{year}{2016}\natexlab{}.
\newblock \showarticletitle{Personal storytelling: Using Natural Language Generation for children with complex communication needs, in the wild…}.
\newblock \bibinfo{journal}{\emph{International Journal of Human-Computer Studies}}  \bibinfo{volume}{92-93} (\bibinfo{year}{2016}), \bibinfo{pages}{1--16}.
\newblock
\showISSN{1071-5819}
\urldef\tempurl%
\url{https://doi.org/10.1016/j.ijhcs.2016.04.005}
\showDOI{\tempurl}


\bibitem[Trnka et~al\mbox{.}(2008)]%
        {Trnka2008}
\bibfield{author}{\bibinfo{person}{Keith Trnka}, \bibinfo{person}{John McCaw}, \bibinfo{person}{Debra Yarrington}, \bibinfo{person}{Kathleen~F. McCoy}, {and} \bibinfo{person}{Christopher Pennington}.} \bibinfo{year}{2008}\natexlab{}.
\newblock \showarticletitle{Word prediction and communication rate in AAC}. In \bibinfo{booktitle}{\emph{Proceedings of the IASTED International Conference on Telehealth/Assistive Technologies}} (Baltimore, Maryland) \emph{(\bibinfo{series}{Telehealth/AT '08})}. \bibinfo{publisher}{ACTA Press}, \bibinfo{address}{USA}, \bibinfo{pages}{19–24}.
\newblock
\showISBNx{9780889867406}


\bibitem[Valencia et~al\mbox{.}(2023)]%
        {Valencia2023}
\bibfield{author}{\bibinfo{person}{Stephanie Valencia}, \bibinfo{person}{Richard Cave}, \bibinfo{person}{Krystal Kallarackal}, \bibinfo{person}{Katie Seaver}, \bibinfo{person}{Michael Terry}, {and} \bibinfo{person}{Shaun~K. Kane}.} \bibinfo{year}{2023}\natexlab{}.
\newblock \showarticletitle{“The less I type, the better”: How AI Language Models can Enhance or Impede Communication for AAC Users}. In \bibinfo{booktitle}{\emph{Proceedings of the 2023 CHI Conference on Human Factors in Computing Systems}} (Hamburg, Germany) \emph{(\bibinfo{series}{CHI '23})}. \bibinfo{publisher}{Association for Computing Machinery}, \bibinfo{address}{New York, NY, USA}, Article \bibinfo{articleno}{830}, \bibinfo{numpages}{14}~pages.
\newblock
\showISBNx{9781450394215}
\urldef\tempurl%
\url{https://doi.org/10.1145/3544548.3581560}
\showDOI{\tempurl}


\bibitem[Valencia et~al\mbox{.}(2024)]%
        {Valencia2024}
\bibfield{author}{\bibinfo{person}{Stephanie Valencia}, \bibinfo{person}{Jessica Huynh}, \bibinfo{person}{Emma~Y Jiang}, \bibinfo{person}{Yufei Wu}, \bibinfo{person}{Teresa Wan}, \bibinfo{person}{Zixuan Zheng}, \bibinfo{person}{Henny Admoni}, \bibinfo{person}{Jeffrey~P Bigham}, {and} \bibinfo{person}{Amy Pavel}.} \bibinfo{year}{2024}\natexlab{}.
\newblock \showarticletitle{COMPA: Using Conversation Context to Achieve Common Ground in AAC}. In \bibinfo{booktitle}{\emph{Proceedings of the CHI Conference on Human Factors in Computing Systems}} (Honolulu, HI, USA) \emph{(\bibinfo{series}{CHI '24})}. \bibinfo{publisher}{Association for Computing Machinery}, \bibinfo{address}{New York, NY, USA}, Article \bibinfo{articleno}{915}, \bibinfo{numpages}{18}~pages.
\newblock
\showISBNx{9798400703300}
\urldef\tempurl%
\url{https://doi.org/10.1145/3613904.3642762}
\showDOI{\tempurl}


\bibitem[van Grunsven and Roeser(2022)]%
        {Janna2022}
\bibfield{author}{\bibinfo{person}{Janna van Grunsven} {and} \bibinfo{person}{Sabine Roeser}.} \bibinfo{year}{2022}\natexlab{}.
\newblock \showarticletitle{AAC Technology, Autism, and the Empathic Turn}.
\newblock \bibinfo{journal}{\emph{Social Epistemology}} \bibinfo{volume}{36}, \bibinfo{number}{1} (\bibinfo{year}{2022}), \bibinfo{pages}{95--110}.
\newblock
\urldef\tempurl%
\url{https://doi.org/10.1080/02691728.2021.1897189}
\showDOI{\tempurl}
\showeprint{https://doi.org/10.1080/02691728.2021.1897189}


\bibitem[Venkatesh and Bala(2008)]%
        {Venkatesh2008}
\bibfield{author}{\bibinfo{person}{Viswanath Venkatesh} {and} \bibinfo{person}{Hillol Bala}.} \bibinfo{year}{2008}\natexlab{}.
\newblock \showarticletitle{Technology Acceptance Model 3 and a Research Agenda on Interventions}.
\newblock \bibinfo{journal}{\emph{Decision Sciences}} \bibinfo{volume}{39}, \bibinfo{number}{2} (\bibinfo{year}{2008}), \bibinfo{pages}{273--315}.
\newblock
\urldef\tempurl%
\url{https://doi.org/10.1111/j.1540-5915.2008.00192.x}
\showDOI{\tempurl}
\showeprint{https://onlinelibrary.wiley.com/doi/pdf/10.1111/j.1540-5915.2008.00192.x}


\bibitem[Vidal et~al\mbox{.}(2020)]%
        {Verónica2020}
\bibfield{author}{\bibinfo{person}{Ver{\'o}nica Vidal}, \bibinfo{person}{Anita McAllister}, {and} \bibinfo{person}{Laura DeThorne}.} \bibinfo{year}{2020}\natexlab{}.
\newblock \showarticletitle{Communication profile of a minimally verbal school-age autistic child: A case study}.
\newblock \bibinfo{journal}{\emph{Language, Speech, and Hearing Services in Schools}} \bibinfo{volume}{51}, \bibinfo{number}{3} (\bibinfo{year}{2020}), \bibinfo{pages}{671--686}.
\newblock


\bibitem[Voros et~al\mbox{.}(2014)]%
        {voros2014}
\bibfield{author}{\bibinfo{person}{Gyula Voros}, \bibinfo{person}{P{\'e}ter Rabi}, \bibinfo{person}{Balazs Pinter}, \bibinfo{person}{Andras Sarkany}, \bibinfo{person}{Daniel Sonntag}, {and} \bibinfo{person}{Andras Lorincz}.} \bibinfo{year}{2014}\natexlab{}.
\newblock \showarticletitle{Recommending Missing Symbols of Augmentative and Alternative Communication by Means of Explicit Semantic Analysis}. In \bibinfo{booktitle}{\emph{2014 AAAI Fall Symposium Series}}.
\newblock


\bibitem[Wang et~al\mbox{.}(2023)]%
        {Wang2023}
\bibfield{author}{\bibinfo{person}{Weizhi Wang}, \bibinfo{person}{Li Dong}, \bibinfo{person}{Hao Cheng}, \bibinfo{person}{Xiaodong Liu}, \bibinfo{person}{Xifeng Yan}, \bibinfo{person}{Jianfeng Gao}, {and} \bibinfo{person}{Furu Wei}.} \bibinfo{year}{2023}\natexlab{}.
\newblock \showarticletitle{Augmenting Language Models with Long-Term Memory}. In \bibinfo{booktitle}{\emph{Advances in Neural Information Processing Systems}}, \bibfield{editor}{\bibinfo{person}{A.~Oh}, \bibinfo{person}{T.~Naumann}, \bibinfo{person}{A.~Globerson}, \bibinfo{person}{K.~Saenko}, \bibinfo{person}{M.~Hardt}, {and} \bibinfo{person}{S.~Levine}} (Eds.), Vol.~\bibinfo{volume}{36}. \bibinfo{publisher}{Curran Associates, Inc.}, \bibinfo{pages}{74530--74543}.
\newblock
\urldef\tempurl%
\url{https://proceedings.neurips.cc/paper_files/paper/2023/file/ebd82705f44793b6f9ade5a669d0f0bf-Paper-Conference.pdf}
\showURL{%
\tempurl}


\bibitem[Watzlawick et~al\mbox{.}(2011)]%
        {watzlawick2011}
\bibfield{author}{\bibinfo{person}{Paul Watzlawick}, \bibinfo{person}{Janet~Beavin Bavelas}, {and} \bibinfo{person}{Don~D Jackson}.} \bibinfo{year}{2011}\natexlab{}.
\newblock \bibinfo{booktitle}{\emph{Pragmatics of human communication: A study of interactional patterns, pathologies and paradoxes}}.
\newblock \bibinfo{publisher}{WW Norton \& Company}.
\newblock


\bibitem[Weitzman(2013)]%
        {Elaine2013}
\bibfield{author}{\bibinfo{person}{Elaine Weitzman}.} \bibinfo{year}{2013}\natexlab{}.
\newblock \showarticletitle{More Than Words—The Hanen Program for Parents of Children with Autism Spectrum Disorder: A Teaching Model for Parent-implemented Language Intervention}.
\newblock \bibinfo{journal}{\emph{Perspectives on Language Learning and Education}} \bibinfo{volume}{20}, \bibinfo{number}{3} (\bibinfo{year}{2013}), \bibinfo{pages}{96--111}.
\newblock
\urldef\tempurl%
\url{https://doi.org/10.1044/lle20.3.86}
\showDOI{\tempurl}
\showeprint{https://pubs.asha.org/doi/pdf/10.1044/lle20.3.86}


\bibitem[Whiteley et~al\mbox{.}(2010)]%
        {whiteley2010gender}
\bibfield{author}{\bibinfo{person}{Paul Whiteley}, \bibinfo{person}{Lynda Todd}, \bibinfo{person}{Kevin Carr}, {and} \bibinfo{person}{Paul Shattock}.} \bibinfo{year}{2010}\natexlab{}.
\newblock \showarticletitle{Gender ratios in autism, Asperger syndrome and autism spectrum disorder}.
\newblock \bibinfo{journal}{\emph{Autism Insights}}  \bibinfo{volume}{2} (\bibinfo{year}{2010}), \bibinfo{pages}{17}.
\newblock


\bibitem[Williams et~al\mbox{.}(2006)]%
        {williams2006}
\bibfield{author}{\bibinfo{person}{Diane~L Williams}, \bibinfo{person}{Gerald Goldstein}, {and} \bibinfo{person}{Nancy~J Minshew}.} \bibinfo{year}{2006}\natexlab{}.
\newblock \showarticletitle{The profile of memory function in children with autism.}
\newblock \bibinfo{journal}{\emph{Neuropsychology}} \bibinfo{volume}{20}, \bibinfo{number}{1} (\bibinfo{year}{2006}), \bibinfo{pages}{21}.
\newblock


\bibitem[Williams(1997)]%
        {williams1997}
\bibfield{author}{\bibinfo{person}{Kathleen~T Williams}.} \bibinfo{year}{1997}\natexlab{}.
\newblock \showarticletitle{Expressive vocabulary test second edition (EVT™ 2)}.
\newblock \bibinfo{journal}{\emph{J. Am. Acad. Child Adolesc. Psychiatry}}  \bibinfo{volume}{42} (\bibinfo{year}{1997}), \bibinfo{pages}{864--872}.
\newblock


\bibitem[Wilson et~al\mbox{.}(2018)]%
        {Wilson2018}
\bibfield{author}{\bibinfo{person}{Cara Wilson}, \bibinfo{person}{Margot Brereton}, \bibinfo{person}{Bernd Ploderer}, {and} \bibinfo{person}{Laurianne Sitbon}.} \bibinfo{year}{2018}\natexlab{}.
\newblock \showarticletitle{MyWord: enhancing engagement, interaction and self-expression with minimally-verbal children on the autism spectrum through a personal audio-visual dictionary}. In \bibinfo{booktitle}{\emph{Proceedings of the 17th ACM Conference on Interaction Design and Children}} (Trondheim, Norway) \emph{(\bibinfo{series}{IDC '18})}. \bibinfo{publisher}{Association for Computing Machinery}, \bibinfo{address}{New York, NY, USA}, \bibinfo{pages}{106–118}.
\newblock
\showISBNx{9781450351522}
\urldef\tempurl%
\url{https://doi.org/10.1145/3202185.3202755}
\showDOI{\tempurl}


\bibitem[Wilson et~al\mbox{.}(2019)]%
        {Wilson2019}
\bibfield{author}{\bibinfo{person}{Cara Wilson}, \bibinfo{person}{Margot Brereton}, \bibinfo{person}{Bernd Ploderer}, {and} \bibinfo{person}{Laurianne Sitbon}.} \bibinfo{year}{2019}\natexlab{}.
\newblock \showarticletitle{Co-Design Beyond Words: 'Moments of Interaction' with Minimally-Verbal Children on the Autism Spectrum}. In \bibinfo{booktitle}{\emph{Proceedings of the 2019 CHI Conference on Human Factors in Computing Systems}} (Glasgow, Scotland Uk) \emph{(\bibinfo{series}{CHI '19})}. \bibinfo{publisher}{Association for Computing Machinery}, \bibinfo{address}{New York, NY, USA}, \bibinfo{pages}{1–15}.
\newblock
\showISBNx{9781450359702}
\urldef\tempurl%
\url{https://doi.org/10.1145/3290605.3300251}
\showDOI{\tempurl}


\bibitem[Wilson et~al\mbox{.}(2020)]%
        {Wilson2020}
\bibfield{author}{\bibinfo{person}{Cara Wilson}, \bibinfo{person}{Laurianne Sitbon}, \bibinfo{person}{Bernd Ploderer}, \bibinfo{person}{Jeremy Opie}, {and} \bibinfo{person}{Margot Brereton}.} \bibinfo{year}{2020}\natexlab{}.
\newblock \showarticletitle{Self-Expression by Design: Co-Designing the ExpressiBall with Minimally-Verbal Children on the Autism Spectrum}. In \bibinfo{booktitle}{\emph{Proceedings of the 2020 CHI Conference on Human Factors in Computing Systems}} (Honolulu, HI, USA) \emph{(\bibinfo{series}{CHI '20})}. \bibinfo{publisher}{Association for Computing Machinery}, \bibinfo{address}{New York, NY, USA}, \bibinfo{pages}{1–13}.
\newblock
\showISBNx{9781450367080}
\urldef\tempurl%
\url{https://doi.org/10.1145/3313831.3376171}
\showDOI{\tempurl}


\bibitem[Winter-Messiers(2007)]%
        {Mary2007}
\bibfield{author}{\bibinfo{person}{Mary~Ann Winter-Messiers}.} \bibinfo{year}{2007}\natexlab{}.
\newblock \showarticletitle{From Tarantulas to Toilet Brushes: Understanding the Special Interest Areas of Children and Youth With Asperger Syndrome}.
\newblock \bibinfo{journal}{\emph{Remedial and Special Education}} \bibinfo{volume}{28}, \bibinfo{number}{3} (\bibinfo{year}{2007}), \bibinfo{pages}{140--152}.
\newblock
\urldef\tempurl%
\url{https://doi.org/10.1177/07419325070280030301}
\showDOI{\tempurl}
\showeprint{https://doi.org/10.1177/07419325070280030301}


\bibitem[Wisenburn and Higginbotham(2008)]%
        {Bruce2008}
\bibfield{author}{\bibinfo{person}{Bruce Wisenburn} {and} \bibinfo{person}{D.~Jeffery Higginbotham}.} \bibinfo{year}{2008}\natexlab{}.
\newblock \showarticletitle{An AAC Application Using Speaking Partner Speech Recognition to Automatically Produce Contextually Relevant Utterances: Objective Results}.
\newblock \bibinfo{journal}{\emph{Augmentative and Alternative Communication}} \bibinfo{volume}{24}, \bibinfo{number}{2} (\bibinfo{year}{2008}), \bibinfo{pages}{100--109}.
\newblock
\urldef\tempurl%
\url{https://doi.org/10.1080/07434610701740448}
\showDOI{\tempurl}
\showeprint{https://doi.org/10.1080/07434610701740448}
\newblock
\shownote{PMID: 18465364}.


\bibitem[Wisenburn and Higginbotham(2009)]%
        {Bruce2009}
\bibfield{author}{\bibinfo{person}{Bruce Wisenburn} {and} \bibinfo{person}{D.~Jeffery Higginbotham}.} \bibinfo{year}{2009}\natexlab{}.
\newblock \showarticletitle{Participant Evaluations of Rate and Communication Efficacy of an AAC Application Using Natural Language Processing}.
\newblock \bibinfo{journal}{\emph{Augmentative and Alternative Communication}} \bibinfo{volume}{25}, \bibinfo{number}{2} (\bibinfo{year}{2009}), \bibinfo{pages}{78--89}.
\newblock
\urldef\tempurl%
\url{https://doi.org/10.1080/07434610902739876}
\showDOI{\tempurl}
\showeprint{https://doi.org/10.1080/07434610902739876}
\newblock
\shownote{PMID: 19444679}.


\bibitem[Wobbrock(2017)]%
        {Wobbrock2017}
\bibfield{author}{\bibinfo{person}{Jacob~O. Wobbrock}.} \bibinfo{year}{2017}\natexlab{}.
\newblock \showarticletitle{SIGCHI Social Impact Award Talk -- Ability-Based Design: Elevating Ability over Disability in Accessible Computing}. In \bibinfo{booktitle}{\emph{Proceedings of the 2017 CHI Conference Extended Abstracts on Human Factors in Computing Systems}} (Denver, Colorado, USA) \emph{(\bibinfo{series}{CHI EA '17})}. \bibinfo{publisher}{Association for Computing Machinery}, \bibinfo{address}{New York, NY, USA}, \bibinfo{pages}{5–7}.
\newblock
\showISBNx{9781450346566}
\urldef\tempurl%
\url{https://doi.org/10.1145/3027063.3058588}
\showDOI{\tempurl}


\bibitem[Wobbrock et~al\mbox{.}(2011)]%
        {Wobbrock2011}
\bibfield{author}{\bibinfo{person}{Jacob~O. Wobbrock}, \bibinfo{person}{Shaun~K. Kane}, \bibinfo{person}{Krzysztof~Z. Gajos}, \bibinfo{person}{Susumu Harada}, {and} \bibinfo{person}{Jon Froehlich}.} \bibinfo{year}{2011}\natexlab{}.
\newblock \showarticletitle{Ability-Based Design: Concept, Principles and Examples}.
\newblock \bibinfo{journal}{\emph{ACM Trans. Access. Comput.}} \bibinfo{volume}{3}, \bibinfo{number}{3}, Article \bibinfo{articleno}{9} (\bibinfo{date}{apr} \bibinfo{year}{2011}), \bibinfo{numpages}{27}~pages.
\newblock
\showISSN{1936-7228}
\urldef\tempurl%
\url{https://doi.org/10.1145/1952383.1952384}
\showDOI{\tempurl}


\bibitem[Zhang et~al\mbox{.}(2023)]%
        {zhang2023prompting}
\bibfield{author}{\bibinfo{person}{Biao Zhang}, \bibinfo{person}{Barry Haddow}, {and} \bibinfo{person}{Alexandra Birch}.} \bibinfo{year}{2023}\natexlab{}.
\newblock \showarticletitle{Prompting Large Language Model for Machine Translation: A Case Study}. In \bibinfo{booktitle}{\emph{Proceedings of the 40th International Conference on Machine Learning}} \emph{(\bibinfo{series}{Proceedings of Machine Learning Research}, Vol.~\bibinfo{volume}{202})}, \bibfield{editor}{\bibinfo{person}{Andreas Krause}, \bibinfo{person}{Emma Brunskill}, \bibinfo{person}{Kyunghyun Cho}, \bibinfo{person}{Barbara Engelhardt}, \bibinfo{person}{Sivan Sabato}, {and} \bibinfo{person}{Jonathan Scarlett}} (Eds.). \bibinfo{publisher}{PMLR}, \bibinfo{pages}{41092--41110}.
\newblock
\urldef\tempurl%
\url{https://proceedings.mlr.press/v202/zhang23m.html}
\showURL{%
\tempurl}


\bibitem[Zolyomi and Snyder(2023)]%
        {Zolyomi2023}
\bibfield{author}{\bibinfo{person}{Annuska Zolyomi} {and} \bibinfo{person}{Jaime Snyder}.} \bibinfo{year}{2023}\natexlab{}.
\newblock \showarticletitle{Designing for Common Ground: Visually Representing Conversation Dynamics of Neurodiverse Dyads}.
\newblock \bibinfo{journal}{\emph{Proc. ACM Hum.-Comput. Interact.}} \bibinfo{volume}{7}, \bibinfo{number}{CSCW2}, Article \bibinfo{articleno}{266} (\bibinfo{date}{oct} \bibinfo{year}{2023}), \bibinfo{numpages}{33}~pages.
\newblock
\urldef\tempurl%
\url{https://doi.org/10.1145/3610057}
\showDOI{\tempurl}


\end{thebibliography}


\end{document}